\documentclass[12pt]{article}
\usepackage{amsmath}
\usepackage{amsfonts}
\usepackage{latexsym}
\usepackage{epsfig}
\usepackage{verbatim}
\newtheorem{theorem}{Theorem}

\newtheorem{lemma}{Lemma}

\begin{document}

\title{Quasiperiodic spin--orbit motion and spin tunes \\ 
in storage rings \footnote{An extended version of 
 Phys. Rev. ST Accel. Beams
{\bf 7}(12), 124002 (2004). }}

\author{ D.~P.~Barber$^a$, J.~A. ~Ellison$^b$ and K.~Heinemann$^a$ \\
$^a$ \small{Deutsches~Elektronen--Synchrotron, DESY, ~22603 ~Hamburg,~Germany} 
\\
$^b$ \small{Department of Mathematics and Statistics, The University of New 
Mexico,} \\  
\small{Albuquerque, New Mexico 87131, U.S.A. }}

\date{ }

\maketitle

\renewcommand{\baselinestretch}{1.0}

\normalsize

\begin{abstract}
  
We present an in--depth analysis of the concept of spin precession
frequency for integrable orbital motion in storage rings.  Spin
motion on the periodic closed orbit of a storage ring can be
analyzed in terms of the Floquet theorem for equations of motion
with periodic parameters and a spin precession frequency emerges in
a Floquet exponent as an additional frequency of the system.  To
define a spin precession frequency on nonperiodic synchro--betatron
orbits we exploit the important concept of quasiperiodicity.  This
allows a generalization of the Floquet theorem so that a spin
precession frequency can be defined in this case too. This frequency
appears in a Floquet--like exponent as an additional frequency in
the system in analogy with the case of motion on the closed orbit.
These circumstances lead naturally to the definition of the uniform
precession rate and a definition of spin tune.  A spin tune is a
uniform precession rate obtained when certain conditions are
fulfilled.  Having defined spin tune we define spin--orbit resonance
on synchro--betatron orbits and examine its consequences.  We give
conditions for the existence of uniform precession rates and spin
tunes (e.g.\ where small divisors are controlled by applying a
Diophantine condition) and illustrate the various aspects of our
description with several examples. The formalism also suggests the
use of spectral analysis to ``measure'' spin tune during computer
simulations of spin motion on synchro--betatron orbits.

\end{abstract}

\newpage

\tableofcontents

\section{Introduction}

This paper provides a rigorous discussion of the concept of spin precession 
frequency 
on synchro--betatron orbits in storage rings.
To set the scene we begin by introducing some key
physical ideas via the equations of orbit and spin motion and the notion of 
spin--orbit equilibrium.

The spin expectation value $\bf S$ (``the spin'') in the rest frame of, 
for example, a proton,  an electron or a muon moving 
in electric and magnetic fields under the influence of the Lorentz force 
precesses  according to the
Thomas--Bargmann--Michel--Telegdi (T--BMT) equation \cite{jackson}
\begin{eqnarray}
&&  d {\bf {S}}/{d t}= {\bf {\tilde \Omega}}({\bf E},{\bf B},{\bf v})
 \times \bf {S}   \; , 
\label{eq:1.1}
\end{eqnarray}
where the precession vector $\bf {\tilde \Omega}$
depends on ${\bf E},~{\bf B}$ and ${\bf v}$ 
which are respectively the electric and
magnetic fields, and the velocity.
Particle motion with respect to the synchronous closed, i.e.\ periodic, orbit 
is described in 
terms  of three pairs of canonical variables which we combine into a vector 
$u$ with six
components. For example, two of the pairs can describe transverse motion and 
one pair
can describe longitudinal (synchrotron) motion within a bunch.  
Since we are dealing with storage rings we take the orbital motion to be 
bounded. In this paper we ignore radiation, interparticle interactions and 
interactions with 
the vacuum system. 
In (\ref{eq:1.1}) the independent variable is the time $t$.
However, since the electric and magnetic guide fields in particle accelerators
and 
storage rings are fixed in space it is more 
convenient to adopt the standard practice of replacing $t$ with 
the angular position around the ring, 
the azimuth $\theta = 2 \pi s /L$, where $s$ is the distance around the ring 
and $L$ is the circumference.  

Since  ${\bf E},{\bf B}$ and ${\bf v}$ depend on $u$ and $\theta$
we can now rewrite (\ref{eq:1.1}) in the form
\begin{eqnarray}
d {\bf {S}}/{d \theta}= {\bf {\Omega}}(\theta,{u(\theta)}) \times \bf {S}   
\label{eq:1.2}
\end{eqnarray}
where ${\bf {\Omega}}$  is the precession vector obtained from 
${\bf {\tilde \Omega}}$
by rescaling with $d t/d \theta$ and transforming to  machine coordinates 
\cite{bhr1}.

If the beam is in equilibrium,  i.e.\ if the 
the phase space density $\rho (\theta, u)$ is $2\pi$--periodic in $\theta$,
we then write it as $\rho_{\rm eq}$ 
so that  $\rho_{\rm eq}(\theta, u)= \rho_{\rm eq}(\theta + 2\pi, u)$.
We normalize it to unity: ${\int} d u \rho_{\rm eq} =1$.
The necessary condition for this kind of equilibrium,  namely that at 
a fixed $u$ the fields  are $2\pi$--periodic in 
$\theta$,  is automatically fulfilled in a storage ring.
But of course the boundedness of the motion is also required.
Conditions necessary for beam equilibrium and a way of calculating $\rho_{\rm eq}$
using ergodic theory and the concept of ``stroboscopic averaging''
are described in detail in \cite{eh2004}.

The statistical properties of the spins are  encoded in the 
quantum mechanical spin density matrix. But for spin 1/2 particles this
can be completely parametrized by the polarization  vector 
\cite{gottfried}. For particle beams we need the 
local polarization  ${\bf P}_{\rm loc}(\theta, u)$ at each point in phase space
$u$. ${\bf P}_{\rm loc}(\theta, u)$ is the average of the normalized 
spin vectors, ${\bf S}/|{\bf S}|$,  at $u$, where $|\cdot|$ denotes the
Euclidean norm.
We define the polarization of the whole beam, the ``beam polarization'', 
at a given azimuth as ${\int} d u \rho_{\rm eq} {\bf  P}_{\rm loc}$. 

Since the T--BMT
equation (\ref{eq:1.2}) is linear in $\bf S$ and since
the particles at $(\theta,u)$
all see the same ${\bf{\Omega}} (\theta, u)$,~
${\bf P}_{\rm loc}(\theta, u(\theta))$ also obeys the T--BMT
equation \cite{dbkh98}.
~Furthermore, the length  of  ~${\bf P}_{\rm loc}(\theta, u)$ is
constant along a phase space trajectory. For a storage ring at fixed energy,
$\bf {\Omega}$ is $2\pi$--periodic in $\theta$
at a fixed position in phase space $u$ so that
${\bf {\Omega}}(\theta, {u}) = {\bf {\Omega}}(\theta + 2\pi, {u})$.
This opens up the possibility of a spin distribution that is the
same from turn to turn, i.e.\ in equilibrium. Then 
${\bf P}_{\rm loc}(\theta, u(\theta))$ not only obeys the 
T--BMT equation, but 
${\bf P}_{\rm loc}(\theta, u)$ is
$2\pi$--periodic in $\theta$ for fixed $u$ and we then write it as 
${\bf P}_{\rm eq}$ so that
${\bf P}_{\rm eq}(\theta, u) = {\bf P}_{\rm eq}(\theta + 2\pi, u)$.
We denote the unit vector along ${\bf P}_{\rm eq}(\theta, u)$
by ${\bf n}(\theta, u)$.
This also obeys the T--BMT
equation along orbits 
and is $2\pi$--periodic in $\theta$:
${\bf n}(\theta, {u}) = {\bf n}(\theta + 2\pi, {u})$. 

The method described in \cite{eh2004} for constructing $\rho_{\rm eq}$ can
be extended as in \cite{hh96} for constructing ${\bf n}$  and from the 
treatments in 
\cite{eh2004} it is clear that the existence of $\rho_{\rm eq}$ and ${\bf n}$ do 
not
require that the orbital motion be integrable.  But, of course, the conclusions
of \cite{eh2004} are still valid if the motion {\em is} integrable. Moreover, 
particle motion in storage rings is usually close enough to integrability 
to allow the motion to be characterized in terms of well defined betatron and 
synchrotron frequencies. This, in turn,  allows predictions to be made
about beam stability via the concept of orbital resonance. Thus, in the 
remainder
of this paper we will assume that the orbital motion is integrable. 
Then, as we shall see, the stability of spin motion can also be discussed in 
terms
of resonance, namely ``spin--orbit resonance''. 
Of course, integrable orbital motion and spin--orbit equilibrium are
idealizations. Nevertheless, these idealizations often provide useful starting
points for 
calculations. 

For integrable particle motion the position 
of a particle in phase space is represented by three pairs of action--angle
variables $(J_i, \phi_i, ~i=1,2,3)$ 
and is determined by a Hamiltonian $H(J)$. 
Thus the orbital phase space is partitioned into disjoint tori, each 
of which is characterized by a unique set of $J${\small s}. 
We now define 
$u: = ({\phi}_1,~{\phi}_2,~{\phi}_3,~J_1,~J_2,~J_3) \equiv (\phi,J)$.
The actions are constants of the motion and for fixed $J$ the constant rate 
of advance of each $\phi_i$, 
~$\omega_i (J): = \partial H/ \partial J_i = d \phi_i/d \theta$,  is called 
an {\em orbital tune}. These frequencies are the number of oscillations per 
turn around the ring. In beam physics such  frequencies are often referred 
to as tunes and we 
have  adopted that usage. We will only consider storage rings running
at fixed nominal energy. 

For integrable motion, the $2\pi$--periodicity in $\theta$  of 
${\bf {\Omega}}$, ${\bf P}_{\rm eq}$ and $\bf n$ is accompanied by
$2\pi$--periodicity in $\phi_1, \phi_2$ and $\phi_3$. 
So as well as being a solution to the T--BMT
equation along orbits $u(\theta)$, $\bf n$ satisfies nontrivial periodicity 
conditions. 
In our later discussions on quasiperiodicity we will require that it also
depend sufficiently regularly on the azimuth and the orbital angles, 
i.e.\ $\bf n$ must be ``smooth'' in the sense defined in the main text.
This corresponds to the expectation that ${\bf P}_{\rm eq}$ also be smooth.
The equilibrium density $\rho_{\rm eq}$  
is also $2\pi$--periodic in $\theta$ and off orbital resonance it just depends on $J$.

Since for every $J$ the field ${\bf  n}(\theta, \phi, J)$ is 
invariant from turn to turn, it is now often
called the {\em invariant spin field} (ISF).  The ISF is a central
object in the theory of polarization in storage rings
\cite{mont98,epac98,spin98,mane87}.  For example, for an ISF sufficiently regular in $J$,
off orbital resonance and away from the spin--orbit resonances to be defined below, an upper limit to the equilibrium beam
polarization at a particular $\theta$ is
${\int} d J \rho_{\rm eq}(J) |\int d \phi ~{\bf n}(\theta, \phi, J)|$ and it is reached 
only when 
the $\int d \phi ~{\bf n}$ are parallel.
This is easy to see by noting
that if $|{\bf P}_{\rm eq}(\theta, \phi, J)|$ were to vary over a torus, the beam
polarization would vary from turn to turn. So equilibrium implies that
$|{\bf P}_{\rm eq}(\theta, \phi, J)|$ is constant over a torus.  
The maximum equilibrium polarization on each torus
is reached when $|{\bf P}_{\rm eq}(\theta, \phi, J)| = 1$.
Note that a zero value for ${\int} d J \rho_{\rm eq}(J) |\int d \phi ~{\bf n}|$ at some 
$\theta$ does not mean
that the beam is depolarized. It could well be that the beam is fully polarized at
each point in phase space but that the geometry of ${\bf  n}$ causes the integral
to vanish. Then, if a change of parameters were to change the geometry of 
${\bf  n}$ so that the integral were to become nonzero, and if the change were carried out 
adiabatically, the beam polarization would reappear.
Furthermore, the fact that the integral vanishes at one position $\theta$ does not
mean that it vanishes at other positions.
We prefer to reserve the term ``depolarization'' for a {\em definitely irreversible} 
loss of  polarization such as occurs in the presence of noise or for an 
{\em effectively  irreversible} loss of the kind that can occur when spin--orbit 
resonances are  crossed \cite{hv2004}.  See Sec. X too.
Although we have introduced ${\bf n}$ via the notion of equilibrium,
the integral ${\int} d J \rho_{\rm eq}(J) |\int d \phi ~{\bf n}|$ also contains useful
information when the spin distribution is not in equilibrium: 
${\int} d J \rho_{\rm eq}(J) |\int d \phi ~{\bf n}|$ gives
an upper limit for the time averaged polarization 
away from spin--orbit resonances. The maximum is 
reached on each torus when the polarization is in equilibrium and with $|{\bf }P_{\rm eq}(\theta, \phi, J)| = 1$. See \cite[Section 2.2.8]{gh2000} and \cite[Section 4.4]{mv2000}.
An example of the origin and behavior of
nonequilibrium beam polarization is given in Figure 9 in
\cite{hh96} where large oscillations are evident.  However,
polarimeters and particle detectors cannot collect data quickly enough
to make such oscillations observable. Instead, only the time averaged
polarization can be observed or exploited. But as we have just seen, we
can still estimate its maximum value. That depends only on the geometry 
of ${\bf n}$ and it is reached for each torus when the spread of ${\bf n}$ is
minimized.
The ISF also provides a perfect tool for estimating the 
long term effects on the beam polarization of small perturbations such as 
radiation \cite{mont98} or electric and magnetic fields which cause 
nonintegrable orbital motion. In particular, one 
begins with a spin--orbit system which is invariant from turn to turn,
i.e.\ with an equilibrium orbital distribution and with  
spins set parallel to the ${\bf n}(\theta, u)$. Then, since the system is initially
in equilibrium, the effects of the perturbations cannot be masked by the natural, potentially
large,  variations of the beam polarization of the kind depicted in Figure 9 in \cite{hh96}. 
An ability to construct ${\bf n}(\theta, u)$ for integrable orbital
motion and understand its behavior is then indispensable.

~For our integrable orbital motion 
of electrons and protons and up to energies of a few  GeV, 
an approximate $\bf n$  can be  calculated in a 
first order perturbation theory by an extension of the code SLIM \cite{chao81,
br99}, and in higher order perturbation theory by the codes, 
SMILE \cite{mane87}, Forget--me--Not \cite{bg98} and SpinLie \cite{ey95}.
However, for the high magnetic fields characteristic of proton rings running 
at energies of hundreds of GeV, perturbative methods are inadequate. 
Then the method of stroboscopic averaging as in the code SPRINT \cite{hh96}
should be used. This is a numerical, nonperturbative algorithm and yields
high accuracy for real rings even  when all modes of orbit oscillation are included 
simultaneously.
One can also use Fourier methods as in the codes SODOM2 \cite{ky99} or MILES \cite{mane20022}.
SODOM2 has been very useful for orbital motion restricted to one plane \cite{mv2000}. 
MILES gives explicit formulae which are applicable to some simple models.
So far, the only practical general way to calculate the invariant spin field is to use 
stroboscopic averaging. 

As for any dynamical system we hope to understand more about spin
motion by studying its spectrum of frequencies. Various quantities,
which seem at first sight to be related to spin frequencies, can be
found in the literature and we will mention some in Section 10.  But a
true component of a spectrum quantifies long term behavior. Thus any
definition of spin precession frequency should reflect that
stipulation. The choice can be further narrowed by requiring that the
spectrum give useful clues about the behavior of {\em sets} of spins,
and in particular about the beam polarization. After all, the experimenters using  
the beams in storage rings are just interested in the beam polarization, not individual spins.

Experience has shown that the best choice for characterizing 
spin motion in storage rings is the traditional one \cite{dk72,dk73,ky86},
namely the so--called {\em amplitude dependent spin tune} 
(briefly ``spin tune''), which we usually denote by $\omega_s$.
Assuming ${\bf n}$ exists, the spin tune measures the number of spin 
precessions
around ${\bf n}(\theta, u)$, per turn around the ring, for a particle on the
orbit $u(\theta)$ 
and  it provides a way to quantify the correlation 
between the spin motion and the orbital motion which ``drives'' it, and 
thereby forecast
a qualitative aspect of spin motion, namely the degree of regularity of the
spin motion. In particular,
the spin motion can in general become very
erratic when a spin tune is near a low order resonance condition
\begin{eqnarray}
\omega_s = m_{0} + m_{1} ~\omega_{1} + m_{2}~ \omega_{2} + m_{3}~ \omega_{3}
 =m\cdot(1,\omega),
\label{eq:1.5}
\end{eqnarray}
where $m$ is a vector  of integers and
the quantity 
$|m_{1}| + |m_{2}| +  |m_{3}|$ is usually called the 
{\em order} of the resonance. 
Correspondingly, close to spin--orbit resonance 
${\bf n}(\theta, u)$ can become a very sensitive function of $u$.
This sensitivity has immediate consequences for work with polarized beams. 
For example, 
the maximum attainable equilibrium beam polarization of a stored high 
energy proton beam can be unacceptably low or the rate
of depolarization, due to synchrotron radiation, of a stored electron beam can
be unacceptably high \cite{mont98}. Note, however, that the 
$|{\int} d \phi~ {\bf  n}|/{(2\pi)^3}$
on a torus can sometimes be small away from  spin--orbit resonance and that proximity to a  
spin--orbit resonance, especially one of high order, does not automatically imply that the
$|{\int} d \phi~ {\bf  n}|/{(2\pi)^3}$ on a torus is low.  The resonance might be very weak. 
Another feature of our definition of spin frequency is that, as we
shall see, it is this quantity whose spectrum one obtains in a
straightforward spectral analysis of spin motion during spin--orbit
tracking simulations. In other words: in an ideal world with 
technology which could select particles on a torus at a fixed $J$, 
it could be {\em measured}.

Right at the resonance condition (\ref{eq:1.5}), ${\bf n}$ is 
in general nonunique. However, as we shall see, our spin--orbit systems
exhibit a tendency to avoid exact spin--orbit resonance.
Since ${\bf \Omega}$ in general depends on $J$ and the particle energy on the 
closed orbit,  $\omega_s$ usually varies with $J$
(hence ``amplitude dependent spin tune''),
and the particle energy on the closed orbit.  
We will call the latter the ``beam energy''.
We emphasize that ${\bf n}$ is a field over the {\em six} dimensional
phase space so that synchrotron motion is built in from the start.
Thus although $\omega_s$ varies with the beam energy and $J_1, J_2$
and $J_3$, it does {\em not} change during a period of synchrotron
motion.  If ${\bf n}$ were defined on four dimensional transverse
phase space and the energy oscillations due to synchrotron motion were
added as an afterthought, it would not be useful for describing {\em
  equilibrium} polarization. Instead, we would have to characterize the
beam polarization using time averages.  We return to this theme in
Section 10. On the closed orbit, i.e.\ for $J_1 = J_2 = J_3 = 0$, an ${\bf n}$
exists
which is independent of $\phi$. We denote it by
${\bf n_0}$ \cite{br99}.

The calculation of spin tune on the closed orbit presents no problem:
it can, as we shall see, be extracted from an eigenvalue of the
1--turn spin map.  But the definition of spin tune for $J \ne 0$,
i.e.\ on synchro--betatron orbits is much more subtle.  Moreover, it
requires precision.  Notions of spin frequency for
synchro--betatron orbits appearing in the literature are often not
precisely presented
and  some appear to possess no capacity for
predicting the qualitative aspects of spin motion in storage rings.

This brings us to the purpose of this paper. This is to provide a
rigorous discussion of the concept of spin precession frequency on
synchro--betatron orbits and thereby consolidate a framework for
systematizing and classifying spin motion in storage rings.
For this we make a careful mathematical study of 
the consequences for spin motion of the periodicities of 
$\bf \Omega$ in $\theta, \phi_1, \phi_2$ and $\phi_3$, using precise 
definitions
and carefully formulated theorems and we make use of
the ISF and other
concepts which we distill from the literature and ``folklore'' on spin 
dynamics in  storage rings 
\cite{hh96,mont98,epac98,spin98,gh2000,mv2000,bg98,ky99,dk72,dk73,ky86,
spin2000,bhr92,hvb99}.
~For example,  we will show that under the appropriate conditions, the 
existence of the ISF with the above mentioned periodicities implies
that the $\theta$ dependence of the general solutions of (\ref{eq:1.2}) will 
contain five frequencies.
~Four of them are the orbital tunes $\omega_1, \omega_2,\omega_3$ and
the circulation tune $\omega_c = 1$, i.e.\ the frequency associated
with the $2\pi$--periodicity in $\theta$. A fifth tune emerges
which, under circumstances to be described, is a spin tune 
$\omega_s$. The general solutions will then be found to be
quasiperiodic with the tunes $1, \omega_1,\omega_2,\omega_3, \omega_s$. 
Moreover  the results obtained here can be viewed as a generalization of Floquet 
theory.
Given the confusion surrounding definitions of spin precession frequencies,
the treatment of the kind that we provide here seems to be very necessary.
Our assumptions about ${\bf \Omega}$ are weak enough to cover the 
situations of most interest for storage rings, namely typical integrable
synchro--betatron motion. 
Several of our theorems assume the existence of $\bf n$ 
but although we have ways to find approximate $\bf n$,
the determination
of complete conditions for its existence is an outstanding mathematical issue. 
This question can, for example, be investigated using ergodic theory 
\cite{si89,CFS} and the method of stroboscopic averaging.
See \cite{eh2004}.
Moreover, simulations indicate that approximate ISF{\small s} {\em do} exist.
This means that one obtains objects which, at least approximately, 
behave like an ISF. Moreover in some instances approximations even lead 
to an $\bf \Omega$ for which an exact ISF can be found, 
e.g.\ as in the single resonance model - see Section 7.

Although we have introduced the vector $\bf n$ by studying spin--orbit equilibrium,
it was first discussed by Derbenev and Kondratenko
as a vehicle for constructing joint action--angle variables for spin
and orbital motion from their semiclassical spin--orbit Hamiltonian
\cite{dk73,ky86,bhr1,mont98,bg98}.  This Hamiltonian is
derived from the Dirac Hamiltonian by a Foldy--Wouthuysen
transformation taken to first order in $\hbar$ \cite{mont98}.  In that
picture the spin tune emerges as the rate of advance of a spin phase
\cite{dk73,ky86,bhr1}.
The terms at first order in $\hbar$ in the Derbenev--Kondratenko Hamiltonian
are those containing spin and these terms imply a force of the
Stern--Gerlach (S--G) type \cite{gottfried}.  
The S--G forces on trajectories appear at first order in
$\hbar$ and  a ``back reaction'' on the spin 
of the S--G perturbation to the orbit would involve an 
addition to the spin precession rate of order $\hbar$. However,
in this paper the effect of S--G forces on spin and orbit motion
is neglected and we just operate with the Lorentz force and the T--BMT
equation and for the Lorentz force and $\bf \Omega$ just include, as is usual, 
the terms 
of zeroth order in $\hbar$.  There are several  reasons for this approach.
First, it is far from clear what form the S--G forces should take. In
fact there is considerable ambiguity in the choice of the S--G forces.
This is covered in detail in \cite{h96}. See \cite{khrip2000} too and
the bibliography in \cite{h96}.
The second ground has to do with the size of the S--G forces.  Since
the S--G forces are of first order in $\hbar$ they are extremely small
compared to the Lorentz forces which are of zeroth order in $\hbar$
\footnote{For example, at the r.m.s. radius of the  920 GeV/c proton beam in a 
quadrupole magnet in the HERA proton ring \cite{gh2000,mv2000},  
a rough estimate for the transverse S--G force from the Derbenev--Kondratenko Hamiltonian 
gives a S--G force which is about $10^{-12}$ of the Lorentz force on a proton. At a fixed 
radius this ratio is essentially independent of the beam energy.
The S--G energy at that radius is of the order of $10^{-21}$ of the kinetic
energy. The S--G energy in a HERA dipole magnet is of the order of  $10^{-19}$ of the 
kinetic energy.}.
They are also small compared to typical spurious perturbations to
trajectories like noise and collective effects.  So S--G forces would not 
cause changes of
practical significance to the results that we present. In particular, in
practical situations in a storage ring there would be no significant
change in the phenomenology of spin--orbit resonances even if the S--G
forces were to cause tiny changes in the orbital tunes.
The third ground is that it is far from clear that it makes sense to
treat an essentially quantum mechanical system with a classical ``over
interpretation'' of the influence of the S--G force on the spin.
An example of an effect which is not taken into account by a naive application
of classical S--G forces, is given in \cite{derb90}. It is implied there that long term 
shifts of an orbit due to S--G forces will be nullified when the spin undergoes a 
quantum flip and  the S--G force then acts in the reverse direction. See \cite[p.137]{bhr1}
for a classical perspective on this. 
In summary, we believe that a too literal interpretation of the S--G--like forces in the 
semiclassical spin--orbit Hamiltonian could lead to manipulations and conclusions 
of little  relevance and utility for illuminating the core phenomenology of spin motion in 
typical storage rings \cite{mane20032,mane20031}.
We believe that the first priority is to begin with just the Lorentz force and the T--BMT
equation. Then, as mentioned earlier, once the equilibrium state of the system has been 
defined, other influences such as  nonlinear fields, noise, collective effects,
synchrotron radiation and the very small S--G--like effects
can be included as perturbations. 

The paper is structured as follows.
In Section 2 we begin by discussing some important consequences of 
(\ref{eq:1.2}). Here we introduce the central concept of a 
{\em uniform precession frame} (UPF)  and the associated 
{\em uniform precession rate} (UPR).
The UPF provides a coordinate system for spin.
Then in Section 3 we give a detailed discussion of spin motion on the closed orbit 
where $\bf \Omega$ is independent of $\phi$  so that the 
Floquet 
theorem applies. 
Sections 2 and 3 contain standard results but we 
present them in
forms which motivate their extension in later sections.
Section 4 contains the definition of a quasiperiodic function and collects some
properties useful for the discussion following.  
In particular it defines a Diophantine condition needed for handling a 
problem with small divisors. The key ideas are formalized in Lemmas 4.3, 4.7 
and 4.8. 
Section 5 uses the concept of a UPF, quasiperiodic with
orbital frequencies, to define the proper UPR, the spin tune and  
spin--orbit resonance.  The main theorem in Section 5 is Theorem 5.3 which
allows us to define equivalence classes of spin tunes. 
The presentation in Sections 2, 3, 4 and 5 is deliberately rather general 
and abstract.  
Then in Section 6 we 
introduce a {\em field} 
called the {\em invariant frame field} (IFF) which is used to construct 
UPF{\small s}. 
There we consider the angular phase space as a whole
to prove theorems about the concepts introduced in 
Section 5. We also connect the abstract ideas introduced earlier to a
familiar physical idea, namely that if the orbital tune were off
orbital resonance  ($m\cdot(1,\omega)=0$ only with the vector of integers $m=0$),
the existence of 
a nonunique ISF would
imply that the system were on spin--orbit resonance.
The main theorems in Section 6 are Theorems 6.3 - 5. 
Theorem 6.3 is used in the proof of Theorem 6.5 and it is generally our main 
tool for showing
that a torus is ``well--tuned''. The proof of Theorem 6.3 relies on Theorem 
5.3. 
Some examples of the formalism for model $\bf \Omega$\,{\small s} are 
presented in Sections 6, 7 and 8.  
Note that, except for some examples, we allow the number of action--angle
pairs, $d$, to be arbitrary (but $\geq 1$) although for spin motion in storage
rings, 
the case $d = 3$ is the most important.
To  aid the reader we mark the key equations
with a $\star\!\star\!\star$ on the left.

As a byproduct of the quasiperiodic structure of the solutions we suggest
using spectral analysis as a way of ``measuring'' the spin tune during 
spin--orbit tracking simulations and thereby complementing other methods 
already  
in use  \cite{epac98,spin98,ky99,gh2000,mv2000,spin2000,spin2_2000}.
Spectral analysis may also lead to a practical method for deciding whether 
an invariant spin field exists. These ideas are presented in Section 9 and 
formalized
in Theorems 9.1 and 9.2.

The paper is summarized in Section 10 where our concepts are also related to 
simulations and used to discuss some popular notions.

~For the rest of the paper, apart from Section 10, we will now adopt a more 
efficient
notation whereby we use  the symbols  $J = (J_1,...,J_d)$,
$\omega = (\omega_1,...,\omega_d)$  and $\phi = (\phi_1,...,\phi_d)$, 
($J, \omega, \phi \in {\mathbb R}^d$) to mean 
respectively  the list of orbital actions, orbital tunes and orbital angles.
~From now on we will also adopt the frame dependent abbreviations
$S = (S_1,S_2, S_3)$,
$\Omega = (\Omega_1,\Omega_2,\Omega_3)$ and $n = (n_1,n_2, n_3)$.
Generally, if $\phi$ appears as an independent variable in a 
function, the function will be $2\pi$--periodic  in 
$\phi_1,...,\phi_d$. In that case we say for brevity that the function
is $2\pi$--periodic in $\phi$.

In terms of the new notation,
the T--BMT
equation and the equations of orbital motion are
\begin{eqnarray}
\dot{{S}} &=& {\cal A}(\theta, \phi, J){S} \,,
\label{eq:1.9} \\
\dot{J} &=& 0,\quad \dot{\phi} = \omega(J) ,
\label{eq:1.10}
\end{eqnarray}
where $\cal A$ is a real 
skew--symmetric $3 \times 3$ matrix with nonzero elements
${\cal A}_{12} = -{\cal A}_{21} =   - \Omega_3, 
 ~{\cal A}_{13} = -{\cal A}_{31} =     \Omega_2$ 
and 
${\cal A}_{23} = -{\cal A}_{32} =    -\Omega_1$. 
The dot over a symbol denotes differentiation w.r.t. $\theta$. 
Because the $J$ dependence is only parametric,
we will often suppress the symbol $J$ in $\omega$ and $\cal A$, 
e.g.\ as in ${\cal A}(\theta, \phi)$.
Clearly,  ${\cal A}$ is $2\pi$--periodic in  
$\theta$ and in $\phi$. For brevity 
we just say that functions with such periodicity are $2\pi$--periodic.  
On the closed orbit, i.e.\ for $J = 0$, $\cal A$ is independent of 
$\phi$. 
Note that on the torus $J = 0$ the 
angular variables $\phi$ play a largely artificial role
because here $\cal A$ is independent of $\phi$.
But their inclusion is very convenient as it allows one to treat   
all tori on the same basis. Then all definitions, e.g.\ that of the ISF, 
apply to all tori.

A function is called $C^r$, if the function together with all of its
partial derivatives up to and including those of order $r$ are continuous.
In this paper we will assume that, for fixed $J$,
${\cal A}$ is a $C^1$ function of $(\theta, \phi)$. 
A $C^1$ function will be called {\em smooth}.
The smoothness of ${\cal A}$ corresponds to the fact that in real storage 
rings, the magnetic and electric fields are smooth functions of space and time.
The labels for the definitions, propositions, theorems and lemmas are chosen in a way which
indicates their relative positions in the text.

\section{General Properties of the Spin Motion}
\setcounter{equation}{0}
We begin by establishing some basic components of our formalism.

Clearly (\ref{eq:1.10}) gives $J(\theta)=J_0$ and
$\phi(\theta)=\omega(J_0)\theta+\phi_0$
where $J_0$ and $\phi_0$ are the actions and phases at $\theta = 0$.
Thus an orbit is labeled by $(\phi_0,J_0)$.  
But if we consider a fixed torus $J_0$ we often suppress the symbol $J_0$. 
By ``a fixed torus $J_0$'' we mean that the orbital
tune has the value $\omega(J_0)$ and that the spin motion is
characterized by the function ${\cal A}(\theta, \phi, J_0)$ of $\theta$ and 
$\phi$.
Equation (\ref{eq:1.9}) thus becomes
\begin{eqnarray}
\!\star\!\star\!\star &&    \qquad \qquad  \qquad \qquad \dot{S} = 
A(\theta; \phi_0,\omega){  S}\,,\quad {  S}(0; \phi_0,\omega) = {  S}_0\,,
\qquad \qquad \qquad \qquad \qquad \qquad
\label{eq:2.1}
\end{eqnarray}
where the real skew--symmetric $A$ is defined by 
\begin{eqnarray}
A(\theta; \phi_0,\omega):= {\cal A}(\theta, \omega \theta + \phi_0) \; .
\label{eq:2.17}
\end{eqnarray}
As will become clear from Definition 4.1 in Section 4, 
$A(\theta; \phi_0,\omega)$ is a quasiperiodic function of $\theta$ with the 
tunes (frequencies) $1, \omega_1,...,\omega_d$.
The solutions of (\ref{eq:2.1}) can be written as
${  S}(\theta; \phi_0,\omega)=\Phi(\theta; \phi_0,\omega) {  S}(0; \phi_0,
\omega)$
in terms of the {\em principal solution matrix} at $\phi_0$ which is
the $3 \times 3$ matrix
$\Phi(\theta; \phi_0,\omega)$, defined uniquely by the initial value
problem
\begin{eqnarray}
\frac{\partial \Phi(\theta; \phi_0,\omega)}{\partial \theta}
&=& A(\theta; \phi_0,\omega)\Phi \; ,
\label{eq:2.2} \\
\Phi(0;\phi_0,\omega) &=& I .
\label{eq:2.0}
\end{eqnarray}
Thus the principal solution matrix
at $\phi_0$
is the spin transport matrix 
from the azimuth $0$ to the azimuth $\theta$.
Occasionally we call a solution ${  S}(\theta;\phi_0)$ of (\ref{eq:2.1})
a ``spin trajectory'' at $\phi_0$.
The choice $\theta = 0$ for the starting azimuth does not imply a loss of 
generality as can be seen by considering the general initial value problem
$S(\theta_0)=S_0,\phi(\theta_0)=\phi_0$ for (\ref{eq:1.9}) and 
(\ref{eq:1.10}).

Note that $A(\theta;\phi_0)$ and $\Phi(\theta;\phi_0)$ are 
$2\pi$--periodic in $\phi_0$  and that by
the smoothness of $A$ the
principal solution matrix
is a smooth function of $(\theta,\phi_0)$ \cite{Amann,Hale}.
We will sometimes suppress the symbols 
$\phi_0$ and $\omega$
in  ${  S}$, $A$ and $\Phi$.
The key property of $\Phi(\theta)$ is that it belongs to $SO(3)$, i.e.
\begin{equation}
\Phi^T(\theta) \Phi(\theta) = I = \Phi (\theta) \Phi^T(\theta) \qquad
{\rm and}\qquad \det(\Phi(\theta))=1 \; ,
\label{eq:2.3}
\end{equation}
as is easily proved using (\ref{eq:2.2}) and (\ref{eq:2.0}).

Let ${  S}^1_0$ and ${  S}^2_0$ be two initial conditions for (\ref{eq:2.1}) 
and 
let $a \cdot b
= a_1b_1 + a_2b_2 + a_3b_3$ be the real inner product.  Then
${  S}^1(\theta) \cdot {  S}^2(\theta) 
= (\Phi(\theta){  S}^1_0) \cdot (\Phi(\theta){  S}^2_0) = 
(\Phi^T(\theta)\Phi(\theta) {  S}^1_0) \cdot {  S}^2_0 = 
{  S}^1_0 \cdot {S}^2_0$ 
so that  the inner product of any two
solutions of (\ref{eq:2.1}) is conserved.  In particular, the length of a spin
vector and the angle between any two spin trajectories at the same $\phi_0$ is 
conserved.  
In addition, it is easy to show that the cross product of two solutions 
is a solution.
In the remainder of
this paper we will, for convenience, allow spins $S$ to have
arbitrary length.

An interesting  property of (\ref{eq:2.1}) is that 
knowledge of one solution completely determines $\Phi$ by a simple
integration.  It is a standard result for linear systems that knowledge of
one solution can be used to reduce the dimension by one.  Here it reduces
the dimension by two because of the special structure of $\Phi$ in 
(\ref{eq:2.3}) as we will now demonstrate.

Let ${  v}^3(\theta)$ be a solution of (\ref{eq:2.1}), i.e.\
a spin trajectory at $\phi_0$, and let ${  v}^3$ be 
of norm 1.
Choose ${  v}^1(\theta)$ and 
${  v}^2(\theta)$ so that
\begin{equation}
V(\theta):=\left[ {  v}^1(\theta), 
{  v}^2(\theta), {  v}^3(\theta) \right]
\label{eq:2.4}
\end{equation}
is a $SO(3)$ matrix.
One can for example require ${  v}^1$ and ${  v}^2$ to be solutions of
$\dot{v}^k=(v^3\times\dot{v}^3)\times v^k$ (k=1,2), whence we can assume
that $V(\theta)$ is smooth, i.e.\ a $C^1$ function.  

Next we make a transformation $\Phi \rightarrow \Psi$ on
(\ref{eq:2.2}) and (\ref{eq:2.0})  defined by 
\begin{equation}
\Phi = V(\theta) \Psi.
\label{eq:2.5}
\end{equation}
This gives
\begin{equation}
\dot{\Psi} = C(\theta) \Psi\,,\quad \Psi(0) = V^T(0) \; ,
\label{eq:2.6}
\end{equation}
where
\begin{equation}
C(\theta) = V^T(\theta)\left(A(\theta)V(\theta) - \dot{V}(\theta)\right).
\label{eq:2.7}
\end{equation}
Since $V(\theta) \in SO(3)$, $\dot{V}^T(\theta) V(\theta) = 
- V^T(\theta) \dot{V}(\theta)$ and
thus $C^T = V^T(A^TV + \dot{V}) = -C$ by the skew  symmetry of $A$.
Therefore $C$ is skew--symmetric as expected for rotations. 
The third column of $C$ is
\begin{eqnarray}
&& C \left( \begin{array}{c} 0 \\ 0 \\ 1 \end{array}\right)
= V^T \left( A{  v}^3 - \dot{  v}^3 \right) = 0,
\nonumber
\end{eqnarray}
since ${  v}^3$ is a solution of (\ref{eq:2.1}),  
and the  skew  symmetry of $C$ yields
\begin{equation}
C(\theta) = c_{_{V}}(\theta) \left( \begin{array}{ccc}0 & -1 & 0\\ 1 & 0 & 0\\ 
0 & 0 &
0 \end{array}\right) =: c_{_{V}}(\theta){\cal J},
\label{eq:2.9}
\end{equation}
where (\ref{eq:2.9}) also serves to define $\cal J$. 
Therefore
\begin{eqnarray}
\Psi(\theta) = 
\exp \left({\cal J}\int\limits^{\theta}_0 c_{_{V}}(\theta')d \theta' \right)
V^T(0)
\nonumber
\end{eqnarray}
as is easily checked by differentiation. Finally,
\begin{eqnarray}
\!\star\!\star\!\star &&    \qquad    \qquad\Phi(\theta;\phi_0) = 
V(\theta) \exp \left({\cal J}
\int\limits^{\theta}_0 c_{_{V}}(\theta')d \theta' 
\right)
V^T(0). \qquad  \qquad \qquad \qquad  \qquad \qquad
\label{eq:2.11}
\end{eqnarray}
The exponentials in these equations can be evaluated by noting that
\begin{equation}
\exp ({\cal J}\tau) = \left( \begin{array}{ccc} \cos \tau & - \sin \tau &
0\\ \sin \tau & \cos \tau & 0\\ 0 & 0 & 1 \end{array} \right).
\label{eq:2.12}
\end{equation}

So we have constructed the complete principal solution matrix
by starting from just one solution
of (\ref{eq:2.1}) and assuming the existence of a smooth $V$.
This is the result we were aiming for. 

Note that by (\ref{eq:2.7}),(\ref{eq:2.9}) we have
\begin{equation}
\dot{V}(\theta)=A(\theta;\phi_0)V(\theta)-c_{_{V}}(\theta)
V(\theta){\cal J} \; .
\label{eq:2.8}
\end{equation}
It follows that $\dot{v}^1=A v^1 -c_{_{V}} v^2$ from which we deduce
$c_{_{V}} =v^2 \cdot (A{  v}^1 - \dot{  v}^1 ) = 
{  v}^2 \cdot (\Omega
\times {  v}^1) - {  v}^2 \cdot {\dot{  v}}^{1} =
 \Omega \cdot {  v}^3 +{  v}^1 \cdot
\dot{  v}^2$. Thus we obtain the useful formula
\begin{equation}
c_{_{V}}(\theta) = 
\Omega(\theta, \omega \theta + \phi_0) \cdot {  v}^3(\theta) 
+ {  v}^1(\theta) \cdot  \dot{  v}^2(\theta).
\label{eq:2.13}
\end{equation}
In addition, since $Tr\lbrack{\cal J}^2\rbrack=-2$, it follows from
(\ref{eq:2.8}) that 
$c_{_{V}}=-(1/2)\; Tr\lbrack\,{\cal J}V^T( A V -\dot{V})\rbrack$.

\vspace*{.15in}

\noindent{\bf Remarks:}
\begin{itemize}
\item[(1)] Equation (\ref{eq:2.5}) is equivalent to a change of basis for 
spin whereby $S$ is expressed as
$S=V(\theta)\hat{S}
=\hat{S}_1 v^1+\hat{S}_2 v^2+\hat{S}_3 v^3$ so that $\hat{S}$ is the spin
in the rotating frame represented by the matrix $V$. 
Moreover $\dot{\hat{S}}=c_{_{V}}(\theta)
{\cal J}\hat{S}$ so that $\hat{S}(\theta)=
\exp \left({\cal J}\int\limits^{\theta}_0 
 c_{_{V}}(\theta')d \theta'\right)\hat{S}(0)$. Thus
$\hat{S}_3$ is constant and  $\hat{S}$ precesses
around $(0, 0, 1)$ at a nonconstant rate $c_{_{V}}(\theta)$. 
~From  (\ref{eq:2.13}) the rate $c_{_{V}}$
is, as one would expect, just 
a combination 
of the projection 
of ${  \Omega}$ onto ${  v}^3$ and the rate,
${  v}^1 \cdot  \dot{  v}^2$, of rotation of ${  v}^1$ and 
${  v}^2$ around ${  v}^3$.
In our discussion
of spin tune in later sections it will be useful to define a frame in which
the spin precesses {\em uniformly}. From  (\ref{eq:2.11}) we have 
\begin{eqnarray}
&& S(\theta) =V(\theta)
\exp \left({\cal J}\int\limits^{\theta}_0 
(c_{_{V}}(\theta')-\nu)
d \theta'\right)
\exp \left({\cal J}\nu\theta\right) V^T(0)S(0)
\; ,
\nonumber
\end{eqnarray}
where $\nu$ is an arbitrary constant.
Thus with the change of basis 
\begin{eqnarray}
S=U(\theta)\hat{S} \; , \qquad
U(\theta)=V(\theta)
\exp \left({\cal J}\int\limits^{\theta}_0 
(c_{_{V}}(\theta')-\nu)
d \theta'\right) \; ,
\nonumber
\end{eqnarray}
we obtain $\dot{\hat{S}}=\nu
{\cal J}\hat{S}$ and we have defined a frame $U$ whose third column is a spin 
trajectory and in which spin has a 
constant precession rate $\nu$.
In the following we will be interested
in the case where the mean  
$\bar{c}_{_{V}} =\lim_{T\rightarrow\infty}
(1/T)\int\limits^{T}_0 c_{_{V}}(\theta)d \theta$ of ${c}_{_{V}}$  
exists and we will 
choose $\nu=\nu_{_{V}}$ where
$\nu_{_{V}}=\bar{c}_{_{V}}$ mod $1$ and is in [0,1).
Thus we can write 
$c_{_{V}}(\theta) = \tilde{c}_{_{V}}(\theta) + \nu_{_{V}}
 +k_{_{V}}$
where $\tilde{c}_{_{V}}$ represents the fluctuating part 
of $c_{_{V}}$ with zero mean and where the integer $k_{_{V}}$
is chosen such that $\nu_{_{V}} \in[0,1)$.
\item[(2)] The ideas in Remark 1 lead to some precise definitions.
On a given torus, let 
$U(\theta)\in SO(3)$ be such that the principal solution matrix
at $\phi_0$ can be written as
\begin{eqnarray}
\Phi(\theta;\phi_0) = U(\theta) 
\exp ({\cal J}\nu\theta) U^T(0) \; ,
\label{eq:2.20}
\end{eqnarray}
where $\nu$ is constant and in $[0,1)$. Then $U$ is
called a {\em uniform precession frame} (UPF) at $\phi_0$ and $\nu$,
which is uniquely determined by $U$, 
is called the {\em uniform precession rate} (UPR) for $U$
and is denoted by $\nu_s(U)$. 
We then call (\ref{eq:2.20}) a {\em standard form} of the 
principal solution matrix.

Under certain conditions which will be described in Definition 5.5, $\nu_s(U)$ 
will be called a spin tune.
Note that, due to (\ref{eq:2.20}), $U(\theta)$ is smooth in 
$\theta$ and satisfies the ordinary differential equation:
\begin{eqnarray}
&& \dot{U}(\theta)=A(\theta;\phi_0)U(\theta)-
\nu U(\theta){\cal J} \; ,
\label{eq:2.30} 
\end{eqnarray}
where $\nu=\nu_s(U)$. In particular by (\ref{eq:2.30}) the vector 
described by the third column of $U$ obeys (\ref{eq:2.1})
so that it is a spin trajectory with unit length.
Moreover,  for every constant $\nu$
the initial value problem defined by (\ref{eq:2.30}) and the arbitrary
initial matrix  $U(0) \in SO(3)$ has the unique solution
$U(\theta) =  \Phi(\theta;\phi_0) U(0)
\exp (-{\cal J}\nu\theta)$.
Thus every solution $U(\theta)$ of (\ref{eq:2.30})
with $U(0)\in SO(3)$ and $\nu\in[0,1)$ is a UPF at 
$\phi_0$ and its UPR $\nu_s(U)$ equals $\nu$, i.e.\
\begin{eqnarray}
&& \dot{U}(\theta)=A(\theta;\phi_0)U(\theta)-
\nu_s(U) U(\theta){\cal J} \; .
\label{eq:2.32} 
\end{eqnarray}
Because $Tr\lbrack\,{\cal J}^2\,\rbrack=-2$, 
for every UPF $U$ one has the useful formula 
\begin{eqnarray}
 \nu_s(U)=-(1/2)\; Tr\lbrack\,{\cal J}\,(U^T A U
-U^T \dot{U})\rbrack \; ,
\label{eq:2.31}
\end{eqnarray}
which follows from (\ref{eq:2.32}).
Note that the interval $[0,1)$ is just a matter of choice -- any
convenient half open interval of length 1 could be chosen, e.g.\
$(0,1\rbrack$. 
\item[(3)]
In this section and in the rest of this paper, the concepts of orthonormal 
reference frame and 
$SO(3)$  matrix are interchangeable. Moreover the elements of the columns of 
such a
matrix  are just the components of the unit coordinate vectors of the 
corresponding frame, 
as for example in (\ref{eq:2.4}). Thus  we will often identify the columns 
with such vectors.  
\item[(4)] It can be shown that if $V:
{\mathbb R}\rightarrow SO(3)$ with $V=:[v^1,v^2,v^3]$, is a smooth and 
$2\pi$--periodic
function (e.g.\ a $2\pi$--periodic UPF) 
then the fractional part of  \\
$\chi:=(1/2\pi)\int_0^{2\pi}\;d\theta
v^1(\theta)\cdot\dot{v}^2(\theta)$,
is independent of $v^1$ and $v^2$, i.e.\
the fractional part of $\chi$ only depends on $v^3$.
If $v^3$ were represented in the ``spinor formalism'' \cite{mv2000} then
it would be found that the fractional part of
$\chi$ is the {\em geometrical phase} of $v^3$ in the sense of
\cite{AA}.
\end{itemize}
~Finally, we summarize the basic eigenstructure of an $SO(3)$
matrix $R$ and its exponential representation in terms of a skew--symmetric
matrix $B$ in the following lemma.

\vspace*{.15in}

\begin{lemma}
a) Let $R$ be a $3 \times 3$ matrix in $SO(3)$.
Then a real number $\mu\in[0,2\pi)$ and a $SO(3)$
matrix $W$ exist such that the spectrum of 
$R$, $\lambda(R)$, is the set 
$\lbrace e^{i \mu}, e^{-i \mu},1 \rbrace$
(whence the eigenvalues are on the unit circle)
and such that
\begin{eqnarray}
RW = W \left(\begin{array}{ccc} \cos \mu & - \sin \mu & 0\\ \sin \mu &
\cos \mu & 0\\ 0 & 0 & 1 \end{array}\right) = W\exp({\cal J}\mu) \; ,
\label{eq:2.14}
\end{eqnarray}
where in the second equality we used (\ref{eq:2.12}).
~Furthermore, any such $W= \left[ {  w}^1,{  w}^2,{  w}^3 \right]$ satisfies  
the relations
$R ({  w}^1 \pm i {  w}^2) = \exp ( \mp i \mu) ({  w}^1 \pm i {  w}^2)$ and 
$R {  w}^3 = {  w}^3$.
Also, by (\ref{eq:2.14}), $R = e^{B \mu}$ 
where the matrix $B = W{\cal J}W^T$ 
is skew--symmetric. 

\noindent b) Conversely, if
$B$ is a real skew--symmetric matrix and if
$\lambda(B)=\lbrace i,-i,0 \rbrace$, then a $SO(3)$
matrix $W$ exists such that $B = W{\cal J}W^T$.
 \end{lemma}
The proof is elementary. See for example \cite{Frank,Gold}.
Note that the relation $R {  w}^3 = {  w}^3$ simply means that ${w}^3$ lies 
along the ``axis of rotation'' for $R$.

\vspace*{.15in}

\section{Spin Motion on the Closed Orbit}
\setcounter{equation}{0}
In this section we consider the case $J = 0$, so that 
the $A$ of (\ref{eq:2.17})
has no $\phi_0$ or $\omega$ dependence and is $2\pi$--periodic and 
smooth in $\theta$.
This case corresponds 
to the $2\pi$--periodic motion of the particle on the closed orbit and
the following theorem applies.

\vspace*{.15in}

\noindent{\bf Floquet Theorem:}\quad {\em For $J = 0$, there exists a 
$\hat{\nu}\in[0,1)$ and $\hat{W}\in SO(3)$ such that
the principal solution matrix defined by 
{\rm (\ref{eq:2.2})} and {\rm (\ref{eq:2.0})}  
is independent of $\phi_0$ and can be decomposed as}
\begin{eqnarray}
\!\star\!\star\!\star &&    \qquad \qquad  \qquad \qquad \qquad \Phi(\theta) 
= \hat{p}(\theta) \hat{W} \exp({\cal J} \hat{\nu} \theta) 
\hat{W}^T \; , \qquad \qquad \qquad \qquad \qquad \qquad \qquad
\label{eq:3.1}
\end{eqnarray}
{\em where the matrix $\hat{p}(\theta)\in SO(3)$ is 
$2\pi$--periodic and smooth  and where
$\hat{p}(0)=I$. Moreover, $\lambda(\Phi(2 \pi)) 
= \lbrace e^{i \hat{\nu} 2 \pi},\, e^{-i \hat{\nu} 2 \pi},\,1 \rbrace$.}
\vspace*{.15in}

\noindent{\em Proof:}\quad 
Since, at $J = 0$, ${\cal A}(\theta, \phi)$ is independent of $\phi$,
$\Phi(\theta; \phi_0)$ is independent of $\phi_0$.
From Lemma 2.1a, we know that there exist
$\mu \in[0, 2\pi)$ and $\hat{W}\in SO(3)$ such that
$\lambda(\Phi(2 \pi)) 
= \lbrace e^{i \mu},\, e^{-i \mu},\, 1 \rbrace$ and $\Phi(2 \pi) =
e^{\hat{B} \mu}$ where $\hat{B} = \hat{W} {\cal J} \hat{W}^T$.
Then with $\mu = 2 \pi \hat{\nu}$, $\hat{\nu} \in [0,1)$,
 $\lambda(\Phi(2 \pi)) 
= \lbrace e^{i \hat{\nu} 2 \pi},\, e^{-i \hat{\nu} 2 \pi},\,1 \rbrace$ and
$\Phi(2 \pi) = e^{\hat{B} \hat{\nu} 2 \pi}$.
A key property of the principal solution matrix is that

\begin{equation}
\Phi(\theta + 2 \pi) = \Phi(\theta) \Phi(2 \pi) \; ,
\label{eq:3.2}
\end{equation}
which we can see by  noting that
the l.h.s. of (\ref{eq:3.2}) is a solution matrix of (\ref{eq:2.2}) by the 
$2\pi$--periodicity of $A$
and the r.h.s. is a solution matrix of  (\ref{eq:2.2}) since $\Phi$ is. 
They are equal, by 
the uniqueness of solutions to the initial value problem for
(\ref{eq:2.2}) and (\ref{eq:2.0}), 
since they are equal at  $\theta=0$.

Define $\hat{p}$ by $\hat{p}(\theta):=\Phi(\theta)e^{-\hat{B} \hat{\nu} 
\theta}$. Clearly
$\hat{p}(0)=I$ and $\hat{p}(\theta)\in SO(3)$ since $\Phi\in SO(3)$ and 
$\hat{B}$ is 
skew--symmetric. The periodicity of $\hat{p}$ is clear because 
\begin{gather*}
\begin{split}
\hat{p}(\theta + 2 \pi) & = \Phi(\theta + 2 \pi) \exp(-\hat{B} 
\hat{\nu}(\theta+2 \pi))
= \Phi(\theta) \Phi(2 \pi) \exp(-\hat{B} \hat{\nu} 2 \pi) \exp(-\hat{B} 
\hat{\nu} \theta)
= \hat{p}(\theta) \; ,
\end{split}
\end{gather*}
where (\ref{eq:3.2}) is used at the second equality.
\hfill $\Box$

\vspace*{.15in}

Using the Floquet theorem we now make several remarks concerning the 
principal solution matrix
defined by (\ref{eq:2.2}),(\ref{eq:2.0}) when $J = 0$, i.e.\ when  $A$ is 
independent of $\phi_0$ 
and $2\pi$--periodic in $\theta$. 
The ``$~\hat{ }~$'' symbol which was specific to the theorem is not needed in 
the 
following remarks. 

\vspace*{.15in}

\noindent{\bf Remarks:}
\begin{itemize}
\item[(1)] If $\Phi(\theta) = p(\theta) 
\exp (B\nu\theta)$, where $\nu \in [0,1)$, where the real
skew--symmetric
matrix $B$ has the spectrum $\lambda(B) 
= \lbrace i,\, -i,\, 0 \rbrace$ and where the matrix $p(\theta)$ is
$2\pi$--periodic, then $p,B,\nu$ will be called {\em Floquet parameters}. 
In particular $\nu$ is called a {\em Floquet frequency}. 
Thus the Floquet
theorem states that Floquet parameters exist and it implies that 
the principal solution matrix 
depends on two frequencies where the Floquet frequency
emerges in addition to the
circulation tune $\omega_c = 1$. Note that the Floquet parameter $p(\theta)$ 
is a smooth element of 
$SO(3)$ with $p(0)=I$. 
\item[(2)] If $p,B,\nu$ are Floquet parameters as defined in Remark 1, 
then from Lemma 2.1b
a $W\in SO(3)$ exists such that
$W{\cal J}W^T=B$, whence  
$\Phi(\theta) = p(\theta) W e^{{\cal J} \nu \theta} (p(0)W)^T$. Thus 
at every $\phi_0$, $p(\theta)W$ is a $2\pi$--periodic UPF with UPR $\nu$, as 
defined in
Section 2. We conclude that every
~Floquet frequency is a UPR of a $2\pi$--periodic UPF
and that (recall Remark 2 of Section 2) a $2\pi$--periodic
unit--length function of $\theta$ exists, which is  
a spin trajectory at every $\phi_0$. This is the $n_0$ mentioned in the  
Introduction.
\par Conversely, if $U$ is a $2\pi$--periodic UPF,
then by (\ref{eq:2.20}) 
\begin{eqnarray}
\Phi(\theta) = U(\theta) 
\exp ({\cal J}\nu_s(U)\theta) U^T(0) =:p(\theta) 
\exp (B\nu\theta)
\; ,
\label{eq:3.6}
\end{eqnarray}
where $p,B,\nu$, defined by
\begin{eqnarray}
&& p(\theta):= U(\theta) U^T(0) \; , \qquad 
B:= U(0) {\cal J}U^T(0) \; , \qquad  \nu  = \nu_s(U) \; ,
\nonumber
\end{eqnarray}
fulfill all conditions of
~Floquet parameters. Thus the UPR of every $2\pi$--periodic UPF is
a Floquet frequency, i.e.\ for
$2\pi$--periodic UPF{\small s} the UPR emerges as a Floquet frequency
and thus as an extra frequency of the system.
We conclude that the set of Floquet frequencies is identical with the
set of UPRs which correspond to $2\pi$--periodic UPF{\small s}.
\item[(3)] To study the set of Floquet frequencies in more detail, we first
consider two sets $p,B,\nu$ and $\tilde{p},\tilde{B},\tilde{\nu}$ of
~Floquet parameters, i.e.\
\begin{eqnarray}
\Phi(\theta) = p(\theta) \exp(B\nu\theta) = 
\tilde{p}(\theta) \exp(\tilde{B}\tilde{\nu}\theta)
\; .
\label{eq:3.7}
\end{eqnarray}
~From (\ref{eq:3.7}) at $\theta = 2 \pi$, we obtain
$\Phi(2\pi) = \exp(2\pi B\nu) =  \exp(2\pi\tilde{B}\tilde{\nu})$,
so that $\lambda(\Phi(2 \pi))= 
\lbrace e^{2\pi i \nu},\, e^{-2\pi i \nu},\, 1 \rbrace=
\lbrace e^{2\pi i \tilde{\nu}},\, e^{-2\pi i \tilde{\nu}},\, 1 \rbrace$. 
Thus the set of Floquet frequencies has at most two elements and (due to
the Floquet theorem) at least one element. In particular,
the set of Floquet frequencies has either one element (which
then is equal to $0$) or it has two elements $\nu,1-\nu$, both of them 
positive.
Note that a Floquet frequency which is in $[0,1/2]$ always exists. 
Moreover, as we will see when we introduce the general concept of spin tune 
with  Definition 5.5, the set of spin tunes contains this
set of Floquet frequencies (see Remark 4 of Section 6).
This in turn allows us to select one of the Floquet frequencies 
as a ``preferred'' spin tune which we denote by $\nu_0$. 
In fact this corresponds to the 
customary choice in which $e^{i {\nu_0} 2 \pi}$ is an eigenvalue of 
$\Phi (2 \pi)$ \cite{chao81,br99}.
\item[(4)] By the definition of
~Floquet parameters, $\nu=0$ is a Floquet frequency
iff $\Phi(2 \pi) = I$. 
~From Remark 3 it follows that if $0$ is a Floquet frequency,
then it is the only Floquet frequency.
~Furthermore all solutions of (\ref{eq:2.1}) are
$2\pi$--periodic iff $\nu=0$ is a Floquet frequency.
It follows by Remark 2 that for the case $\nu=0$ 
every $2\pi$--periodic UPF has a  zero UPR.
See also Remark 4 in Section 6  and Theorem 6.4.
\end{itemize}

The Floquet theorem does not give a method for 
constructing the principal solution matrix
at $J = 0$ since  $\hat{p}$ is defined by $\Phi,W$ and 
$\hat{\nu}$ 
--- it is a theorem about its properties.
We now construct, by
the method pointed out in Remark 1 of Section 2,
a standard form for the principal solution matrix
in order to compute Floquet parameters
used in, or implied by,
other treatments of spin motion on the closed orbit  \cite{br99,bhr92,bmrr}.
We first show that the matrix (\ref{eq:2.4}) exists and 
can be chosen to  be 
$2\pi$--periodic in $\theta$.

Let ${  \xi}^3$ with length 1 denote the eigenvector 
for eigenvalue 1 of the $SO(3)$ matrix $\Phi(2 \pi)$. 
Then the solution of (\ref{eq:2.1})
with ${S}_0 = {  \xi}^3$ is $2\pi$--periodic since 
${S}(\theta) = \Phi(\theta){  \xi}^3 = 
\Phi(\theta)\Phi(2 \pi){  \xi}^3 =
\Phi(\theta+2 \pi){  \xi}^3 = {S}(\theta+2 \pi)$.
We now define ${  v}^3(\theta) := \Phi(\theta){  \xi}^3$. To complete
the construction of the matrix  $V$  we must now
construct the $2\pi$--periodic vectors ${  v}^1$ and ${  v}^2$.
To do this we can, for example, assume 
that a constant unit vector $e$ exists such that
${  v}^3(\theta)\times e$ has no zeros. Then we can define 
${  v}^1(\theta):={  v}^3(\theta)\times e/|{  v}^3(\theta)\times e|$
and ${  v}^2(\theta):={  v}^3(\theta)\times{  v}^1(\theta)$.
In any case the Floquet theorem ensures the existence of a smooth and 
$2\pi$--periodic $SO(3)$ matrix $V(\theta)$, whose third column is 
${  v}^3(\theta)$: just choose a $SO(3)$ matrix $V(0)$ whose third
column is ${  \xi}^3$ and then choose $V(\theta) = \hat{p}(\theta) V(0)$.

We now consider an arbitrary smooth $2\pi$--periodic $SO(3)$ matrix 
$V$ whose third column is a 
solution of (\ref{eq:2.1}) and  we construct  a $2\pi$--periodic UPF and
the corresponding standard form for the 
principal solution matrix.
~From (\ref{eq:2.7}) and (\ref{eq:2.9}) $c_{_{V}}(\theta)$ is 
$2\pi$--periodic and can be written as $c_{_{V}}=\bar{c}_{_{V}}+\tilde{c}_{_{V}}$
where, as before,  $\bar{c}_{_{V}}$ and $\tilde{c}_{_{V}}$ are the mean and 
zero--mean parts of
$c_{_{V}}$. From Remark 1 of Section 2 we can write
\begin{equation}
U(\theta)= 
V(\theta)\exp({\cal J}(\alpha(\theta) + k_{_{V}}\theta))
\label{eq:3.5}
\end{equation}
where $d\alpha/d\theta=\tilde{c}_{_{V}}$ and where
$\nu_{_{V}}\in[0,1)$ and the integer $k_{_{V}}$ are chosen such that
$\nu_{_{V}}+k_{_{V}}=\bar{c}_{_{V}}$. The principal solution matrix
is then given by (\ref{eq:3.6}) with
$\nu_s(U)=\nu_{_{V}}$.
Because the mean of $\tilde{c}_{_{V}}$ vanishes, 
the integral
$\int^{\theta}_0 \, \tilde{c}_{_{V}}(\theta') d \theta'$  is 
$2\pi$--periodic in $\theta$ and thus
$\exp ({\cal J} \alpha(\theta))$ and $U(\theta)$ are 
$2\pi$--periodic in $\theta$.

The standard  form for the principal solution matrix
that we promised is given by 
the first equality in (\ref{eq:3.6}), where $U$ is defined by
(\ref{eq:3.5}), and the UPR corresponding to $U$ is given by 
$\nu_s(U)=\bar{c}_{_{V}}-k_{_{V}}$.
In (\ref{eq:3.5}) we have a procedure to calculate 
$U(\theta)$ and $\nu_s(U)$ and
thus $\Phi(\theta)$,  knowing a periodic solution of (\ref{eq:2.1}).
~From $U(\theta)$ and $\nu_s(U)$ one can construct Floquet parameters
as in Remark 2.

\vspace*{.15in}

\noindent{\bf Remarks:}
\begin{itemize}
\item[(5)]
~From Remark 2 it is clear that the unit--length $2\pi$--periodic spin 
trajectory $n_0$ exists and from Remark 4   
it is clear that the direction of $n_0$
is not unique iff $\nu_0 = 0$.
\item[(6)]
The  above construction for ${v}^3(\theta)$ is easily done numerically.
~First integrate (\ref{eq:2.2}) from $\theta=0$ to  $\theta=2 \pi$
numerically to determine $\Phi(2 \pi)$ and then solve the linear system
$\Phi(2 \pi)\xi^3=\xi^3$.
Then ${  v}^3$ can be found on a grid of points in $[0,2\pi]$ by numerically
integrating (\ref{eq:2.1}) with $S_0=\xi^3$.
This is the way that $n_0(\theta)$
is constructed in SLIM \cite{chao81} and other related spin codes.
\item[(7)]
Since ${  v}^3$ is a $2\pi$--periodic solution of (\ref{eq:2.1}) 
so is $-{  v}^3$. 
One can therefore replace 
~$V = [{  v}^1,~{  v}^2, ~{  v}^3]$ ~by ~$V = [{  v}^1,~-{  v}^2, ~-{  v}^3]$.
By (\ref{eq:2.13}) it follows that with this replacement 
${\bar c}_{_{V}}$ becomes $-{\bar c}_{_{V}}$.
Thus if  ~$[{  v}^1,~{  v}^2, ~{  v}^3]$ ~leads to 
$\nu_{_{V}} \in (\frac{1}{2},1)$ then
~$[{  v}^1,~-{  v}^2, ~-{  v}^3]$ leads to $\nu_{_{V}} \in (0,\frac{1}{2})$,  
so that one can choose $V$ in (\ref{eq:3.5})
so as to put 
the UPR of $U$ in $[0,\frac{1}{2}]$.  
\end{itemize}
\section{Quasiperiodicity and a Diophantine Condition}
\setcounter{equation}{0}
In the previous section we set up the Floquet form for the 
principal solution matrix for the
 case $J = 0$ where $A(\theta)$ is $2\pi$--periodic. 
However, one of our aims is to
obtain an analogous form when $J \ne 0$, i.e.\ when 
$A$ contains the frequency vector $\omega$ in addition to the
circulation tune $1$.
~For this and other purposes we need to introduce the concept of 
quasiperiodicity.

A periodic function has one basic frequency whereas a quasiperiodic
function has a finite number of basic frequencies (tunes) denoted by $\nu =
(\nu_1,\nu_2, \dots, \nu_k)$.  We give the following definition.

\vspace*{.15in}

\noindent{\bf Definition 4.1:} 
a) A function $f: {\mathbb R}
\rightarrow {\mathbb R}$ is  said to be {\em quasiperiodic} 
with tune vector $\nu$ in
${\mathbb R}^k$ 
if a continuous and $2\pi$--periodic function $F:{\mathbb R}^k
\rightarrow {\mathbb R}$ exists and
\begin{eqnarray}
\!\star\!\star\!\star &&  \qquad \qquad \qquad \qquad \qquad \qquad \qquad 
f(\theta) = F(\nu\theta)\; .
\label{eq:4.0} \qquad \qquad \qquad \qquad \qquad \qquad \qquad \qquad
\end{eqnarray}
A function $f: {\mathbb R}
\rightarrow {\mathbb C}$ is  said to be {\em quasiperiodic} 
with tune vector $\nu$ in
${\mathbb R}^k$ 
if its real and imaginary parts are quasiperiodic with tune vector $\nu$ in
${\mathbb R}^k$. 
A real or complex matrix valued function is said to be {\em quasiperiodic}
with tune vector $\nu$ in
${\mathbb R}^k$
if its components are. 
\newline\noindent
b) The {\em spectrum} of a quasiperiodic $f$ is defined by $\Lambda(f):=
\lbrace \lambda\in{\mathbb R}: a(f,\lambda)\neq 0\rbrace$, where  
\begin{equation}
a(f,\lambda):= \lim_{T \rightarrow \infty} \frac{1}{T} \int_0^T 
f(\theta)\exp(-i\lambda \theta)
 d\theta \; .
\label{eq:4.02}
\end{equation}
c) The {\em mean} of $f$ is defined by
\begin{equation}
\bar{f} := a(f,0) \; ,
\label{eq:4.01}
\end{equation}
and the zero--mean part of $f$ is denoted by $\tilde{f}$ and defined by
$f=:\bar{f}+\tilde{f}$. \hfill $\Box$

\vspace*{.15in}

The class of functions so defined will be denoted by ${\cal Q}(\nu;k)$,
where we refer to $\nu$ as the tune vector and $k$ as the order.
If either  $\nu$ or $k$ are obvious from the context 
or not relevant we may omit either or both.  
We prove that $a(f,\lambda)$ exists in Lemma 4.3d. 
Note that for $k = 1$ all functions in ${\cal Q}$ are periodic.
A tune vector $\nu$ is said to be {\em nonresonant} if the equation
$$m \cdot \nu:= m_1 \nu_1 + \cdots + m_k \nu_k = 0,$$
where $m \in {\mathbb Z}^k$ (i.e.\ $m$ is a $k$ vector of integers), 
 has $m = 0$ as the only solution. 
If there are nontrivial solutions then $\nu$ is said to be 
{\em resonant}.
For some applications, e.g.\ as in Lemma 4.3d, a tune vector
may be assumed to be nonresonant, since
otherwise the order can be reduced, as the following lemma shows.

\vspace*{.15in}

\setcounter{lemma}{1}
\begin{lemma} 
If $f$ is in ${\cal Q}(\nu;k)$, then a 
nonresonant $\hat{\nu}$ exists such that $f$ is in 
${\cal Q}(\hat{\nu};\hat{k})$, where $\hat{k}\leq k$. 
\end{lemma}
{\em Proof:} 
Let $f \in {\cal Q}(\nu;k)$ and let $f:{\mathbb R}\rightarrow {\mathbb R}$.
Then $f(\theta)=F(\nu\theta)$ where ${F}:{\mathbb R}^k
\rightarrow {\mathbb R}$ is continuous and $2\pi$--periodic.
If $\nu$ is nonresonant, then nothing has to be proved. If $\nu=0$, then
$f$ is constant, so that $f\in{\cal Q}(1;1)$. If
$\nu$ is resonant and nonzero then $k\geq 2$ and there exists
nonzero $m \in {\mathbb Z}^k$ such that
$m \cdot \nu= 0$. If $m_k\neq 0$, then 
$\hat{F}:{\mathbb R}^{k-1}
\rightarrow {\mathbb R}$, defined by $\hat{F}(z_1,...,z_{k-1}):=
~F(-m_k z_1,...,-m_k z_{k-1},m_1 z_1+...+m_{k-1} z_{k-1})$, is continuous
and $2\pi$--periodic and
$f(\theta)=F(\nu\theta)=\hat F(\hat \nu\theta)$ where
$\hat{\nu}:=-(1/m_k)(\nu_1,...,\nu_{k-1})$. 
The case $m_k= 0$ can be dealt with analogously because $m$ is nonzero.
We have therefore shown that $f\in {\cal Q}(\hat\nu;k-1)$, 
where $\hat\nu\neq 0$. If $\hat\nu$ is resonant,
we can repeat the above procedure until
a nonresonant tune vector is obtained. For complex valued functions and
matrix valued functions the proof proceeds analogously.
\hfill $\Box$

\vspace*{.15in}

\noindent 
A different proof of Lemma 4.2 can be found in \cite[Appendix 3.8]{Loch}.

The order in every $\cal Q$ can also be increased since the $F$ in 
(\ref{eq:4.0}) can be viewed as a function with domain ${\mathbb R}^{k+l}$. 
Thus for an arbitrary real vector $\mu$ in ${\mathbb R}^l$,
${\cal Q}(\nu;k)\subset{\cal Q}(\nu,\mu;k+l)$. 
~For later use we note that if $a$ is an integer, $\exp({\cal J} a \theta)$ is
$2\pi$--periodic in $\theta$ and therefore in ${\cal Q}(1;1)$. 

The class of functions ${\cal Q}(\nu)$ obviously depends on $\nu$,
but two different $\nu$ can generate the same class.
~For example,  ${\cal Q}$ is
unaltered if $\nu$ is replaced by $\hat \nu = Q \nu$ where
$Q$ is a matrix with integer entries and determinant $\pm 1$,
i.e.\ $f \in {\cal Q}(\hat \nu;k)$ iff $f \in {\cal Q}(\nu;k)$.
To see this let $f \in {\cal Q}(\nu;k)$. 
Then there exists a function $F$ 
such that (\ref{eq:4.0}) holds.
Let $H(z):= F(Q^{-1} z)$. Then $H$ is $2\pi$--periodic 
since $Q^{-1}$ has integer entries, and $f \in {\cal Q}(\hat\nu;k)$
since $f(\theta) = H(\hat \nu \theta)$.
The converse is proved in the same way.

The periodic function $F$ in Definition 4.1
has the Fourier series 
\begin{equation}
\sum\limits_{m \in {\mathbb Z}^k} F_m
\exp(im \cdot z),
\label{eq:4.1}
\end{equation}
 where the Fourier coefficients of $F$  are defined by
\begin{eqnarray}
&& F_{m}:=\frac{1}{(2\pi)^k}
\int_0^{2\pi} \cdots\int_0^{2\pi}
  F(z) \exp(-i m\cdot z) dz_1 \cdots dz_k 
\label{eq:4.11}
\end{eqnarray}
and where $m\cdot z: = m_1z_1 + \cdots + m_kz_k$.
If $F$ is of class $C^k$ and if we define
\begin{equation}
S_N(z):= 
\sum_{m \in {\mathbb Z}^k\atop ||m||\leq N} 
~F_m \exp(im \cdot z) \; ,
\label{eq:4.600}
\end{equation}
then the sequence $\{S_N\}$ converges uniformly to $F$ on ${\mathbb R}^k$
\cite[p.411]{Koe}.
~Furthermore, the sequence $\{f_N\}$, defined by 
$f_N(\theta):=S_N(\nu\theta)$, converges uniformly to $f$.
Thus we have the representation
\begin{equation}
f(\theta) = \lim_{N \rightarrow \infty}f_N (\theta)
 =\lim_{N \rightarrow \infty}  \sum_{m \in {\mathbb Z}^k\atop ||m||\leq N} 
~F_m \exp(i(m \cdot \nu)\theta),
\label{eq:4.2}
\end{equation}
for $f$ in terms of the Fourier coefficients of $F$.
Here the norm $||m||$ of the integer vector $m$ is the max norm, i.e.\
$||m||:=max\lbrace|m_1|,...,|m_k|\rbrace$.

~For our definition of quasiperiodicity $F$ is only required to be in $C^0$,
not in $C^k$. Thus in general (\ref{eq:4.2}) does not apply and 
 we need the following lemma.

\vspace*{.15in}

\setcounter{lemma}{2}
\begin{lemma} 
a) Let $F$ be a continuous and $2\pi$--periodic real or complex matrix valued 
function
on ${\mathbb R}^k$. 
Then the sequence $\{\Sigma_N\}$, defined by
\begin{eqnarray}
&& \Sigma_N(z):=\sum_{m \in {\mathbb Z}^k\atop ||m||\leq N}\; A_{N,m}
~F_{m}\exp(i m \cdot z) \; ,
\label{eq:4.3}
\end{eqnarray}
converges uniformly to $F$ on ${\mathbb R}^k$, where
\begin{eqnarray}
&& A_{N,m}:=\prod_{n=1}^k\;\frac{N+1-|m_n|}{N+1}\; .
\nonumber
\end{eqnarray}
b) If $f\in{\cal Q}(\nu;k)$ and if $F$ is a continuous and $2\pi$--periodic 
function on ${\mathbb R}^k$ such that $f(\theta)=F(\nu\theta)$, then
\begin{equation}
f(\theta) = \lim_{N \rightarrow \infty} \Sigma_N (\nu\theta)
 =\lim_{N \rightarrow \infty}  \sum_{m \in {\mathbb Z}^k\atop ||m||\leq N} 
A_{N,m} F_m \exp(i(m \cdot \nu)\theta),
\label{eq:4.8}
\end{equation}
where the convergence is uniform. Thus {\rm (\ref{eq:4.8})}
generalizes {\rm (\ref{eq:4.2})} to this less smooth case.\\
c) Let $\nu$ be nonresonant under the conditions of Lemma 4.3b. 
Then the mean of $f$, defined by {\rm (\ref{eq:4.01})}, exists and
is given by
\begin{eqnarray}
&&\bar{f}   = F_0 := 
\frac{1}{(2\pi)^k}
\int_0^{2\pi} \cdots\int_0^{2\pi} F(z)dz_1 \cdots dz_k 
\end{eqnarray}
and the Fourier coefficients $F_m$ of $F$ in {\rm (\ref{eq:4.11})}
satisfy the relation 
\begin{eqnarray}
&&  F_{m} = \lim_{T \rightarrow \infty} \frac{1}{T} 
\int_0^T ~f(\theta) e^{-i(m \cdot \nu) \theta} d\theta 
= \lim_{T \rightarrow \infty} \frac{1}{T} 
\int_0^T ~F(\nu\theta) 
e^{-i(m \cdot \nu) \theta} d\theta 
\; . 
\label{eq:4.4}
\end{eqnarray}
In particular, $f=0$ implies $F=0$.

\noindent
d) Let $f$ be in ${\cal Q}(\nu;k)$.
Then $a(f,\lambda)$ exists for all $\lambda$. \\
e) Let $f$ be in ${\cal Q}(\nu;k)$.
Then $\Lambda(f)\subset \lbrace m\cdot\nu :m\in {\mathbb Z}^k\rbrace$. 
\end{lemma}
{\em Proof of Lemma 4.3a:} See for example \cite{Maa}. 
\hfill $\Box$
\newline\noindent
{\em Proof of Lemma 4.3b:} A simple consequence of Lemma 4.3a. 
\hfill $\Box$
\newline\noindent
{\em Proof of Lemma 4.3c:}
That $\bar{f}$ exists and equals $F_0$ is proved for example in
\cite[Section 10.3]{Arn}. The proof of (\ref{eq:4.4}) is similar. 
\hfill $\Box$
\newline\noindent
{\em Proof of Lemma 4.3d:} Let $f\in {\cal Q}(\nu;k)$. 
Then by Lemma 4.2, $f\exp(-i\lambda \, \cdot)
\in {\cal Q}(\hat{\nu};\hat{k})$, where
$\hat{\nu}$ is nonresonant and where $f\exp(-i\lambda \, \cdot)$  
denotes the function which maps $\theta$ to $f(\theta)\exp(-i\lambda \, 
\theta)$.
Applying Lemma 4.3c we conclude that
$a(f,\lambda)$ exists for all $\lambda$. 
\hfill $\Box$  
\newline\noindent
{\em Proof of Lemma 4.3e:} Since $a(\Sigma_N (\nu \, \cdot),\lambda)$ and
$a(f,\lambda)$ exist by Lemma 4.3d we have
\begin{eqnarray}
&& |a(\Sigma_N (\nu \, \cdot),\lambda)-a(f,\lambda)|
=|a(\Sigma_N (\nu \, \cdot)-f,\lambda)|
\nonumber\\
&&
=|\lim_{T \rightarrow \infty} \frac{1}{T} 
\int_0^T ~(\Sigma_N (\nu\theta)-f(\theta)) \exp(-i\lambda\theta)d\theta|
\leq \sup_{\theta}|\Sigma_N (\nu\theta)-f(\theta)| \; ,
\label{eq:4.10}
\end{eqnarray}
where we note that $\Sigma_N (\nu \, \cdot)$ and $f$ are bounded
functions and where $|\cdot|$ denotes the Euclidean norm. 
It follows that
\begin{equation}
\lim_{N \rightarrow \infty} a(\Sigma_N (\nu \, \cdot),\lambda)
=a(f,\lambda) \; ,
\label{eq:4.13}
\end{equation}
since the convergence in (\ref{eq:4.8}) is uniform.
If $\lambda\not\in
\lbrace m\cdot\nu :m\in {\mathbb Z}^k\rbrace$, then 
$a(\Sigma_N (\nu \, \cdot),\lambda) = 0$ and
(\ref{eq:4.13}) gives $a(f,\lambda)=0$. Thus if $\lambda\not\in
\lbrace m\cdot\nu :m\in {\mathbb Z}^k\rbrace$, then
$\lambda\not\in\Lambda(f)$, so that
$\Lambda(f)\subset \lbrace m\cdot\nu :m\in {\mathbb Z}^k\rbrace$. 
\hfill $\Box$ 

\vspace*{.15in}

\noindent{\bf Remarks:}
\begin{itemize}
\item[(1)] Lemma 4.3a is a multidimensional generalization 
of Fej\'er's theorem
and it shows that a continuous function can be recovered from its Fourier 
coefficients. 
~For $k=1$, $\Sigma_N$ in (\ref{eq:4.3}) is the ${\rm Ces\grave{a}ro}$ sum of 
the sequence $\lbrace S_N\rbrace$. For $k>1$ it is a natural generalization
of ${\rm Ces\grave{a}ro}$ summability and is one of many techniques
for summation.
The claim in Lemma 4.3c
that $\bar{f} = F_0$, is a ``flow'' version of Weyl's 
equidistribution theorem which is one of the main theorems of ergodic theory
\cite{si89},\cite[Chapter 3]{CFS},\cite{Arn}. Of course, if $F$ is in $C^k$
then $S_N$ and $\Sigma_N$ converge to the same function $F$.
\item[(2)] Definition 4.1 gives $f$ in terms of $F$, 
and Lemma 4.3a
and (\ref{eq:4.4}) show
how to recover $F$ given $f$ with nonresonant $\nu$. Furthermore Lemma 4.3b 
gives a representation for $f$ even when (\ref{eq:4.2}) is not valid.
If $\nu$ is resonant then Lemma 4.3 does not allow $F$ to be recovered.  
However we can find
$\hat{F}$ for the nonresonant tune vector $\hat{\nu}$ of Lemma 4.2.
\item[(3)] If $f$ is quasiperiodic, then by Theorem 4.3d 
$\bar{f}$ exists, whence $\tilde{f}:=f-\bar{f}$ is
quasiperiodic and the mean of $\tilde{f}$ exists. In particular
$\bar{\tilde{f}}=\overline{f-\bar{f}}=0$ which justifies calling
$\tilde{f}$ the zero--mean part of $f$.
\item[(4)] From (\ref{eq:4.3}) and (\ref{eq:4.4}) we see that $F$, for
nonresonant $\nu$, is determined
by its values on the curve $D':=\lbrace\nu\theta:\theta\in{\mathbb R}\rbrace$.
That this is to be expected can be shown as follows. 
Because $F$ is $2\pi$--periodic, the knowledge of $F$ at the points of $D'$  
implies that $F$ is even known at the points of the larger set  
$\tilde D:=
\lbrace\nu\theta + 2\pi M:\theta\in{\mathbb R},M\in{\mathbb Z}^k\rbrace$.
But $\tilde D$ is dense in ${\mathbb R}^k$ since 
$\nu$ is nonresonant and 
any continuous function is determined by its values on a dense set.
\end{itemize}
The following definition will be convenient in some of our proofs:

\vspace*{.15in}

\noindent{\bf Definition 4.4:}\quad A function $f: {\mathbb R}
\rightarrow {\mathbb R}$ is  said to be {\em almost periodic} if a
complex sequence $f_n$ and a real sequence
$\lambda_n$ exist such that
the sequence of functions $\sum_{n=1}^N f_n \exp(i\lambda_n\theta)$
converges uniformly to $f$ as $N\rightarrow\infty$.
A function $f: {\mathbb R}
\rightarrow {\mathbb C}$ is  said to be {\em almost periodic} if its real
and imaginary parts are.
A real or complex matrix valued function is said to be {\em almost periodic}
if it is
almost periodic in each component. 
\hfill $\Box$

\vspace*{.15in}

\noindent{\bf Remark:}
\begin{itemize}
\item[(5)] 
Due to Lemma 4.3b, the 
quasiperiodic functions form a subset of the {\em almost periodic
functions}. In contrast to the quasiperiodic functions,
the almost periodic functions generally have no finite set of
basic frequencies. The ~Fourier representation of such a function is given by
$\sum\limits^\infty_{n=1} f_n \exp(i \lambda_n \theta)$,
where $\lbrace \lambda_n \rbrace$ is the set of frequencies.  A
quasiperiodic function can be written in this way, but
(\ref{eq:4.2}) is often more  convenient for analysis.  
Almost periodic functions were
introduced by H. Bohr and a standard modern treatment can be found 
in \cite{Fink},
and in summarized form in the appendix of \cite{Hale}.  
\end{itemize}

We have now provided the basic machinery needed to deal with the 
quasiperiodic functions appearing in the following sections. 
In the remainder of this section
we treat a special aspect of quasiperiodicity, namely the so called
small divisor problem and we show how one solves this problem 
by using sufficiently differentiable functions. 
This material will only be needed in Theorems 6.5c-d. 

Integration as in the definition of $\alpha(\theta)$ in Section 3 (see
(\ref{eq:3.5})) is, of
course, a smoothing operation and the integral of a  periodic
function with zero mean is periodic. 
It is perhaps surprising then that the  integral of
a  quasiperiodic function with zero mean 
is not necessarily quasiperiodic.
This is an example of the so--called small divisor problem.
To see this let $\nu$ be nonresonant and let
$f$ be in ${\cal Q} (\nu;k)$ with zero mean 
with its associated $F$ in $C^k$. Define
 $g(\theta) = \int_{0}^{\theta}f(\theta')\;d\theta'$ for $f$ given by
(\ref{eq:4.2}). Because of the 
uniform convergence in (\ref{eq:4.2}) 
we can interchange limit and sum giving 
\begin{equation}
g(\theta) = 
\lim_{N \rightarrow \infty}\sum_{m \in {\mathbb Z}^k\atop 
0<||m||\leq N}
 -i(m \cdot\nu)^{-1}F_{m}
 \biggl(\exp(i(m \cdot\nu)\theta)  - 1\biggr).
\end{equation}
However, although $g$ is well defined, it may not be quasiperiodic 
if $k \ge 2$.
The source of the problem is that the divisor $m\cdot\nu$, while not zero,
 can be arbitrarily small and this can lead to an unbounded $g$, thereby  contradicting 
(4.1). 
We give such an example in Section 8.
To find a sufficient condition for quasiperiodicity we define
\begin{eqnarray}
G_N(z):= \sum_{m \in {\mathbb Z}^k\atop 0<||m||\leq N} 
-i(m \cdot\nu)^{-1} F_{m}
\biggl(\exp(im \cdot z)  - 1\biggr) \; ,
\label{eq:4.5}
\end{eqnarray}
 and note that $g(\theta) = \lim_{N \rightarrow \infty} G_N(\nu\theta)$ 
so that
$G_N$ converges on the lines $z=\nu\theta+2\pi l$ in ${\mathbb R}^k$
where $l$ is an integer.
Now suppose that $G_N$ 
converges pointwise in ${\mathbb R}^k$ to a 
function $G$. Then $G$ is 
$2\pi$--periodic and $g(\theta) = G(\nu\theta)$.
Since $g$ is continuous, $G$ is continuous along the lines 
$z=\nu\theta+2\pi l$, 
but Definition 4.1 requires that
 $G$ be continuous  everywhere on ${\mathbb R}^k$.
So we need a condition that ensures both pointwise convergence of $G_N$ and 
the continuity of the limit. 
Uniform convergence guarantees 
both, and a sufficient condition for this is a so called
``Diophantine condition'' which places a condition on how close 
$m \cdot\nu$ can be to zero. 

In the following we will be concerned with the tune vector $\nu=(1,\omega)$
where $1$ is the circulation tune and $\omega\in{\mathbb R}^d$ is the orbital 
tune. Thus we formulate the Diophantine condition and the convergence
of $G_N$ in terms of $\nu=(1,\omega)$ and $k=d+1$. 
See \cite[Appendix 4]{Loch} for some details on Diophantine conditions and
 \cite{Dumas} for further discussion and its use in another beam dynamics 
 context.

\vspace*{.15in}

\noindent{\bf Definition 4.5:}\quad
The tune vector $\omega\in {\mathbb R}^d$  is said
to satisfy a {\em Diophantine condition} if, for positive $\tau$,
$\omega$ is in the {\em Diophantine set}
${\Omega}(\tau):=\bigcup_{\gamma \in (0,1]}{\Omega}(\tau, \gamma)$
where ${\Omega}(\tau, \gamma)$ is the set
$$\lbrace \omega \in {\mathbb R}^d : | m\cdot(1, \omega) | \geq \gamma
||m||^{-\tau} , m \in {\mathbb Z}^{d+1},\; m \neq 0\rbrace$$.
\hfill $\Box$

\vspace*{.15in}

\noindent Note that the symbol $\Omega$ is also used in the Introduction for 
the precession vector. However its meaning should be clear from the context. 

It follows from the definition that ${\Omega}^c(\tau,\gamma) = 
\bigcup_{m\in{\mathbb Z}^{d+1}\setminus\lbrace 0\rbrace}\; 
 {\mathcal Z}(\tau,\gamma,m)$, where ${\mathcal Z}(\tau,\gamma,m):=
 \lbrace \omega\in {\mathbb R}^d:|m\cdot(1,\omega)|
<\gamma ||m||^{-\tau} \rbrace \; $.  
For fixed $(\tau,\gamma)$ and for each $m$ the ``resonant zone'' 
 ${\mathcal Z}(\tau,\gamma,m)$
 is either empty or 
a thickened $d-1$ dimensional plane centered on the resonant plane,
 $m\cdot (1,\omega) = m_0 + m_1\omega_1 +\cdots
 +m_d\omega_d =0$, with thickness proportional to $\gamma$.  
For example, when $d=1,\|m\|=1$ and $m_1\neq 0$ 
the corresponding zones are intervals centered on the three points
in  ${\mathbb R}$: $\omega_1=\eta$
 where $\eta \in \lbrace-1,0,1\rbrace$, each with thickness $2\gamma$.
When $d=2,\|m\|=1$ and $(m_1,m_2)\neq 0$  
the corresponding zones are centered on the twelve lines
in  ${\mathbb R}^2$: $\omega_1=\eta,\omega_2=\eta,\omega_2=\pm\omega_1+\eta$
where $\eta \in \lbrace-1,0,1\rbrace$ with thickness either $2\gamma$ or
$\sqrt{2}\gamma$.
More generally, if $m=(m_0,\hat{m})$ with $m_0\in{\mathbb Z}$ and 
$\hat{m}\in({\mathbb Z}^d\setminus\lbrace 0\rbrace)$ then 
one can show by using rotations in ${\mathbb R}^d$ that
${\mathcal Z}(\tau,\gamma,m)$ can be rotated into the set
$\lbrace \omega\in {\mathbb R}^d:
|\frac{m_0}{|\hat{m}|}+\omega_1|
<\frac{\gamma||m||^{-\tau}}{|\hat{m}|} \rbrace$.
Thus the thickened $d-1$ dimensional plane
${\mathcal Z}(\tau,\gamma,m)$ has thickness 
$2\gamma||m||^{-\tau}/|\hat{m}|$.
For $m$ such that $\hat{m}=0$ the resonance condition cannot 
be satisfied and we have  ${\mathcal Z}(\tau,\gamma,m)=\emptyset$.
Note also that ${\mathcal Z}(\tau,\gamma,m)$ is undefined for $m=0$. 

\vspace*{.15in}

\noindent{\bf Definition 4.6} (orbital resonance):\quad 
~We say that the torus at $J$ is off orbital resonance if $(1,\omega(J))$ is 
nonresonant (otherwise we say that it is on orbital resonance)
 and this is certainly the case if $\omega(J) \in \Omega(\tau)$.
Usually, {\rm (\ref{eq:1.10})} is said to be resonant if $\omega$ is resonant.
Our usage is different because our basic system is {\rm (\ref{eq:2.1})},
 which includes the circulation tune, 1. \hfill $\Box$
  
\vspace*{.15in}

We can now interpret $\Omega(\tau,\gamma)$ as the closed set in 
${\mathbb R}^d$
 constructed by successively removing the open resonance zones,
 corresponding to the resonance planes $m\cdot(1,\omega)=0$, with increasing 
$\|m\|$.
Thus its construction is similar to the construction of a Cantor set.
The resonance planes are dense in ${\mathbb R}^d$ and thus 
$\Omega(\tau,\gamma)$   
 is small in the sense that it has an empty interior. 
However, we will show in the proof of Lemma 4.8 
that it is large in the sense that for $\tau > d+1$ the Lebesgue
measure
 of its complement relative to ${\mathbb R}^d$ is proportional to $\gamma$ 
(in the sense of (\ref{eq:4.23})) which can
 be arbitrarily small.
We could take our Diophantine set to be  $\Omega(\tau,\gamma)$ for $\gamma$ 
small as 
 in \cite{Dumas}. Here we take the larger set $\Omega(\tau)$.

Now if $\omega\in {\Omega}(\tau,\gamma)$ then $|(m \cdot(1, \omega))^{-1} 
~F_{m} |\leq \gamma^{-1} ||m||^\tau |F_{m}|$ 
and thus $G_N$ converges uniformly if
$\sum ||m||^\tau | F_{m} |$ converges.  
Thus the Diophantine condition 
leads to a simple sufficient condition for the quasiperiodicity of $g$. 
In this context we now state and prove the following lemma which 
addresses the differentiability of $F$.

\vspace*{.15in}

\setcounter{lemma}{6}
\begin{lemma} Let $F:{\mathbb R}^{d+1}
\rightarrow {\mathbb R}$ be of class $C^n$
and $2\pi$--periodic and let $F_0=0$.
Let $\omega\in\Omega(\tau)$ where
$0<\tau<n-d-2$. Then $G_N:{\mathbb R}^{d+1}
\rightarrow {\mathbb R}$, given by
{\rm (\ref{eq:4.5})} with $\nu=(1,\omega)$, 
converges uniformly on ${\mathbb R}^{d+1}$ 
to a smooth
function $G$ which is $2\pi$--periodic.
Moreover, $\nabla G(z)\cdot (1,\omega)=F(z)$.
\end{lemma}
{\em Proof:} 
Because a constant $M\geq 0$ exists
such that $||m||^n |F_{m}|\leq M$ (see \cite[p.409]{Koe}), we have,
for every $\tau>0$,
\begin{eqnarray}
&& \sum_{m \in {\mathbb Z}^{d+1}\atop 
0<||m||\leq N} ||m||^\tau |F_{m}|
\leq M\sum_{m \in {\mathbb Z}^{d+1}\atop 
0<||m||\leq N}  ||m||^{\tau-n}
= M\sum_{j=1}^N  j^{\tau-n}
\sum_{m \in {\mathbb Z}^{d+1}\atop ||m||=j} 1 \; .
\label{eq:4.6}
\end{eqnarray}
A combinatorial argument gives
\begin{eqnarray}
&& \sum_{m \in {\mathbb Z}^{d+1}\atop 
||m||=j} 1 \leq 3^{d+1} (d+1) j^d  \; .
\label{eq:4.15}
\end{eqnarray}
Moreover, if $\omega \in {\Omega}(\tau,\gamma)$, then
$|(m \cdot(1, \omega))^{-1}|\leq \gamma^{-1} ||m||^\tau$ and 
(\ref{eq:4.6}) and (\ref{eq:4.15}) give
\begin{eqnarray}
&& \sum_{m \in {\mathbb Z}^{d+1}\atop 
0<||m||\leq N} |(m \cdot(1, \omega))^{-1} F_{m} |
\leq  3^{d+1} (d+1) M \gamma^{-1}
\sum_{j=1}^N  j^{\tau-n+d} \; .
\label{eq:4.9}
\end{eqnarray}
Since $\tau<n-d-2$ we conclude that the l.h.s. of (\ref{eq:4.9})
converges as $N\rightarrow\infty$ for every
$\omega\in {\Omega}(\tau)$. It follows that $G_N$ converges uniformly to a
continuous function $G$ which is $2\pi$--periodic.

\noindent
~From (\ref{eq:4.5}) we have
\begin{eqnarray}
&& \nabla G_N(z)=  \sum_{m \in {\mathbb Z}^{d+1}\atop 
0<||m||\leq N} (m \cdot (1,\omega))^{-1} F_{m} \exp(im \cdot z) ~m \; .
\label{eq:4.18}
\end{eqnarray}
Then if $\omega \in {\Omega}(\tau,\gamma)$ and repeating the above 
argumentation we find
\begin{eqnarray}
&& |\nabla G_N(z)| \leq  
\sum_{m \in {\mathbb Z}^{d+1}\atop 
0<||m||\leq N}
|( m\cdot (1,\omega))^{-1} F_{m} |  ~|m|
 \leq  \sqrt{d+1}
\sum_{m \in {\mathbb Z}^{d+1}\atop 
0<||m||\leq N}
\; |(m \cdot (1,\omega))^{-1} F_{m} |  ~||m||
\nonumber\\
&&
\leq  3^{d+1} (d+1)^{3/2} M \gamma^{-1}
\sum_{j=1}^N j^{\tau+1-n+d} \; . \qquad
\label{eq:4.22}
\end{eqnarray}
Thus, if $\omega \in {\Omega}(\tau)$,
$\nabla G_N$ converges uniformly and a standard result (see
\cite[8.6.3]{Di60},\cite[p.117]{Lang}) means that 
$G$ is $C^1$ and
\begin{eqnarray}
&& \nabla G(z)=\lim_{N\rightarrow\infty}
\sum_{m \in {\mathbb Z}^{d+1}\atop 
0<||m||\leq N}
  (m \cdot (1,\omega))^{-1} F_{m} \exp(im \cdot z) ~m\; . 
\label{eq:4.25}
\end{eqnarray}
It follows that $\nabla G(z)\cdot(1,\omega)=\lim_{N\rightarrow\infty}
\nabla G_N(z)\cdot(1,\omega)=F(z)$ since, with $n \geq d+1$, the $S_N$
in (\ref{eq:4.600}) converge. 
\hfill$\Box$  



\noindent {\bf Remark:}
\begin{itemize}
\item[(6)] To prove (\ref{eq:4.15}) we define the sets
$s:=\lbrace m \in {\mathbb Z}^{d+1}: ||m||=j\rbrace$ and 
$s_i:=\lbrace m \in s: |m_i|=j\rbrace$. It follows that
$s=\bigcup_{i=1}^{d+1} s_i$ and that $s_i$ contains $2(2j+1)^d$ elements.
Then $s$ contains no more than
$2(d+1)(2j+1)^d$ elements and because
\begin{eqnarray}
&& 2(2j+1)^d \leq  3(2j+1)^d \leq 3(3j)^d =
 3^{d+1}j^d \; , \nonumber
\end{eqnarray}
we conclude that $s$ contains no more than
$3^{d+1}(d+1)j^d$ elements, thus proving (\ref{eq:4.15}).
\end{itemize}

Lemma 4.7 provides the basic framework that we need for
discussing the uniform convergence of the sequence $G_N$.  
In particular it shows that as
$n$ increases beyond $\tau+d+2$ the small divisor problem
loses much of its potency.  This comes as no surprise because
the inequality $||m||^n |F_{m}|\leq M$ implies that the
~Fourier coefficients decrease with increasing $||m||$ more rapidly as $n$ 
increases. Then with growing $n$ the small divisor in (\ref{eq:4.5}) 
can come closer to zero without destroying the convergence.

However, although Lemma 4.7 takes the mystery out of the
working of the Diophantine condition, it puts the burden on
determining which $\omega$ are in the set ${\Omega}(\tau)$ and it is
not so easy to decide, off orbital resonance, if $\omega$ is in 
${\Omega}(\tau)$.
But some relief  comes from the following lemma, which 
shows that if $\tau > d + 1$ then the complement,  ${\Omega}^c(\tau)$,
of ${\Omega}(\tau)$ is a small set in terms of Lebesgue measure.
If in addition  $\tau < n-d-2$, the sequence $G_N$ converges 
uniformly for almost every $\omega$.
~For these two conditions to be consistent we thus need $n>2d+3$.
These results will be central to the statement and proof of Theorems 6.5c-d.

\vspace*{.15in}

\setcounter{lemma}{7}
\begin{lemma} If $\tau>d+1$,
then $\mu({\Omega}^c(\tau))=0$, where
$\mu$ denotes the Lebesgue measure.
\end{lemma}
{\em Proof:} 
~Let $B(R):=\lbrace \omega \in {\mathbb R}^d: |\omega| \leq R\rbrace$ and
define ${\mathfrak S}(R,m):= {\mathcal Z}(\tau,\gamma,m) \cap B(R)$.
Then from Definition 4.5, ${\Omega}^c(\tau,\gamma)\cap B(R) = 
\bigcup_{m\in{\mathbb Z}^{d+1}\setminus\lbrace 0\rbrace}\; 
{\mathfrak S}(R,m)$. Note that ${\mathcal Z}(\tau,\gamma,m)$ (hence 
${\mathfrak S}(R,m)$) is undefined for $m=0$. 

\noindent 
We will show below that
\begin{eqnarray}
&&\mu({\mathfrak S}(R,m))
\leq  2\gamma R^{d-1} \alpha(d-1)
||m||^{-\tau} \; ,
\label{eq:4.223}
\end{eqnarray}
where $\alpha(n):=\pi^{n/2}/\Gamma(n/2+1)$ (note that $\alpha(d)$ is the 
volume of $B(1)$). 

\noindent
Assuming the validity of (\ref{eq:4.223}),
\begin{eqnarray}
&&   \mu\biggl( {\Omega}^c(\tau,\gamma)\cap B(R)\biggr) 
\leq  2\gamma R^{d-1} \alpha(d-1)
\sum_{m\in{\mathbb Z}^{d+1}\setminus\lbrace 0\rbrace}\; 
||m||^{-\tau} \; .
\nonumber
\end{eqnarray}
As before (see (\ref{eq:4.15}))
\begin{eqnarray}
&& \sum_{m\in{\mathbb Z}^{d+1}\setminus\lbrace 0\rbrace}\; 
||m||^{-\tau} = \sum_{j=1}^\infty  j^{-\tau}
 \sum_{m\in {\mathbb Z}^{d+1}\atop ||m||=j} 1 
\leq 3^{d+1} (d+1)\sum_{j=1}^\infty  j^{d-\tau} 
\; .
\nonumber
\end{eqnarray}
Therefore
\begin{eqnarray}
&&   \mu\biggl( {\Omega}^c(\tau,\gamma)\cap B(R)\biggr)
\leq  2\gamma R^{d-1} \alpha(d-1)3^{d+1}(d+1)
\sum_{j=1}^\infty  j^{d-\tau} 
\; ,
\label{eq:4.23}
\end{eqnarray}
and for $\tau>d+1$ the series converges. Now
${\Omega}^c(\tau,1/p)$ decreases monotonically to
${\Omega}^c(\tau)$ as the positive integer $p\rightarrow\infty$. Therefore
\begin{eqnarray}
&&   \mu\biggl( {\Omega}^c(\tau)\cap B(R)\biggr)
=\lim_{p\rightarrow\infty}
 \mu\biggl( {\Omega}^c(\tau,1/p)\cap B(R)\biggr) = 0 
\nonumber
\end{eqnarray}
by continuity and the finiteness of Lebesgue measure restricted to
$B(R)$ where the second equality follows from (\ref{eq:4.23}). 
Since this is true for all $R$ we obtain the required result.

To prove (\ref{eq:4.223}) we first recall that 
${\mathcal Z}(\tau,\gamma,m)$ is empty if
$m=(m_0,0,...,0)$. Thus ${\mathfrak S}(R,m)$ can only be nonempty if
$m=(m_0,\hat{m})$ with $m_0\in{\mathbb Z}$ and 
$\hat{m}\in({\mathbb Z}^d\setminus\lbrace 0\rbrace)$ so that we only have 
to consider this case. Using the fact that volumes are
invariant under rotations it follows from the remarks after Definition 4.5
that
\begin{eqnarray}
&& \mu({\mathfrak S}(R,m)) 
\leq \mu\biggl( \lbrace \omega\in B(R) :|\frac{m_0}{|\hat{m}|}+\omega_1|
<\gamma||m||^{-\tau} \rbrace \biggr)  \; ,
\label{eq:4.2230}
\end{eqnarray}
where we also used the fact that $1/|\hat{m}|\leq 1$.
If $d=1$ then (\ref{eq:4.2230}) and the fact that
$\alpha(0)=1$ immediately yield (\ref{eq:4.223}). Furthermore, for $d\geq 2$,
we conclude from (\ref{eq:4.2230}) that
\begin{eqnarray}
&& \mu({\mathfrak S}(R,m)) \leq 
\mu\biggl(\lbrace \omega\in {\mathbb R}^d: \sqrt{\omega_2^2+...+\omega_d^2}
\leq R,
|\frac{m_0}{|\hat{m}|}+\omega_1|
<\gamma||m||^{-\tau} \rbrace\biggr) \nonumber\\
&& =\mu\biggl(\lbrace \omega\in {\mathbb R}^d: 
\sqrt{\omega_2^2+...+\omega_d^2}\leq R,
|\omega_1|<\gamma
||m||^{-\tau}\rbrace\biggr) =
 2\gamma R^{d-1} \alpha(d-1)
||m||^{-\tau} \; ,
\nonumber
\end{eqnarray}
where the r.h.s. is just the volume of the cylinder in
${\mathbb R}^d$ with height
$2\gamma ||m||^{-\tau}$ and radius $R$ and where in the first 
equation we used the fact that volumes are
invariant under translations. This completes the proof.
\hfill$\Box$

\section{Introduction to the Quasiperiodicity of the Spin Motion. \\ 
Definition of the Spin Tune}
\setcounter{equation}{0}
\setcounter{theorem}{0}
We can now continue the study of the 
principal solution matrix $\Phi(\theta;\phi_0,\omega)$
for (\ref{eq:2.1}) which is defined by the initial value problem
\begin{eqnarray}
\frac{\partial \Phi}{\partial \theta}
&=& A(\theta; \phi_0,\omega)\Phi \; , \qquad
\Phi(0;\phi_0,\omega) = I 
\label{eq:5.02}
\end{eqnarray}
where 
\begin{eqnarray}
&& A(\theta; \phi_0,\omega):= {\cal A}(\theta, \omega \theta + \phi_0) \; .
\label{eq:5.03}
\end{eqnarray}
In the language of Section 4
$A(\cdot; \phi_0,\omega)$ is quasiperiodic in ${\cal Q}(1,\omega;d+1)$.
One of the aims of this paper is to define the spin tune for this system.
We are guided by the special case $J = 0$ of 
Section 3 and therefore, and as will become clear below, 
our emphasis in this section is on 
principal solution matrices, $\Phi$,
where $\Phi(\cdot;\phi_0,\omega)$ is quasiperiodic and can be written in the 
standard form, 
\begin{eqnarray}
\!\star\!\star\!\star &&  \qquad \qquad  \qquad \qquad\Phi(\theta;\phi_0,
\omega) = U(\theta) 
\exp ({\cal J}\nu\theta) U^T(0) \; , \qquad \qquad \qquad \qquad \qquad \qquad
\label{eq:5.01}
\end{eqnarray}
where $U$ is a quasiperiodic $SO(3)$ matrix in ${\cal Q}(1,\omega;d+1)$
and $\nu\in[0,1)$ and both may depend on $(\phi_0,\omega)$. 
In that case all solutions of (\ref{eq:2.1})
are in ${\cal Q}(1,\omega,\nu;d+2)$ with a very simple frequency
structure in the tune $\nu$. 
Such principal solution matrices
fulfill a generalized Floquet theorem.

There has been extensive study of the equation $\partial \Phi/ \partial \theta
= A(\theta;y)\Phi$, where
the matrix $A(\cdot;y)$ is almost periodic and $y$ is
a vector of parameters, one goal being to find
conditions under which almost periodic solutions exist
(see for example \cite{Fink,Yos}). 
However our problem (\ref{eq:5.02})
is quite special because of the 
parameter  dependence of $A$ induced by ${\cal A}(\theta,\phi)$ in
(\ref{eq:5.03}). In fact, for every integer $N$ we have
\begin{eqnarray}
&& A(\theta+2\pi N;\phi_0,\omega)=A(\theta;2\pi N\omega+\phi_0,\omega) \; ,
\label{eq:5.04} 
\end{eqnarray}
which follows from (\ref{eq:5.03}) and the $2\pi$--periodicity of
${\cal A}(\theta,\phi)$ in $\theta$. We will return to this in Section 9
where we will see  that 
condition (\ref{eq:5.04}) has useful consequences for the 
spectrum  $\Lambda(\Phi(\cdot;\phi_0,\omega))$ of the 
principal solution matrix.

The case $J = 0$ where $A(\cdot; \phi_0,\omega)$ is $2\pi$--periodic, i.e.\ 
in ${\cal Q}(1;1)$ and independent of $\phi_0$ and $\omega$,
was discussed in Section 3. There we found a solution of (\ref{eq:2.1}) in 
${\cal Q}(1;1)$, namely ${  v}^3(\theta)=\hat{p}(\theta)\xi^3$.
All other linearly independent
solutions are in ${\cal Q}(1,\nu;2)$. The latter follows directly
from the Floquet theorem, but more importantly also from the construction of 
the 
$2\pi$--periodic
UPF using ${  v}^3$ which led, with $U$ in ${\cal Q}(1;1)$, to the 
representation (\ref{eq:5.01}). 
It is this construction that 
points the way 
for a generalization of the Floquet theorem to 
$A\in{\cal Q}(1,\omega;d+1)$. 

We begin with the following proposition.

\vspace*{.15in}

\noindent{\bf Proposition 5.1}
{\em Consider the initial value problem
\begin{equation}
\dot{S} = A(\theta){  S} \; , \quad {  S}(0) = {  S}_0\,,
\label{eq:5.0}
\end{equation}
with a real skew--symmetric $3\times 3$
matrix $A\in {\cal Q}(\nu;k)$  
and make the following assumptions:
\newline\noindent
a) Equation {\rm (\ref{eq:5.0})} has a nonzero solution ${  v}^3$ in 
${\cal Q}(\nu;k)$.
\newline\noindent
b) There exists a smooth $SO(3)$ matrix
$V=\left[ {  v}^1,  {  v}^2, {  v}^3 \right]$ 
in ${\cal Q}(\nu;k)$.
\newline\noindent
c) Let $c:= {  v}^2 \cdot (A{  v}^1 - \dot{  v}^1 )$.
Since $c\in{\cal Q}(\nu;k)$ its mean, $\bar{c}$, and zero-mean part, 
$\tilde{c}$,
 exist (recall Remark 3 of Section 4).
We assume that $\int\limits^{\theta}_0 \tilde{c}(\theta')d \theta'\in
 {\cal Q}(\nu;k)$. 
\newline\noindent
Then all solutions of {\rm (\ref{eq:5.0})} are in ${\cal Q}(\nu,\nu_{k+1};k+1)$
where $\nu_{k+1}=\bar{c}$ mod $1$ and $\nu_{k+1}\in[0,1)$. }

\vspace*{.15in}
\noindent {\em Proof:} From (c) there is an integer $l$ such that
$c(\theta) = \bar{c}+\tilde{c}(\theta)
=\nu_{k+1} + l+\tilde{c}(\theta)$ and by the same calculation that
leads to (\ref{eq:2.11}) the principal solution matrix
for (\ref{eq:5.0}) is
\begin{eqnarray}
&&\Phi(\theta) = 
V(\theta) \exp \left({\cal J}(\int\limits^{\theta}_0 \tilde{c}(\theta')
d \theta'
+l\theta)\right)
 \exp \left({\cal J}\nu_{k+1}\theta \right)V^T(0) \; .
\nonumber
\end{eqnarray}
Clearly, $\Phi(\theta)\in {\cal Q}(\nu,\nu_{k+1};k+1)$ and thus all solutions
are quasiperiodic as stated. \hfill $\Box$ 

\vspace*{.15in}

However as pointed out in Section 4, even if conditions a) and b) are 
fulfilled 
condition c) might still be problematic since the integral of
a zero--mean quasiperiodic function may not be quasiperiodic.

To continue our discussion of the quasiperiodic structure of the 
principal solution matrix defined in (\ref{eq:5.02}) and its potential
representation in (\ref{eq:5.01}) we
elaborate on  the definition of UPF and UPR from Remark 2 in Section 2.

\vspace*{.15in}

\noindent{\bf Definition 5.2} (UPF and UPR): 
Consider (\ref{eq:5.02}) for a fixed $\phi_0$ on a fixed torus $J_0$. If
$\nu\in [0,1)$ and $U \in SO(3)$ exist such 
that the principal solution matrix
at $\phi_0$ can be written as in (\ref{eq:5.01}), then the
orthonormal coordinate system represented by $U$ is called a UPF (uniform
precession frame) at $\phi_0$ and $\nu=:\nu_s(U)$ is called the UPR 
(uniform precession rate) for $U$. 
If $U\in {\cal Q}(1,\omega;d+1)$, 
then it is called {\em proper}.
A UPR is called {\em proper} if it corresponds to a proper UPF.
The set of all proper UPF{\small s} at $\phi_0$ is denoted by
${\mathfrak U}(\phi_0)$ and the set of all proper UPF{\small s} on the torus
is given by ${\mathfrak U}:=\bigcup_{\phi_0\in{\mathbb R}^d}
{\mathfrak U}(\phi_0)$. In addition we denote the set of all proper 
UPR\,{\small s} 
at $\phi_0$ by $\Xi(\phi_0)$ and then the set of all proper UPR\,{\small s} on
the torus is given by $\Xi:=\bigcup_{\phi_0\in{\mathbb R}^d}\Xi(\phi_0)$.
\hfill $\Box$ 

\vspace*{.15in}

\noindent
~For further background material see Remarks 1 and  2 in Section 2. 
Recall that in Definition 5.2 ``on a fixed torus $J_0$'' means that the orbital
tune has the value $\omega(J_0)$ and that the spin motion is
characterized by the function ${\cal A}(\cdot, \cdot, J_0)$.
Recall also that we often suppress the symbol $J_0$.
~From the discussion in Section 2 we know that UPF{\small s} always exist. 
However the existence of a {\em proper} UPF imposes additional constraints.
~For example Proposition 5.1 gives rise to the existence
of a proper UPF for (\ref{eq:5.02}) (and a proper UPR) but the conditions are
rather severe.
If a proper UPF $U$ at $\phi_0$ exists, then clearly
$\Phi(\cdot;\phi_0)\in{\cal Q}(1,\omega,\nu_s(U);d+2)$. However it is 
unknown if the converse holds although this is a reasonable conjecture.
Note that for each proper UPR there is an infinite number of proper 
UPF{\small s}:
if $U$ is a proper UPF at $\phi_0$ then 
$U \exp({\cal J} y)$ is a proper UPF at $\phi_0$ if $y$ is a real constant.

It is easy to see that if a proper UPF $U$ exists at a given $\phi_0$, 
then for all
$\nu\in[0,1)$ such that 
$\nu = \pm \nu_s(U) +  j_0 + j \cdot \omega$, where
$(j_0,j)\in {\mathbb Z}\times
{\mathbb Z}^d$, $\tilde U(\theta):= 
\left[ {  u}^1(\theta), \pm{  u}^2(\theta), 
\pm{  u}^3(\theta) \right] \\
\cdot \exp(-{\cal J} (j_0 +j \cdot\omega)\theta)$ is 
a proper UPF at that $\phi_0$ with $\nu_s(\tilde U)=\nu$. 
This motivates the definition of an equivalence relation where 
$\nu_1$ and $\nu_2$ in $[0,1)$ are said to be equivalent - and
we write $\nu_1\sim \nu_2$ - iff there exist
$(\varepsilon, j_0,j)\in\lbrace -1,1\rbrace\times{\mathbb Z}\times
{\mathbb Z}^d$ such that $\nu_2 = 
\varepsilon \nu_1 +  j_0 + j \cdot \omega$.
The equivalence relation partitions the interval into equivalence
classes $\lbrack\nu\rbrack:=\lbrace\mu\in[0,1): 
\mu\sim\nu\rbrace$ such that
$\lbrack\nu_1\rbrack=\lbrack \nu_2\rbrack$ iff $\nu_1\sim \nu_2$ and 
$\lbrack \nu_1\rbrack\cap \lbrack \nu_2\rbrack=\emptyset$ iff 
$\nu_1\not\sim \nu_2$.
We note that if $\omega$ has one irrational component
then each $\lbrack \nu\rbrack$ is a dense subset of $[0,1)$.
To see this suppose that $\omega_1$ is irrational.
Then $\varepsilon \nu +  j_1\omega_1\; mod \;1$ is dense as $j_1$ varies over
${\mathbb Z}$ so that $\mu = \varepsilon \nu +  j_0 + j\cdot\omega$ is dense
in $[0,1)$ as $j_0$ and $j$ vary.

~From the above motivation for the  definition of equivalence it is clear that
if $\nu_1\in\Xi(\phi_0)$ then the equivalence $\nu_1\sim \nu_2$ implies  
$\nu_2\in\Xi(\phi_0)$, i.e.\ $\lbrack \nu_1\rbrack\subset\Xi(\phi_0)$.
Now suppose that $U_1$ and $U_2$ are in
${\mathfrak U}(\phi_0)$ and that $\nu_s(U_1)= \nu_1$. Then
$\exp ({\cal J}\nu_s(U_2)\theta)=
U_2^T(\theta)U_1(\theta)\exp ({\cal J}\nu_s(U_1)\theta) \\
\cdot U_1^T(0) U_2(0)$
so that the l.h.s. is in ${\cal Q}(1,\omega,\nu_s(U_1);d+2)$. Thus 
$\nu_s(U_2)=j_0 + j\cdot\omega+j_{d+2}\nu_s(U_1)$ by Lemma 4.3e.
It is plausible that $j_{d+2}=\pm 1$. If that is so, then 
$\nu_1 = \nu_s(U_1)\sim\nu_s(U_2)$. Then  since
$U_2$ is an arbitrary member of ${\mathfrak U}(\phi_0)$,
$\Xi(\phi_0) \subset  \lbrack \nu_1\rbrack$.
In fact this is the case, and the joint 
conditions $\lbrack \nu_1\rbrack\subset\Xi(\phi_0)$ and 
$\Xi(\phi_0) \subset  \lbrack \nu_1\rbrack$ 
can be embodied in a theorem.

\vspace*{.15in}

\setcounter{theorem}{2}
\begin{theorem} If ~$\Xi(\phi_0)$ is nonempty with an element $\nu$, then
$\lbrack \nu\rbrack=\Xi(\phi_0)$.
\end{theorem}

Although all proper UPR\,{\small s} are equivalent at a given $\phi_0$ on the
torus $J_0$, it is not true in general that all proper UPR\,{\small s}
on the torus are equivalent. But if they are, we then say that the torus is 
well--tuned
and we call any $\nu\in\Xi$ a spin tune. Since this situation is central to 
this paper, we delay the proof of Theorem 5.3 in order to formalize
this definition.

\vspace*{.15in}

\noindent{\bf Definition 5.4} (well--tuned): 
A torus is said to be {\em well--tuned} iff the following two conditions hold:
\newline\noindent
a) $\Phi(\cdot;\phi_0)$ has a proper UPF
for each $\phi_0\in{\mathbb R}^d$, i.e.\ each 
${\mathfrak U}(\phi_0)$ is nonempty.
\newline\noindent
b) Let $U$ and $\hat{U}$ be in ${\mathfrak U}$, then $\nu_s(U)\sim
\nu_s(\hat{U})$. 
Thus for every $\nu\in\Xi$, $\lbrack\nu\rbrack \supset \Xi$
~and, by Theorem 5.3, $\lbrack\nu\rbrack \subset \Xi$ so that then,
$\lbrack\nu\rbrack = \Xi$.

A torus that is not well--tuned is called  {\em ill--tuned}. 
\hfill $\Box$ 

\vspace*{.15in}

\noindent
Note that by Theorem 5.3 a torus is well--tuned iff the
$\Xi(\phi_0)$ have an element in common. This criterion is very convenient
and we will use it for example in the proof of Theorem 6.3a.
Note also that $\Xi(\phi_0)$  always contains at most countably many elements.
In particular, if a torus is well--tuned then $\Xi$ contains at most
countably many elements.

\vspace*{.15in}

\noindent{\bf Definition 5.5} (spin tune): 
Let a torus $J_0$ be well--tuned, then each element of 
$\Xi$ is called a {\em spin tune}. Thus for each spin tune $\nu$, 
$\lbrack\nu\rbrack = \Xi$.
\hfill $\Box$ 

\vspace*{.15in}

To prove Theorem 5.3 we need the following lemma.

\vspace*{.15in}

\setcounter{lemma}{5}
\begin{lemma} 
Let $\exp(ic_0\theta)$, where $c_0$ is a real constant, be in 
${\cal Q}(\nu;k)$ with $k\geq 1$.
Then there exists $n\in{\mathbb Z}^k$
such that $c_0 = n\cdot\nu$. Moreover $n$ is unique if $\nu$ is nonresonant.
\end{lemma}

\vspace*{.15in}
{\em Proof:} First of all we observe that 
\begin{eqnarray}
&& \Lambda(\exp(ic_0\, \cdot))=\lbrace c_0 \rbrace \; .
\label{eq:6.500}
\end{eqnarray}
Also, by Lemma 4.3e we see that $\Lambda(\exp(ic_0\, \cdot))
\subset \lbrace m\cdot\nu :m\in {\mathbb Z}^k\rbrace$.  Hence by
(\ref{eq:6.500}):
\begin{eqnarray}
&& c_0 \in \lbrace m\cdot\nu :m\in {\mathbb Z}^k\rbrace \; .
\label{eq:6.501}
\end{eqnarray}
Thus $n\in{\mathbb Z}^k$ exists such that $c_0 = n\cdot\nu$. Clearly 
$n$ is unique if $\nu$ is nonresonant.
\hfill $\Box$

\vspace*{.15in}
\noindent We can now prove Theorem 5.3.

\vspace*{.15in}
\noindent
{\em Proof of Theorem 5.3:}
Consider 
$U=: \left[{u}^1, ~{u}^2, ~{u}^3 \right]$ 
and 
$\hat{U}=: \left[\hat{u}^1,~\hat{u}^2,~\hat{u}^3 \right]$, both in 
${\mathfrak U}(\phi_0)$, 
and define the smooth functions 
$g_\pm:{\mathbb R}\rightarrow {\mathbb C}$ by
\begin{eqnarray}
&& g_\pm:=({u}^1\pm i{u}^2)\cdot 
(\hat{u}^1+ i \hat{u}^2) \; .
\label{eq:6.401} 
\end{eqnarray}
It follows from (\ref{eq:2.32}) that
\begin{eqnarray}
&& \dot{g}_\pm=i\biggl(\nu_s(\hat{U})\pm \nu_s(U)\biggr)g_\pm \; ,
\nonumber 
\end{eqnarray}
so that
\begin{eqnarray}
&& g_\pm(\theta)=\exp\biggl(i(\nu_s(\hat{U})\pm \nu_s(U))
\theta\biggr)
g_\pm(0) \; .
\label{eq:6.400} 
\end{eqnarray}
Because $U$ and $\hat{U}$ are in
${\cal Q}(1,\omega;d+1)$,  (\ref{eq:6.401}) ensures
that $g_\pm$ are also in ${\cal Q}(1,\omega;d+1)$.
Now, if $g_+$ is not the zero function
then by (\ref{eq:6.400}) $\exp(i(\nu_s(\hat{U}) + \nu_s(U))\theta)$ 
is in ${\cal Q}(1,\omega;d+1)$ and by
Lemma 5.6 
 $\nu_s(\hat{U}) + \nu_s(U) = 
m\cdot(1,\omega)$ for some
$m\in{\mathbb Z}^{d+1}$.
Thus either $g_+=0$ or $m\in{\mathbb Z}^{d+1}$
exist
such that $\nu_s(\hat{U}) + \nu_s(U) =  
m\cdot(1,\omega)$. 
Similarly either $g_-=0$ or 
$m\in{\mathbb Z}^{d+1}$ exist
such that $\nu_s(\hat{U})-\nu_s(U) =  
m\cdot(1,\omega)$.
By definition either $g_+$ or
$g_-$ is different from the zero function, since  
otherwise ${u}^j\cdot \hat{u}^k=0$ for ($j,k=1,2$) which is obviously false. 
Thus $m\in{\mathbb Z}^{d+1}$ 
exist such that either
$\nu_s(\hat{U}) + \nu_s(U) =  m\cdot(1,\omega)$ or
$\nu_s(\hat{U})-\nu_s(U) =  m\cdot(1,\omega)$. 
Hence $\varepsilon\in\lbrace{-1, 1\rbrace}$ exists such that 
$\nu_s(\hat{U}) =\varepsilon \nu_s(U) +  
m \cdot (1,\omega)$, i.e.\ $\Xi(\phi_0)\subset\lbrack\nu_s(U)\rbrack$.
But by the remarks before Theorem 5.3 we also have that
$\Xi(\phi_0)\supset\lbrack\nu_s(U)\rbrack$.
\hfill $\Box$

\vspace*{.15in}

\noindent{\bf Remarks:}
\begin{itemize}
\item[(1)] For the case of $J = 0$ and arbitrary $\phi_0$ each
Floquet frequency  $\nu_0$ is in $\Xi(\phi_0)$ 
(see Remark 2 in Section 3).
We will see in Remark 4 of Section 6 that the torus $J = 0$
is well--tuned so that every Floquet frequency $\nu_0$ is a spin tune.
\item[(2)] As we will see in Remarks 13
and 14 in Section 6, matrices ${\cal A}(\theta,\phi)$
can be found with which
there are tori where ${\Xi}(\phi_0)\neq\emptyset$
for every $\phi_0$ but which are not well--tuned.
\item[(3)] Consider a well--tuned torus. We say that the torus is
on a ``spin--orbit resonance'' 
if $0 \in \Xi$. Thus on  spin--orbit resonance the set of spin tunes
reads as \\
$\Xi=\lbrace \omega_s \in[0,1):\omega_s = 
 j_0 + j \cdot \omega; ~j_0 \in {\mathbb Z},
j \in {\mathbb Z}^d\rbrace$.
~For $d=3$ the spin--orbit resonance condition amounts to 
(\ref{eq:1.5}).
In general the condition takes the form
\begin{eqnarray}
\omega_s = m_{0} + m_{1} ~\omega_{1} + \cdots~  + m_{d}~ \omega_{d}
\label{eq:5.00}
\end{eqnarray}
and the order of the resonance is $|m_{1}| + \cdots +  |m_{d}|$.
{\em Thus on a well--tuned torus, if one $\phi_0$ 
has resonant spin motion
then all $\phi_0$ have resonant spin motion}. 
Of course the same is true for nonresonant spin motion.
This is a key aspect of being well--tuned: 
spin--orbit resonance is not defined in terms of a spin frequency 
$\omega_s$ that 
varies with $\phi_0$ in the sense that $[\omega_s]$ is dependent on $\phi_0$. 
An important consequence of this definition of resonance is  
presented in Theorem 6.4.
Clearly, the order of a resonance 
depends on the chosen $\omega_s$ and that in turn depends on the chosen
UPF. Thus the order of a spin--orbit resonance is frame dependent. 
However, in practice the order is fixed by choosing a ``preferred'' spin 
tune
and the corresponding UPF for every $J$. The way to do this will be explained 
in Section 10 where some
numerical results on spin--orbit resonance are also mentioned.
A numerical method for  computing $\Xi$ is outlined in Section 9.
\item[(4)]
If one component of $\omega$ is irrational and if the torus is well--tuned,
then from the previous
discussion the sets $\Xi$ and $\lbrack 0\rbrack$ are countable dense subsets 
of 
$[0,1)$. Thus even though the complement of $\Xi$ is uncountable and
a number chosen at random from $[0,1)$ is unlikely to be a spin tune,
$\Xi$ contains spin tunes arbitrarily close to any number in this interval.
This simply implies that every spin tune (in particular the preferred 
spin tune) is close to a number in the 
resonance set $\lbrack 0\rbrack$ so that
the system is always arbitrarily near to a high order resonance.
Although the set $\lbrack 0\rbrack$ is a dense subset of $[0,1)$,
the subset of low order spin--orbit resonances is not dense in $[0,1)$.
Thus the closeness of a preferred spin tune $\omega_s$ to a number in 
the set $\lbrack 0\rbrack$
does not mean that $\omega_s$ is near to a low order spin--orbit resonance.
\item[(5)]
The set $\Xi_J$ of proper UPR\,{\small s} on the torus $J$ is a function 
of $J$ which is uniquely determined by the $J$ dependence of $\cal A$ and
the orbital tune $\omega$.
Of course, if in addition, $\cal A$ depends on extra parameters like the 
beam energy, then the set $\Xi_J$ is also a  function of these parameters.
Thus if all tori are well--tuned the spin tunes $\omega_s$ will vary
with $J$ and any extra parameters. 
Therefore the distance 
$|\omega_s - m_{0} - m_{1} ~\omega_{1} - \cdots~  - m_{d}~ \omega_{d}|$
from the spin--orbit resonance condition (\ref{eq:5.00}) 
depends on $J$ and any extra parameters. 
\item[(6)] Consider a torus for which ${\mathfrak U}(\phi_0)\neq\emptyset$
for each $\phi_0$ and for each $\phi_0$ choose a single proper UPF
$U(\cdot;\phi_0)$. The corresponding UPR\,{\small s} $\nu_s(U(\cdot;\phi_0))$
define a function of $\phi_0$. 
If the torus is well--tuned, then $\nu_s(U(\cdot;\phi_0))$
can be chosen to be a constant.
However for an ill--tuned torus,
$\nu_s(U(\cdot;\phi_0))$ usually has to vary with
$\phi_0$. This nonconstant function can nevertheless be
continuous and we present examples in Remarks 13 and 14 in Section 6.
However, because of the freedom to choose the 
proper UPR\,{\small s}, $\nu_s(U(\cdot;\phi_0))$, this function is usually very
irregular and discontinuous.
\item[(7)]
If a torus is ill--tuned, then in domains of $\phi_0$ where 
$\nu_s(U(\cdot;\phi_0))$ varies with $\phi_0$,  
it is likely that there are some
$\phi_0$\,{\small s} 
for which the resonance condition $\nu_s(U(\cdot;\phi_0)) = 
m_0+m\cdot\omega$ holds.
\item[(8)] If the principal solution matrix
$\Phi$ can be written as
\begin{eqnarray}
\Phi(\theta;\phi_0) =
 p(\theta;\phi_0)\exp\biggl(B(\phi_0)\nu(\phi_0)\theta\biggr) ,
\label{eq:2.100}
\end{eqnarray}
where $\nu(\phi_0) \in [0,1)$, $B(\phi_0)$ is a real
skew--symmetric matrix with spectrum \\
$\lambda(B(\phi_0)) 
= \lbrace i,\, -i,\, 0 \rbrace$ and the matrix $p(\cdot;\phi_0)$ is
in ${\cal Q}(1,\omega;d+1)$, 
then $p,B,\nu$ will be called {\em generalized Floquet parameters} 
at $\phi_0$.
In particular $\nu(\phi_0)$ will be called a {\em generalized 
~Floquet frequency}. By definition the generalized Floquet parameter 
$p(\theta;\phi_0)$ is smooth
in $\theta$ and 
$p\in SO(3)$,  with $p(0;\phi_0)=I$. 
Because $p(\cdot;\phi_0)$ has the same periodicities as $A(\cdot;\phi_0)$,
the $p,B,\nu$ are indeed  generalizations of the 
~Floquet parameters introduced in Section 3. 
Note that (\ref{eq:2.100}) generates a large class of $A$\,{\small s}, namely
$A(\theta;\phi_0)=(\partial\Phi/\partial\theta)
(\theta;\phi_0)\Phi^T(\theta;\phi_0)$, where
$\Phi$ is given by (\ref{eq:2.100}).

If $p(\theta;\phi_0), B(\phi_0), \nu(\phi_0)$ are generalized 
~Floquet parameters at $\phi_0$, then from Lemma 2.1b
a $W(\phi_0)\in SO(3)$ exists such that
$W(\phi_0){\cal J}W^T(\phi_0)=B(\phi_0)$, whence  
$\Phi(\theta;\phi_0) = p(\theta;\phi_0) W(\phi_0) e^{{\cal J} \nu(\phi_0) 
\theta} (p(0;\phi_0) W(\phi_0))^T$. Thus 
$p(\theta;\phi_0) W(\phi_0)$ is a proper 
UPF at $\phi_0$ with UPR 
$\nu(\phi_0)$. 

Conversely, if $U(\theta;\phi_0)$ is a proper UPF at
$\phi_0$, then by
(\ref{eq:2.20}), $\Phi$ can be written as in (\ref{eq:2.100}) with 
$p,B,\nu$ given by 
\begin{eqnarray}
&& p(\theta;\phi_0) :=  U(\theta;\phi_0)U^T(0;\phi_0) \; ,  \qquad
B(\phi_0) := U(0;\phi_0) {\cal J}U^T(0;\phi_0) \; , \nonumber\\
&&\nu(\phi_0):=\nu_s(U(\cdot;\phi_0)) \; .
\nonumber
\end{eqnarray}
Therefore the UPR corresponding to a proper UPF is a generalized Floquet
frequency and is thus a frequency
additional to those in $A$
in analogy with the Floquet frequency which
emerges in the case where $A$ is 
$2\pi$--periodic.
We conclude that the set of generalized Floquet frequencies at an arbitrary 
$\phi_0$ is identical with the set $\Xi(\phi_0)$.

Note that the condition $\Xi(\phi_0)\neq\emptyset$ at some $\phi_0$ 
implies that generalized Floquet parameters exist at that $\phi_0$.
This can be viewed as a generalized Floquet theorem.
Moreover, if the torus is well--tuned,
then for every $\phi_0$ generalized Floquet parameters exist
and the set of generalized Floquet frequencies is the same at every $\phi_0$
and is identical with the set $\Xi$ of spin tunes.
\item[(9)] 
The spin trajectories at a given $\phi_0$
are quasiperiodic if $\Xi(\phi_0)\neq\emptyset$
but 
there are $A(\cdot;\phi_0)$ in ${\cal Q}(1,\omega;d+1)$
which give rise to nonquasiperiodic  spin trajectories (see Section 8).
\end{itemize} 
\section{The Invariant Spin Field and the Quasiperiodicity of the Spin Motion}
\setcounter{equation}{0}
In this section we present an application of the concepts introduced
in Section 5.  In particular we consider the important situation where
for a given ${\cal A}(\cdot,J)$, a torus admits an invariant spin field, a 
central 
object in the
theory of polarization whose relevance was explained in the
Introduction. It was pointed out there that the invariant spin field
is a solution to the T--BMT
equation along particle orbits and that it
is $2\pi$--periodic in $\theta$ and $\phi$. We now
return to it and show, among other things, that if the torus is off
orbital resonance, the existence of a nonunique invariant spin field
implies that the system is on spin--orbit resonance. We begin by studying
general spin fields.

Let ${\cal S}(\theta,\phi)$ be a field such that
$S(\theta;\phi_0):={\cal S}(\theta,\phi_0+\omega\theta)$
is the spin of the particle which starts with phase  $\phi_0$ at $\theta=0$.
By (\ref{eq:2.1}) we must have 
\begin{eqnarray}
&&\frac{d}{d\theta}S(\theta;\phi_0)=
{\cal D}{\cal S}(\theta,\omega\theta+\phi_0)
={\cal A}(\theta,\omega\theta+\phi_0){\cal S}(\theta,\omega\theta+\phi_0) 
\nonumber
\end{eqnarray}
and thus the evolution of the field is described by the 
partial differential equation
\begin{eqnarray}
\!\star\!\star\!\star &&  \qquad \qquad \qquad \qquad  \qquad \qquad{\cal D} 
{\cal S}= {\cal A}(\theta,\phi) {\cal S} \; . \qquad  \qquad \qquad  \qquad 
\qquad  \qquad \qquad  \qquad \qquad
\label{eq:6.1}
\end{eqnarray}
Here, ${\cal D}{\cal S}(\theta,\phi) = D_1{\cal S}(\theta,\phi) +
 \omega\cdot\nabla_\phi {\cal S}(\theta,\phi)$, where
$D_k$ will denote the derivative with respect to the $k$-th argument, 
be it scalar or multicomponent.
Consider the {\em generalized  principal solution matrix} 
$\varphi(\theta,\phi)$, defined by
\begin{eqnarray}
{\cal D}\varphi = {\cal A}(\theta,\phi)\varphi \; , \qquad 
\varphi(0,\phi) = I \; ,
\label{eq:6.0001}
\end{eqnarray}
then the general solution of (\ref{eq:6.1}) is given by
\begin{eqnarray}
{\cal S}(\theta,\phi) =\varphi(\theta,\phi){\cal S}(0,\phi-\omega\theta) \; .
\label{eq:6.0000}
\end{eqnarray}
This can be seen directly:  since ${\cal D}$ is a derivation, 
${\cal D}{\cal S}(\theta,\phi)
=\biggl({\cal D}\varphi(\theta,\phi)\biggr){\cal S}(0,\phi-\omega\theta)+
\varphi(\theta,\phi){\cal D}\biggl({\cal S}(0,\phi-\omega\theta)\biggr)$ 
and ${\cal D}\biggl({\cal S}(0,\phi-\omega\theta)\biggr)=0$. 
Equations (\ref{eq:6.1}) and (\ref{eq:6.0001}) are amenable to solution by the
method of characteristics. In this case 
the characteristics for (\ref{eq:6.1}) are just the
spin trajectories $S(\cdot,\phi_0)$ and
the characteristics for (\ref{eq:6.0001}) are just the
$\Phi(\cdot,\phi_0)$.

Now suppose, as in Section 2, that we know one solution of 
(\ref{eq:6.1}). Let ${\cal V}(\theta,\phi)$ be in $SO(3)$ such that its 
third column is this solution and make the 
transformation
$\varphi\rightarrow\psi$ defined by
\begin{equation}
\varphi = {\cal V} \psi.
\label{eq:6.0003}
\end{equation}
Then
\begin{equation}
 {\cal D}\varphi =({\cal D} {\cal V}) \psi +
{\cal V}{\cal D}\psi = {\cal A}{\cal V} \psi 
\nonumber
\end{equation}
so that
\begin{eqnarray}
{\cal D} \psi = {\cal C}_{\cal V}(\theta,\phi){\cal J}\psi \; , \qquad
 \psi(0,\phi)={\cal V}^T(0,\phi) \; ,
\label{eq:6.0004}
\end{eqnarray}
where, in analogy to Section 2, ${\cal C}_{\cal V}$ is defined by
\begin{eqnarray}
{\cal V}^T({\cal A}{\cal V}- {\cal D}{\cal V})={\cal C}_{\cal V}{\cal J} \; ,
\label{eq:6.0005}
\end{eqnarray}
and thus
\begin{eqnarray}
{\cal C}_{\cal V}=-(1/2)\; Tr\lbrack\,{\cal J}\,({\cal V}^T {\cal A} {\cal V}
-{\cal V}^T {\cal D} {\cal V})\rbrack \; ,
\label{eq:6.201}
\end{eqnarray}
\begin{eqnarray}
{\cal D} {\cal V} = {\cal A} {\cal V} -{\cal C}_{\cal V}{\cal V} {\cal J} \; 
\label{eq:6.201b}
\end{eqnarray}
which is analogous to (\ref{eq:2.8}). \\
Equation (\ref{eq:6.0004}) also can be solved by the method of characteristics
giving
\begin{eqnarray}
\psi(\theta,\phi) = 
\exp \left({\cal J}\int\limits^{\theta}_0 {\cal C}_{\cal V}
(\theta',\omega \theta'+\phi-\omega\theta)d\theta' \right)
{\cal V}^T(0,\phi-\omega\theta) \; ,
\label{eq:6.0006}
\end{eqnarray}
as is easily checked by direct substitution in (\ref{eq:6.0004}).
The generalized  principal solution matrix
thus becomes
\begin{eqnarray}
\varphi(\theta,\phi) = {\cal V}(\theta,\phi)
\exp \left({\cal J}\int\limits^{\theta}_0 {\cal C}_{\cal V}
(\theta',\omega \theta'+\phi-\omega\theta)d\theta' \right)
{\cal V}^T(0,\phi-\omega\theta) \; .
\label{eq:6.0007}
\end{eqnarray}
The principal solution matrix of (\ref{eq:5.02}) 
is easily constructed from the generalized  principal solution matrix
as the following theorem
shows.

\vspace*{.15in}

\setcounter{theorem}{0}
\begin{theorem} Let ${\cal V}(\theta,\phi)$ be a smooth
$SO(3)$ matrix on a fixed torus, with third column satisfying 
{\rm (\ref{eq:6.1})}. Then the 
principal solution matrix
at $\phi_0$ on the torus is given by
\begin{eqnarray}
\!\star\!\star\!\star && \qquad \qquad \qquad \Phi(\theta;\phi_0) = 
\varphi(\theta,\omega\theta+\phi_0) = 
{\cal V}(\theta,\omega\theta+\phi_0) \qquad \qquad \qquad \qquad \qquad 
\qquad \nonumber \\
&& \qquad \qquad \qquad \qquad \qquad \cdot \exp \left({\cal J}
\int\limits^{\theta}_0 {\cal C}_{\cal V}
(\theta',\omega \theta'+\phi_0)d\theta' \right)
{\cal V}^T(0,\phi_0) \; .
\label{eq:6.0008}
\end{eqnarray}
\end{theorem}
\noindent{\em Proof:} Clearly $\Phi(0;\phi_0)=I$. Moreover,
$\Phi$ satisfies (\ref{eq:5.02}) because
$(d/d\theta){\cal V}(\theta,\omega\theta+\phi_0)=
{\cal A}{\cal V}-{\cal C}_{\cal V}{\cal V}{\cal J}$
and $(d/d\theta)
\int\limits^{\theta}_0 {\cal C}_{\cal V}
(\theta',\omega \theta'+\phi_0)d\theta' ={\cal C}_{\cal V}
(\theta,\omega\theta+\phi_0)$.
\hfill $\Box$ 

\vspace*{.15in}

\noindent We now give several definitions. In particular we 
give a formal definition of the invariant spin field.

\vspace*{.15in}

\noindent{\bf Definition 6.2}  (ISF,IFF): 
Consider a fixed torus. 

\noindent a) A field ${\cal S}(\theta,\phi)\in{\mathbb R}^3$ 
is said to be a {\em spin field} for (\ref{eq:2.1}) 
iff ${\cal S}$ 
is  smooth in $(\theta, \phi)$, $2\pi$--periodic in $\phi$ and
if it satisfies the partial differential equation (\ref{eq:6.1}).
A spin field ${\cal S}$ is called an {\em invariant spin field} 
(ISF) for (\ref{eq:2.1}) if it is also $2\pi$--periodic in $\theta$
and $|{\cal S}|=1$.

\noindent b) An $SO(3)$ matrix, ${\cal V}(\theta,\phi)$, is called
a {\em frame field} for (\ref{eq:2.1}) iff
${\cal V}$ is smooth in $(\theta, \phi)$, $2\pi$--periodic in $\phi$ and if 
its 
third column is a spin field.
A frame field is called an {\em invariant frame field} (IFF) 
for (\ref{eq:2.1}) iff it is also $2\pi$--periodic in $\theta$. Thus the
third column of an IFF is an ISF.
A {\em uniform} IFF, ${\cal V}$, is such that the function 
${\cal C}_{\cal V}$ as defined by (\ref{eq:6.201}) is
constant (independent of $(\theta,\phi)$) and in $[0,1)$.
\hfill $\Box$ 

\vspace*{.15in}

\noindent Note that in (\ref{eq:6.0007}) both ${\cal V}$ and $\varphi$  
are frame fields.

\vspace*{.15in}

\setcounter{theorem}{2}
\begin{theorem} 
a) Consider a fixed torus. 
If a uniform IFF ${\cal V}$ exists then, for every $\phi_0$,
$U(\cdot;\phi_0)$, defined by
$U(\theta;\phi_0):={\cal V}(\theta,\phi_0+\omega\theta)$, is a proper UPF
at $\phi_0$ with $\nu_s(U(\cdot;\phi_0))={\cal C}_{\cal V}$.
Moreover the torus is well--tuned and ${\cal C}_{\cal V}$ is a spin tune.

\noindent b) Consider a fixed torus. 
If ${\cal V}$ is a frame field
then it satisfies the partial differential equation 
{\rm(\ref{eq:6.201b})}, where ${\cal C}_{\cal V}$ is given by
{\rm (\ref{eq:6.201})}.
\end{theorem}
\noindent{\em Proof of Theorem 6.3a:} 
~From Theorem 6.1 it follows that
$U(\cdot;\phi_0)$ is a UPF at $\phi_0$ with
UPR $\nu_s(U(\cdot;\phi_0))={\cal C}_{\cal V}$. Of course,
$U(\cdot;\phi_0)$ is in ${\cal Q}(1,\omega;d+1)$ so that
the UPF $U(\cdot;\phi_0)$ is proper. Because ${\cal C}_{\cal V}$ is contained 
in each $\Xi(\phi_0)$ we conclude (recall the comment after Definition 5.4)
that the torus is well--tuned and that 
${\cal C}_{\cal V}$ is a spin tune.
\hfill $\Box$ 

\noindent{\em Proof of Theorem 6.3b:} 
With
\begin{eqnarray}
&& {\cal V}=: 
\left[{  \mathfrak v}^1, ~{  \mathfrak v}^2, ~{  \mathfrak v}^3 \right] \; ,
\label{eq:6.3} 
\end{eqnarray}
\noindent 
and
\begin{eqnarray}
&& {\mathfrak w}^j:=({\cal D}-{\cal A}){  \mathfrak v}^j \; ,
\end{eqnarray}
and since $\cal V$ is in SO(3) and ${\cal S}\equiv 
{  \mathfrak v}^3$ satisfies (\ref{eq:6.1}), we have:
\begin{eqnarray}
&&{\mathfrak w}^1 ={ ( \mathfrak w}^1\cdot{  \mathfrak v}^2) {  \mathfrak v}^2
\; , \qquad
{\mathfrak w}^2 =-{ ( \mathfrak w}^1\cdot{  \mathfrak v}^2) {  \mathfrak v}^1
\; , \qquad
{  \mathfrak w}^3 = 0 \; ,
\nonumber
\end{eqnarray}
so that
\begin{eqnarray}
{\cal D} {\cal V} - {\cal A} {\cal V} =({  \mathfrak w}^1\cdot{  \mathfrak v}^2
){\cal V} {\cal J} \; .
\label{eq:6.04}
\end{eqnarray}
Because $Tr\lbrack\,{\cal J}^2\,\rbrack=-2$, it follows from 
(\ref{eq:6.04}) that ${  \mathfrak w}^1\cdot{  \mathfrak v}^2=
(1/2)\; Tr\lbrack\,{\cal J}\,({\cal V}^T {\cal A} {\cal V}
-{\cal V}^T {\cal D} {\cal V})\rbrack$. 
\hspace*{16.2cm}$\Box$ 

\vspace*{.15in}

\noindent
We now make some further remarks on IFF{\small s} and ISF{\small s}.



\noindent{\bf Remarks:}
\begin{itemize}
\item[(1)] The name ``invariant frame field'' (IFF) is chosen so as to
reflect the fact that like the ISF it is  $2\pi$--periodic in $\theta$. 
\item[(2)] If ${\cal V}$ is an IFF, then we call the 
${\mathfrak v}^1,{  \mathfrak v}^2$ in (\ref{eq:6.3})
pseudo--$u^1,u^2$--axes. They are used in the program SPRINT
for the numerical calculation of the spin tune
\cite{hh96,epac98,spin98,gh2000,mv2000,spin2000,hvb99}.
If the IFF is uniform, then ${\mathfrak v}^1,{  \mathfrak v}^2$
are called $u^1,u^2$--axes (see also \cite{ky86,bhr92}).
An example of a  numerical calculation of $u^1,u^2$--axes can be found in  
\cite{mv2000}. See Section 10 also.
\item[(3)] From the proof of Theorem 6.3b it follows for every IFF
${\cal V}$ that (see also \cite{bhr92})
\begin{eqnarray}
{\cal C}_{\cal V}(\theta, \phi)= 
{  \mathfrak v}^1\cdot\biggl( -{\cal A}(\theta,\phi){  \mathfrak v}^2
+{\cal D}{  \mathfrak v}^2\biggr) =
\Omega \cdot {  \mathfrak v}^3
+   {  \mathfrak v}^1 \cdot ({\cal D} {  \mathfrak v}^2) 
\; .
\label{eq:6.4}
\end{eqnarray}
Thus if ${  \mathfrak v}^2$ is an ISF, then 
${\cal C}_{\cal V}=0$ and ${\cal V}$ is a uniform IFF.
\item[(4)] A uniform IFF always exists for the closed orbit so that by 
Theorem 6.3a the torus $J = 0$ is  well--tuned. This follows easily 
from the fact that in this case proper UPF{\small s} are uniform 
IFF{\small s} (see Remark 2 in Section 3).
Thus at $J = 0$ every Floquet frequency  is a spin tune by Remark 1 of 
Section 5.
Note also that for $J = 0$ any $2\pi$--periodic spin trajectory $S$ with
$|S|=1$ is an ISF.
\item[(5)]
Because the coefficients of the partial differential equation (\ref{eq:6.1}) 
are $2\pi$-periodic,
it is plausible that an ISF exists.  But as we signaled  earlier,
it remains as a mathematical challenge to prove it and in Remark 14
we will give an example where no ISF exists.
The coefficients of the partial differential equation (\ref{eq:6.201b}) are 
also $2\pi$-periodic
so that it is again plausible, but again mathematically
challenging to prove, that an IFF exists.
Note that if an ISF ${\cal S}$ exists, then an IFF ${\cal V}$
is easily constructed in analogy with
the construction of $V(\theta)$ in Section 3 if there is a $2\pi$--periodic
smooth unit vector which is nowhere parallel to ${\cal S}$ (see also
\cite{bhr92}). However, if there is no such unit vector
then it can happen that an IFF ${\cal V}$ with the third column
${\cal S}$ does not exist.
\end{itemize}

\noindent If an ISF does exist, it might not be unique. In that case 
the torus is on a  spin--orbit resonance, as shown in the following 
theorem.

\vspace*{.15in}

\setcounter{theorem}{3}
\begin{theorem} 
Consider the 
case of a fixed torus which is off orbital resonance 
and assume that
an ISF, $n$, exists. Let $n$ be nonunique, i.e.\ let
$\hat{n}$ be another ISF such that the vector product of $n$ and $\hat{n}$
is nonzero at some $(\theta,\phi)$. Then a uniform IFF exists (so that the
torus is well--tuned) and the system is on a spin--orbit resonance.
\end{theorem}

\noindent{\em Proof:}
Let $n$ and $\hat{n}$ be ISF{\small s} whose vector product is nonzero at 
some $(\theta,\phi)$ and let 
the torus be off orbital resonance.
We observe that $F:=|n\times\hat{n}|^2$ is a smooth and 
$2\pi$--periodic function
with ${\cal D} F=0$. As will be shown below, it
 follows that off orbital resonance the function $F$ is constant. 
Thus the angle between the 
ISF{\small s} is the same at all points and $m$, defined by
$m:=(n\times\hat{n})/|n\times\hat{n}|$, is an ISF, perpendicular to $n$.
Therefore ${\cal V}:=[m\times n,m,n]$ is an IFF. Due to
Remark 3 the IFF ${\cal V}$ is uniform with ${\cal C}_{\cal V}=0$.
We conclude from Theorem 6.3a that the torus is well--tuned and that
$0$ is a spin tune. In particular the torus is on a spin--orbit resonance
(see also Remark 3 in Section 5).

To complete the proof, we now consider
a smooth and $2\pi$--periodic function
${\cal F}:{\mathbb R}^{d+1}\rightarrow {\mathbb R}$ for which  ${\cal D} 
{\cal F}=0$.
We define ${\cal H}:{\mathbb R}^{d+1}\rightarrow {\mathbb R}$ via 
\begin{eqnarray}
&& {\cal H}(\theta,\phi):={\cal F}(\theta,\phi+\omega\theta) \; .
\label{eq:25} 
\end{eqnarray}
Hence ${\cal H}(\theta,\phi)$ is a smooth function $2\pi$-periodic in $\phi$
 such that $D_1{\cal H} = 0$.
Because $\cal H$ is smooth, we have
\begin{eqnarray}
&& {\cal H}(\theta,\phi)
 = {\cal H}(0,\phi) +\int_0^{\theta}\; D_1 {\cal H}(\theta_0,\phi)d\theta_0 =  
 {\cal H}(0,\phi)\; .
\label{eq:26} 
\end{eqnarray}

Because $\cal F$ is $2\pi$-periodic
in $\theta$, we obtain
\begin{eqnarray}
&& 0= {\cal F}(2\pi,\phi) - {\cal F}(0,\phi) = {\cal H}(2\pi,\phi-2\pi\omega)-
 {\cal H}(0,\phi) 
\; ,
\nonumber
\end{eqnarray}
whence by (\ref{eq:26})
\begin{eqnarray}
&& 0= {\cal H}(0,\phi-2\pi\omega) - {\cal H}(0,\phi) \; .
\label{eq:28} 
\end{eqnarray}
Thus for the Fourier coefficients 
$g_r:=(1/2\pi)^d
\int_0^{2\pi}\cdots\int_0^{2\pi} {\cal H}(0,\phi)\exp(-ir\cdot\phi)d\phi$
of ${\cal H}(0,\cdot)$  it follows  that
\begin{eqnarray}
&& g_r=\exp(2\pi i r\cdot\omega)g_r \; ,
\label{eq:29} 
\end{eqnarray}
where $r\in{\mathbb Z}^d$.
Because $(1, \omega)$ is nonresonant  $g_r$ vanishes for $r\neq 0$. 
Thus by Lemma 4.3a
${\cal H}(0,\phi)$ is constant, i.e.\ independent of $\phi$. 
Therefore by (\ref{eq:26}) ${\cal H}(\theta,\phi)$ is constant, 
i.e.\ independent of 
$\theta,\phi$ and then by (\ref{eq:25}) 
${\cal F}(\theta,\phi)$ is constant, i.e.\ independent of 
$\theta,\phi$. 
\hfill $\Box$

\vspace*{.15in}

Theorem 6.4 addresses the uniqueness 
of the ISF as well as its nonuniqueness. In particular, 
the contrapositive of Theorem 6.4 yields:
if off orbital resonance a uniform IFF  exists and if
the system is not on spin--orbit resonance, then the 
ISF is unique up to a sign.  
This behavior was predicted earlier in \cite{ky86}.

\vspace{3mm}

We now focus on the case where an IFF exists and $C_{\cal V}$ 
 is not constant.
We first define
\begin{eqnarray}
&& c_{\cal V}(\theta;\phi_0) :={\cal C}_{\cal V}(\theta,\omega\theta+\phi_0) 
\; .
\label{eq:6.8}
\end{eqnarray}
Since $c_{\cal V}(\cdot;\phi_0)\in{\cal Q}(1,\omega;d+1)$, its mean,
$\bar{c}_{\cal V}(\phi_0)$, and zero--mean part,
$\tilde{c}_{\cal V}(\theta;\phi_0)$,  exist. 
We thus have an important decomposition of $C_{\cal V}$, namely,
\begin{eqnarray}
{\cal C}_{\cal V}(\theta,\phi)=\bar{c}_{\cal V}(\phi-\omega\theta) + 
\tilde{c}_{\cal V}(\theta;\phi-\omega\theta) \; .
\label{eq:6.6}
\end{eqnarray}
Since the l.h.s. is $2\pi$--periodic, the r.h.s. is too and  in fact 
$\bar{c}_{\cal V}(\phi-\omega\theta)$ and 
$\tilde{c}_{\cal V}(\theta;\phi-\omega\theta)$ are individually 
$2\pi$--periodic as we check in Lemma 6.6a below.
~From (\ref{eq:6.6}) the integral in the exponential of 
(\ref{eq:6.0007}) is
\begin{eqnarray}
\int\limits^{\theta}_0 {\cal C}_{\cal V}
(\theta',\omega \theta'+\phi-\omega\theta)d\theta' =\theta
\bar{c}_{\cal V}(\phi-\omega\theta)
+\beta_{\cal V}(\theta,\phi) \; ,
\label{eq:6.0009}
\end{eqnarray}
where 
\begin{eqnarray}
\beta_{\cal V}(\theta,\phi):=
\int\limits^{\theta}_0 \tilde{c}_{\cal V}
(\theta'; \phi-\omega\theta)d\theta'\; .
\label{eq:6.0010}
\end{eqnarray}
Clearly ${\cal D}\beta_{\cal V}(\theta,\phi)=\tilde{c}_{\cal V}
(\theta; \phi-\omega\theta)$, which leads to consideration of
the partial differential equation
\begin{equation}
{\cal D} \alpha(\theta, \phi)  =  \tilde{c}_{\cal V}
(\theta;\phi-\omega\theta) =: {\cal \check F}(\theta,\phi).
\label{eq:6.0}
\end{equation}
Then for every solution $\alpha$ of (\ref{eq:6.0}) 
\begin{eqnarray}
\beta_{\cal V}
(\theta,\phi)= \alpha(\theta, \phi) - \alpha(0,\phi-\omega\theta) \; .
\label{eq:6.0011}
\end{eqnarray}
The existence of a $2\pi$-periodic $\alpha$ will be important below
 (note that ${\cal \check F}$ is always $2\pi$-periodic).

We now write 
\begin{eqnarray}
&& \bar{c}_{\cal V}(\phi_0) =
\nu_{\cal V}(\phi_0) + k_{\cal V}(\phi_0) \; ,
\label{eq:6.109}
\end{eqnarray}
where the integer $k_{\cal V}(\phi_0)$ is uniquely determined by
the condition $\nu_{\cal V}(\phi_0)\in
[0,1)$. The generalized  principal solution matrix
from (\ref{eq:6.0007}) now becomes
\begin{eqnarray}
&&\varphi(\theta,\phi) 
\nonumber\\
&&= {\cal V}(\theta,\phi)
\exp \biggl({\cal J}(\alpha(\theta, \phi) -\alpha(0,\phi-\omega\theta) 
+k_{\cal V}(\phi-\omega\theta)\theta
+\nu_{\cal V}(\phi-\omega\theta)\theta)\biggr)
{\cal V}^T(0,\phi-\omega\theta) \nonumber\\
&&= {\cal U}(\theta,\phi)
\exp \left({\cal J}\nu_{\cal V}(\phi-\omega\theta)\theta\right)
{\cal U}^T(0,\phi-\omega\theta)
\; ,
\label{eq:6.0012}
\end{eqnarray}
where 
\begin{eqnarray}
&&{\cal U}(\theta,\phi):=
{\cal V}(\theta,\phi)
\exp \biggl({\cal J}(\alpha(\theta, \phi)
+k_{\cal V}(\phi-\omega\theta)\theta)\biggr) \; .
\label{eq:6.0013}
\end{eqnarray}
Then the principal solution matrix
becomes
\begin{eqnarray}
\Phi(\theta;\phi_0) = U(\theta;\phi_0) 
\exp ({\cal J}\nu_{\cal V}(\phi_0) \theta) ~ U^T(0;\phi_0),
\label{eq:6.07}
\end{eqnarray}
where
\begin{eqnarray}
&& U(\theta;\phi_0) := 
{\cal U}(\theta,\omega\theta+\phi_0) \nonumber\\
&&=
{\cal V}(\theta,\omega\theta+\phi_0)
\exp \biggl({\cal J}(\alpha(\theta, \omega\theta+\phi_0) 
 +k_{\cal V}(\phi_0)\theta)\biggr) \; .
\label{eq:6.108}
\end{eqnarray}
We can now state and prove the next basic result of this section. The proof 
will depend,
in part, on Lemma 6.6b which follows later in order not to break the flow.
Theorems 6.5c-d
 use the Diophantine condition. 
\vspace*{.15in}
\setcounter{theorem}{4}
\begin{theorem} 
Consider a fixed torus $J_0$.

\noindent a) If an IFF, $\cal V$, exists
and {\rm (\ref{eq:6.0})}
has a smooth and $2\pi$--periodic solution $\alpha$, then 
$U(\cdot;\phi_0)$, defined by {\rm (\ref{eq:6.108})}, 
is a proper UPF at $\phi_0$ with UPR equal to $\nu_{\cal V}(\phi_0)$.
\newline\noindent
b) If the conditions of Theorem 6.5a hold and if the torus is 
off orbital resonance, then ${\cal U}$, defined by {\rm (\ref{eq:6.0013})}
is a uniform IFF and ${\cal C}_{\cal U}=\nu_{\cal V}$.
Moreover the torus is well--tuned and
$U(\cdot;\phi_0)$, defined for every $\phi_0$ by 
{\rm (\ref{eq:6.108})},
is a proper UPF at $\phi_0$ whose UPR is a spin tune and
$\nu_s(U(\cdot;\phi_0))={\cal C}_{\cal U}=\nu_{\cal V}$.
\newline\noindent c) Let 
$0<\tau<r-d-3$ and let ${\cal A}(\cdot,J_0)$ be in $C^r$.
If an IFF, ${\cal V}$, exists in $C^r$ and if $\omega(J_0) \in \Omega(\tau)$
then a uniform IFF exists (and thus the torus is well-tuned).

\noindent d) Let $r>2d+4$ and let ${\cal A}(\cdot,J_0)$ be in $C^r$.
If an IFF, ${\cal V}$, exists in $C^r$ for every $\omega(J_0)$ in
a Borel subset ${\cal R}$ of ${\mathbb R}^d$ then
a uniform IFF exists 
(and thus the torus is well-tuned)
for $\mu$-almost every $\omega(J_0)$ in ${\cal R}$.
\end{theorem}
\noindent{\em Proof of Theorem 6.5a:}
It is clear from (\ref{eq:6.108}) that
$U(\cdot;\phi_0)$ is in ${\cal Q}(1,\omega;d+1)$ and the result follows from
(\ref{eq:6.07}).
\hfill $\Box$

\noindent{\em Proof of Theorem 6.5b:}
~From (\ref{eq:6.0013}), ${\cal U}$ and ${\cal V}$ have the same third
column and ${\cal U}$ is smooth and $2\pi$--periodic.  Therefore
${\cal U}$ is an IFF. From (\ref{eq:6.201}) an easy calculation gives
${\cal C}_{\cal U}=\nu_{\cal V}$.  Thus by Lemma 6.6b,
${\cal C}_{\cal U}$ is constant, independent of $\phi_0$.  Hence
${\cal U}$ is a uniform IFF.
Therefore by Theorem 6.3a $U(\cdot;\phi_0)$ is a proper UPF at $\phi_0$ with
UPR $\nu_{\cal V}$ which 
is a spin tune and the torus
is well--tuned.
\hfill $\Box$

\noindent{\em Proof of Theorem 6.5c:}
From (\ref{eq:6.201}),  ${\cal C}_{\cal V}$ is in $C^{r-1}$
 since ${\cal V}$ and ${\cal A}$ are in $C^r$.
Using the fact that the torus is off orbital resonance
 (since $\omega(J)\in\Omega(\tau)$),
 we have by Lemma 6.6b that $\bar{c}_{\cal V}$ is a constant. Thus, by 
(\ref{eq:6.6}), ${\cal \check F} = {\cal C}_{\cal V} - \bar{c}_{\cal V}$
 is in $C^{r-1}$.
It follows from the condition $\tau < r-d-3$ and Lemma 4.7 that
\begin{equation}
\alpha(\theta,\phi) = 
\sum\limits_{m \in {\mathbb Z}^{d+1} \backslash \lbrace 0 \rbrace} 
\frac{1}{i m \cdot
(1,\omega)} {\cal \check F}_m \exp(i m \cdot(\theta,\phi)) \; ,
\label{eq:6.2}
\end{equation}
is smooth and $2\pi$--periodic in $\theta$ and $\phi$ and that it
satisfies (\ref{eq:6.0}), where ${\cal \check F}_m$ denotes the $m$--th Fourier coefficient
of $\cal \check F$. The claims now follow from Theorems 6.5b and 6.3a.
\hfill $\Box$

\noindent{\em Proof of Theorem 6.5d:}
The interval $(d+1,r-d-3)$ is not empty so that we pick a $\tau$ in that interval.
Because $\tau < r-d-3$ we have, by Theorem 6.5c, for
$\omega(J_0)\in(\Omega(\tau)\cap{\cal R})$ 
a uniform IFF (and thus a well-tuned torus). Because $\tau>d+1$ we
have by Lemma 4.8 that $\mu$--almost every $\omega(J_0)$ in 
${\cal R}$ is in $\Omega(\tau)\cap{\cal R}$. This proves our claim.
\hfill $\Box$

\vspace*{.15in}

\noindent We now complete the discussion by stating and proving the Lemmas 6.6a and 6.6b 
mentioned earlier in this section.

\vspace*{.15in}

\setcounter{lemma}{5}
\begin{lemma} 
a) If ${\cal V}$ denotes an IFF, then $\bar{c}_{\cal V}(\phi-\omega\theta)$ 
and $\tilde{c}_{\cal V}(\theta;\phi-\omega\theta)$
are $2\pi$--periodic in $\theta$ and in $\phi$.

\noindent b) If ${\cal V}$ denotes an IFF and if the torus is off
orbital resonance, then $\bar{c}_{\cal V},\nu_{\cal V}$ and
$k_{\cal V}$ are constant.
\end{lemma}
{\em Proof of Lemma 6.6a:} 
The periodicities in (\ref{eq:6.6})
can be demonstrated as follows: 
\begin{eqnarray}
&& \bar{c}_{\cal V}(\phi-\omega(\theta+2\pi))=
\lim_{T \rightarrow \infty} \frac{1}{T} \int_0^T 
{\cal C}_{\cal V}(\theta',\phi-\omega(\theta+2\pi)+\omega\theta') d\theta'  
\nonumber\\
&&=
\lim_{T \rightarrow \infty} \frac{1}{T} \int_0^T 
{\cal C}_{\cal V}
(\theta'-2\pi,\phi-\omega(\theta+2\pi)+\omega\theta') d\theta'  
\nonumber\\
&&
=
\lim_{T \rightarrow \infty} \frac{1}{T} \int_{-2\pi}^{T-2\pi} 
{\cal C}_{\cal V}(\theta'',\phi-\omega\theta+\omega\theta'') d\theta''
\nonumber\\
&&=\bar{c}_{\cal V}(\phi-\omega\theta)+
\lim_{T \rightarrow \infty} \frac{1}{T} \biggl(\int_{-2\pi}^0 - 
\int_{T-2\pi}^T\biggr)
{\cal C}_{\cal V}(\theta'',\phi-\omega\theta+\omega\theta'') d\theta''
\nonumber\\
&&
=\bar{c}_{\cal V}(\phi-\omega\theta) \; ,
\label{eq:6.6a}
\end{eqnarray}
where at the second equality we have used the fact that
${\cal C}_{\cal V}
(\cdot,\phi)$ is $2\pi$--periodic and at the last equality we have
used the fact that ${\cal C}_{\cal V}$ is bounded. This shows
that the first term on the r.h.s. of (\ref{eq:6.6}) is
$2\pi$--periodic in $\theta$ and thus that all three terms 
in (\ref{eq:6.6}) have this periodicity property.
That all three terms in (\ref{eq:6.6}) are $2\pi$--periodic
in $\phi$ is trivial.
\hfill $\Box$

\noindent{\em Proof of Lemma 6.6b:} 
Off orbital resonance we find, due to (\ref{eq:6.8}) and by applying
Lemma 4.3c, $\bar{c}_{\cal V}(0) = \bar{c}_{\cal V}(\phi_0) 
= (1/ 2 \pi)^{(d+1)} \int^{2 \pi}_0 \cdots
\int^{2 \pi}_0  ~{\cal C}_{\cal V} d\theta d\phi $.
Thus $\bar{c}_{\cal V},\nu_{\cal V}$ and
$k_{\cal V}$ in (\ref{eq:6.109})
are $\phi_0$-independent in this case. 
\hfill $\Box$

\vspace*{.15in}

\noindent{\bf Remarks:}
\begin{itemize}
\item[(6)]
The $2\pi$--periodicity in $\theta$ and $\phi$ of 
$\bar{c}_{\cal V}(\phi-\omega\theta)$ 
and $\tilde{c}_{\cal V}(\theta;\phi-\omega\theta)$
is suggested by examining the formal Fourier series 
$\sum\limits_{m,n} c_{m,n} 
e^{i(n\theta +m \cdot \phi)}$ of ${\cal C}_{\cal V}$
where $n\in{\mathbb Z},m\in{\mathbb Z}^d$.
Then it is easy to show that 
the resonant module part of this sum, defined by $m\cdot(1,\omega)=0$ for 
$m\in{\mathbb Z}^{d+1}$ corresponds to $\bar{c}_{\cal V}(\phi-\omega\theta)$. 
The remaining part with $m\cdot(1,\omega)\neq 0$ corresponds to
$\tilde{c}_{\cal V}(\theta;\phi-\omega\theta)$.
Their formal Fourier series display the $2\pi$--periodicity in 
$\theta$ and $\phi$. 
Off orbital resonance the relation $n+m\cdot\omega= 0$ 
implies that $n = m = 0$ so that $\bar{c}_{\cal V}(\phi_0)=c_{0,0}$, 
i.e.\ it is independent of $\phi_0$ as in Lemma 6.6b.

\item[(7)] 
Under the conditions of Theorem 6.5a,
$\nu_{\cal V}(\phi_0)$ is the UPR at
$\phi_0$ associated with the UPF defined 
in (\ref{eq:6.108}). However the UPF is not unique since the 
principal solution matrix in
(\ref{eq:6.07}) can be written as
\begin{eqnarray}
\Phi(\theta;\phi_0) = U(\theta;\phi_0) \exp ({\cal J} h(\phi_0)) 
\exp ({\cal J}\nu_{\cal V}(\phi_0) \theta)\exp (-{\cal J} h(\phi_0))
 U^T(0;\phi_0),
\nonumber
\end{eqnarray}
so that $\hat{U}(\theta;\phi_0)=
U(\theta;\phi_0) \exp ({\cal J} h(\phi_0))$ is also a UPF for an arbitrary
smooth $h(\phi_0)$.
\item[(8)] 
Under the conditions of Theorem 6.5a, and using the notation in (\ref{eq:6.3}),
${S}(\theta;\phi_0) = \Phi(\theta;\phi_0){  \mathfrak v}^3(0,\phi_0)$ 
gives a solution of 
(\ref{eq:2.1}) which is in ${\cal Q}(1,\omega;d+1)$.
This is easy to check  by (\ref{eq:6.07}):
since  ${\cal V}^T(0,\phi_0) {  \mathfrak v}^3(0,\phi_0) = (0,0,1)$  is 
the eigenvector of ${\cal J}$ with zero eigenvalue,
${S}(\theta;\phi_0) = {  \mathfrak v}^3(\theta,\phi (\theta))$  and 
$\nu_{\cal V}(\phi_0)$ has dropped  out.
By (\ref{eq:6.07}) all other linearly independent solutions are in 
${\cal Q}(1,\omega,\nu_{\cal V}(\phi_0);d+2)$.
\item[(9)] The  IFF underlying the definition (\ref{eq:6.109})
of $\nu_{\cal V}(\phi_0)$
is, of course, not unique. For example,
if $\nu_{\cal V}(\phi_0) \in (\frac{1}{2},1)$ then by
changing the signs of ${  \mathfrak v}^2$ and ${  \mathfrak v}^3$ we find 
$\nu_{\cal V}(\phi_0) \in
(0,\frac{1}{2})$.  Thus in analogy to the case $J = 0$ in
Remark 7 in Section 3 we can 
choose $\cal V$ such that  $\nu_{\cal V} \in
[0,\frac{1}{2}]$.
\item[(10)] If ${\cal V}$ denotes a uniform IFF, then 
$c_{\cal V},\bar{c}_{\cal V},\nu_{\cal V},k_{\cal V}$
are constant and $\tilde{c}_{\cal V}$ is zero. In particular
$\nu_{\cal V}={\cal C}_{\cal V}$.
\item[(11)] Under the conditions of Theorem 6.5b it is possible,
by rotating an arbitrary IFF into a uniform  IFF, 
to  construct a proper UPF at every point $\phi_0$
of the torus such that the UPR  is independent of $\phi_0$ and is a spin tune. 
\item[(12)] Theorem 6.5a
holds on and off orbital resonance although if $\omega$ does not satisfy
a Diophantine condition, appropriate solutions of (\ref{eq:6.0}) may not exist.
Note also that in Section 8 we will consider an example where the conditions
of Theorem 6.1 hold but where it turns out that
the conditions of Theorem 6.5a cannot hold due to the presence of 
nonquasiperiodic spin motion.
\item[(13)] To illustrate Theorem 6.5,
it is instructive to consider simple, but perhaps unphysical, models. 
Here we will use the model  defined by
${\cal A}(\theta,\phi):=\sqrt{2J}\cos(\phi-\theta){\cal J}$ 
with $d = 1$.
This represents a precession around the vertical and, of course,  such an 
$\cal A$ will not be found in real storage rings. 
Note that ${\cal A}$ is a smooth and $2\pi$--periodic
function and, because
$(0,0,1)$ is an ISF, we can choose the IFF $\cal V$ as 
${\cal V}=I$ so that 
by (\ref{eq:6.201})
${\cal C}_{\cal V}(\theta,\phi)=\sqrt{2J}\cos(\phi-\theta)$.
We first consider a case on orbital resonance, $\omega=1$.
~From (\ref{eq:6.8}), $c_{\cal V}(\theta;\phi_0)=\sqrt{2J}\cos(\phi_0)$, whence
$\bar{c}_{\cal V}(\phi_0)
=c_{\cal V}(\theta;\phi_0),~\tilde{c}_{\cal V}=0$. Clearly (\ref{eq:6.0})
is solved with $\alpha=0$ 
and (\ref{eq:6.108}) gives
$U(\theta;\phi_0)=\exp \left({\cal J}k_{\cal V}(\phi_0)\theta\right)$.
This 
$U(\cdot;\phi_0)$ is a proper UPF at $\phi_0$ with UPR 
$\nu_{\cal V}(\phi_0)$,
where $\nu_{\cal V}(\phi_0)$ 
and the integer $k_{\cal V}(\phi_0)$ are determined by
the condition
\begin{eqnarray}
&& k_{\cal V}(\phi_0)+\nu_{\cal V}(\phi_0) =\sqrt{2J}\cos(\phi_0) \; .
\label{eq:6.110}
\end{eqnarray}
Obviously, if $J > 0$, $\Xi$ has uncountably many elements  so that the 
torus is 
ill--tuned; 
in particular, by Theorem 6.3a, 
no uniform IFF exists.

We now consider the case of irrational $\omega$. 
~From (\ref{eq:6.8}), $c_{\cal V}(\theta;\phi_0)=\sqrt{2J}\cos((\omega-1)\theta
+\phi_0)$, thus $\bar{c}_{\cal V}=0$ and $\tilde{c}_{\cal V}(\theta;\phi_0)=
c_{\cal V}(\theta;\phi_0)$. Since $\bar{c}_{\cal V}=0$, 
$\nu_{\cal V}$ and $k_{\cal V}$ are zero and since 
${\cal D} \alpha(\theta, \phi)  = c_{\cal V}
(\theta;\phi-\omega\theta)$,
we have $\alpha(\theta,\phi)=\sqrt{2J}\sin(\phi-\theta)/(\omega-1)$. 
Then (\ref{eq:6.0013}) gives
${\cal U}(\theta,\phi)=\exp \left({\cal J}(\sqrt{2J}/(\omega-1))
\sin(\phi-\theta)  \right)$. 
One observes that ${\cal U}$ is a uniform IFF and 
${\cal C}_{\cal U}=0$. Hence by Theorem 6.3a the torus
is well--tuned and on a spin--orbit resonance.
\item[(14)] 
In Theorems 6.3--5,
we assumed that an ISF exists and 
for the physically interesting ${\cal A}$\,{\small s} we hope this is true.
It is also clear from Section 3 that for $J = 0$ an
ISF does exist.
Nevertheless, for $J \ne 0$ this assumption in the theorems 
is not superfluous as we 
will now show by constructing, for a fixed torus, an ${\cal A}$ such that 
proper UPF{\small s} 
exist at every $\phi_0$ but an ISF does not. 

We consider the case where $d=1,\omega=1$ and where ${\cal A}(\theta,\phi)$ 
depends only on $\phi-\theta$ and is given by
\begin{eqnarray}
{\cal A}(\theta,\phi)\equiv\hat{\cal A}(\phi-\theta):= 
\left(\begin{array}{ccc} 0 & -\Omega_3(\phi-\theta) 
& \Omega_2(\phi-\theta) \\
\Omega_3(\phi-\theta) & 0 &  -\Omega_1(\phi-\theta) \\
-\Omega_2(\phi-\theta) & \Omega_1(\phi-\theta) & 0
\end{array}\right) \; .
\nonumber
\end{eqnarray}
Here the function $\Omega$ is smooth and $2\pi$--periodic and we assume
$|\Omega| < 1$.
The principal solution matrix
is given by $\Phi(\theta;\phi_0)=\exp (\hat{\cal A}(\phi_0)\theta)$. 
Because $\hat{\cal A}$ is skew--symmetric with spectrum 
$\lambda(\hat{\cal A}(\phi_0))=
\lbrace i|\Omega(\phi_0)|,-i|\Omega(\phi_0)|,0\rbrace$, it follows 
by Lemma 2.1b that a
$SO(3)$-matrix $W(\phi_0)$ exists such that 
\begin{eqnarray}
&&  \hat{\cal A}(\phi_0)=W(\phi_0)|\Omega(\phi_0)|{\cal J}W^T(\phi_0) \; .
\label{eq:6.105a}
\end{eqnarray}
Hence
\begin{eqnarray}
&& \Phi(\theta;\phi_0) = W(\phi_0) 
 \exp ({\cal J}|\Omega(\phi_0)|\theta) W^T(\phi_0) \; ,
\label{eq:6.105b}
\end{eqnarray}
so that $W(\phi_0)$ is a proper UPF at $\phi_0$ 
with UPR $|\Omega(\phi_0)|$. 

We now proceed to construct an $\Omega$ such that an ISF does not exist.
If an ISF $n$ exists then, by (\ref{eq:6.0000}) and  Theorem 6.1,
we have
\begin{eqnarray}
&& \Phi(2\pi;\phi_0)n(0,\phi_0)=n(0,\phi_0+2\pi\omega) \; ,
\nonumber
\end{eqnarray}
for all $\phi_0$.
Then, by (\ref{eq:6.105b}),
\begin{eqnarray}
&&  \exp ({\cal J}|\Omega(\phi_0)|2\pi) W^T(\phi_0)n(0;\phi_0) = 
 W^T(\phi_0)n(0;\phi_0) \; .
\nonumber
\label{eq:6.105c}
\end{eqnarray}
If $|\Omega(\phi_0)|\neq{\rm integer}$, i.e.\ if $\Omega(\phi_0)\neq 0$,
then every eigenvector of $\exp ({\cal J}|\Omega(\phi_0)|2\pi)$ for eigenvalue
$1$ is parallel to $(0,0,1)$. In addition, since
$\hat{\cal A}(\phi_0)\Omega(\phi_0)=0$, (\ref{eq:6.105a}) implies that
\begin{eqnarray}
&&  {\cal J} W^T(\phi_0)\Omega(\phi_0) = 0 \;  .
\nonumber
\label{eq:6.105f}
\end{eqnarray}
It follows that
$W^T(\phi_0)n(0;\phi_0)$ and $W^T(\phi_0)\Omega(\phi_0)$ are parallel to
$(0,0,1)$. Hence $n(0;\phi_0)={\rm const}\times\Omega(\phi_0)$.
We conclude that for $\Omega(\phi_0)\neq 0$
\begin{eqnarray}
&& n(0;\phi_0) =\pm \Omega(\phi_0)/|\Omega(\phi_0)| \; .
\nonumber
\label{eq:6.105g}
\end{eqnarray}
Thus if an ISF $n$ exists, then 
\begin{eqnarray}
 n(0,\phi_0)= 
\left\{ \begin{array}{l}  \pm \Omega(\phi_0)/|\Omega(\phi_0)| 
\qquad  {\rm if} \qquad \Omega(\phi_0)\neq 0     \\
\xi  \qquad  {\rm if} \qquad \Omega(\phi_0)= 0  \; ,
                          \end{array}
                  \right.
\nonumber
\label{eq:6.105h}
\end{eqnarray}
where $\xi$ is arbitrary of norm one.
To obtain our example, we present a smooth $\Omega$ such that
$\Omega(\phi_0)/|\Omega(\phi_0)|$ does not have a continuous extension
to those $\phi_0$, where $\Omega(\phi_0)=0$.
Let $\psi(x):=1/\sin(x)$ and 
\begin{eqnarray}
\Omega(\phi_0):=\left\{ \begin{array}{l} 
(1/2)\sin^3(\phi_0)
(\cos(\psi(\phi_0)),-\sin(\psi(\phi_0)),0)
\qquad  {\rm if} \qquad \sin(\phi_0)\neq 0     \\
  0 \qquad  {\rm if} \qquad \sin(\phi_0)= 0 
 \; .
                          \end{array}
                  \right.
\nonumber
\label{eq:6.105i}
\end{eqnarray}
It is easy to show that $\Omega$ is smooth with $|\Omega| < 1$ so that
${\cal A}(\theta,\phi)$ is smooth and $2\pi$--periodic.
However
\begin{eqnarray}
\Omega(\phi_0)/|\Omega(\phi_0)| = \frac{\sin(\phi_0)}{|\sin(\phi_0)|}
(\cos(\psi(\phi_0)),-\sin(\psi(\phi_0)),0) \; , \qquad \sin(\phi_0)\neq 0
\nonumber
\label{eq:6.105j}
\end{eqnarray}
does not have a continuous extension to all of $\phi_0$. Thus there is
no ISF. 

In summary, we have an example of a torus on orbital resonance for which a 
proper UPF exists at each $\phi_0$ 
but which is ill--tuned because the set 
$\lbrace (1/2) |\sin^3(\phi_0)|: \phi_0 \in {\mathbb R}\rbrace$ 
is a subset of $\Xi$ and has uncountably many elements. Furthermore there is 
no ISF and 
thus no IFF. Of course,
an  $\cal A$ as exotic as that defined here will not  
emerge from the fields of a real storage ring.
\item[(15)] 
Now that IFF{\small s} are available we 
present another generalization of the Floquet Theorem.
Starting with (\ref{eq:6.07}) and assuming that the conditions of Theorem 6.5a
hold, we write
\begin{eqnarray}
&&\Phi(\theta;\phi_0) = U(\theta;\phi_0) 
\exp ({\cal J}\nu_{\cal V}(\phi_0) \theta) U^T(0;\phi_0) \nonumber\\
&& = 
U(\theta;\phi_0)U^T(0;\phi_0) U(0;\phi_0)\exp ({\cal J}
\nu_{\cal V}(\phi_0) \theta) 
U^T(0;\phi_0)
\nonumber\\
&&
=
 p(\theta;\phi_0)\exp\biggl(B(\phi_0) \nu(\phi_0) \theta\biggr) , 
\label{eq:6.17}
\end{eqnarray}
where 
\begin{eqnarray}
&& p(\theta;\phi_0) := U(\theta;\phi_0)U^T(0;\phi_0) \; , \; \;
B(\phi_0) :=  U(0;\phi_0) {\cal J} U^T(0;\phi_0) \; , \; \;
\nu(\phi_0):=\nu_{\cal V}(\phi_0) \; ,
\nonumber
\end{eqnarray}
and where $U$ is given by (\ref{eq:6.108}). Thus $U$ is a proper UPF.
It follows from Remark 8 of Section 5 that 
$p(\cdot;\phi_0),B(\phi_0),\nu(\phi_0)$ are generalized Floquet parameters
at $\phi_0$.
Thus under the conditions of 
Theorem 6.5a we have generalized Floquet parameters and 
have obtained a generalization of the Floquet theorem at every 
$\phi_0$.
Note that $\nu_{\cal V}(\phi_0)$ is given by (\ref{eq:6.109}).
\item[(16)] 
~For speed and practicality, simulations of particle motion in storage rings
are often made with the approximations that the fields at the ends of 
magnets fall to zero abruptly (the ``hard edge approximation'') 
or that a magnet has zero length but the correct field integral 
(the ``thin lens  approximation'').  
Then the ${\cal A}$\,{\small s} are not smooth in
$\theta$ so that the theory of this paper would have to modified.
~For example an ISF would be defined without imposing smoothness
in $\theta$. The IFF would also not be smooth in $\theta$. 
We return to these matters in Section 10.
\end{itemize}
We now prove a simple partial converse of Theorem 6.3a which gives a large
class of well--tuned ${\cal A}$\,{\small s} with a uniform IFF. 

\vspace*{.15in}

\setcounter{theorem}{6}
\begin{theorem} Let $q(\theta,\phi)$ be a $SO(3)$
matrix which is of class $C^2$ and $2\pi$--periodic such
that $q(0,\phi) = I$ and let $b$ be a constant real skew--symmetric matrix. 
Let   $A$ be defined by {\rm (\ref{eq:5.03})}
with ${\cal A}  = ({\cal D}q + q b)q^T$.
Then a uniform IFF exists, so that the torus is well--tuned.
\end{theorem}

\noindent{\em Proof:}\quad Because $b$ is skew--symmetric, by Lemma 2.1b
a constant $SO(3)$ matrix $W$ exists
such that $b=(\nu+m)W{\cal J}W^T$, where  
$\nu\in[0,1)$ and the integer $m$ is
constant. Defining ${\cal V}:=qW\exp({\cal J}m\theta)$, 
we observe that
the smooth $SO(3)$ function ${\cal V}(\theta,\phi)$ is $2\pi$--periodic 
and that ${\cal D}{\cal V}={\cal A}{\cal V}-\nu{\cal V}{\cal J}$.  Hence
${\cal C}_{\cal V}=\nu$.
~For the third column of ${\cal V}$ we have
\begin{eqnarray}
&& {\cal D}{\mathfrak v}^3 ={\cal D}{\cal V}(0,0,1) 
={\cal A}{\cal V} (0,0,1) \; .
\label{eq:6.202}
\end{eqnarray}
So it is an ISF and therefore $\cal V$ is an IFF.
The IFF $\cal V$ is uniform so that by Theorem 6.3a
the torus is well--tuned.
\hfill $\Box$

\vspace*{.15in}

\noindent{\bf Remark:}
\begin{itemize}
\item[(18)] 
We obtain the principal solution matrix
as in (\ref{eq:6.07}), with $\nu_{\cal V}=\nu$ and
\begin{eqnarray}
U(\theta;\phi_0) =q(\theta,\omega\theta+\phi_0) W
 \exp \left({\cal J} m\theta\right) \; .
\label{eq:6.203}
\end{eqnarray}
Combining (\ref{eq:6.07}) and (\ref{eq:6.203}) gives the generalized 
~Floquet form
\begin{eqnarray}
&&\Phi(\theta;\phi_0)= 
q(\theta,\omega\theta+\phi_0)W \exp \left({\cal J}(m+\nu)\,\theta
\right)W^T  =
q(\theta,\omega\theta+\phi_0) \exp(b\theta) .
\nonumber
\end{eqnarray}
\end{itemize}

We now apply some ideas of this section to some more simple models. 
Then in Section 9 we return to the general case and consider the spectrum of 
the principal solution matrix.
\section{The Single Resonance Model}
\setcounter{equation}{0}
In this section we examine 
the so--called ``single resonance model''. This model
provides an approximation for the matrix $\cal A$  and it is 
frequently used in approximate descriptions of spin--orbit resonance
effects both in proton and electron rings \cite{hh96, mane92}.
The reason for its popularity is that it can be solved exactly and 
delivers useful indications about behavior near  spin--orbit resonance even 
though
$\cal A$  is an approximation. Since it can be solved exactly it provides
an example for $\Phi(\theta;\phi_0)$ in (\ref{eq:6.07}) and (\ref{eq:6.108}).

The single resonance model
effectively describes the spin motion for a particle limited to a 
harmonic vertical betatron oscillation around a horizontal circular design 
orbit. Thus there is only one pair of action--angle variables 
$J, \phi$, and
$d = 1$. The matrix $\cal A$ then contains terms due to a constant vertical
magnetic field and terms due to a radial quadrupole field oscillating 
with the tune $\omega = d \phi /d \theta$. 
The radial field can be decomposed into two counter rotating horizontal 
components rotating with the tune $\omega$. Close to resonance it is sufficient
to consider just one of the rotating field components and to neglect the other
\cite{br99,mane92}. The matrix $\cal A$ then takes the form:
\begin{eqnarray}
\!\star\!\star\!\star &&  \qquad \ \qquad {\cal A}(\phi):=
 \left(\begin{array}{ccc} 0 & -\sigma_1  & \sigma_2 \sqrt{2J}\sin(\phi) \\
 \sigma_1 & 0 &  -\sigma_2 \sqrt{2J}\cos(\phi)  \\
 -\sigma_2 \sqrt{2J}\sin(\phi)  &
 \sigma_2 \sqrt{2J}\cos(\phi) & 0
\end{array}\right) \; , \qquad \qquad
\label{eq:7.1}
\end{eqnarray}
where $\sigma_1$ and $\sigma_2$ are real constants describing the
strength of the vertical field and
of the rotating horizontal field respectively.
Note that ${\cal A}$ is a smooth function of $\phi$, 
independent of  $\theta$.

Our first aim is to show that a proper UPF exists at every $\phi_0$.
We first consider the case
$\sigma_1\neq\omega,\sigma_2\sqrt{2J}\neq 0$ and define 
$\sigma:= \sqrt{(\sigma_1-\omega)^2+2\sigma_2^2 J}$.  Then
it is easy to verify that (see also \cite{hh96}) 
\begin{eqnarray}
&& {  \mathfrak v}^3(\phi)=\frac{1}{\sigma}
 \left(\begin{array}{c} \sigma_2 \sqrt{2J}\cos(\phi)  \\ 
 \sigma_2 \sqrt{2J}\sin(\phi)  \\ 
  \sigma_1-\omega \end{array}\right) \; .  
\label{eq:7.2}
\end{eqnarray}
This is $2\pi$--periodic and it satisfies (\ref{eq:6.1}). It is thus an ISF. 
Since 
$(1, 0, 0) \times {  \mathfrak v}^3(\phi)\neq 0$, we can define 
${  \mathfrak v}^2(\phi) \bot {  \mathfrak v}^3(\phi)$ by
\begin{eqnarray}
&& {  \mathfrak v}^2(\phi):=
\frac{(1, 0, 0) \times {  \mathfrak v}^3(\phi)}{|(1, 0, 0) \times 
{  \mathfrak v}^3(\phi)|} =
\frac{1}{\sqrt{\sigma^2-2\sigma_2^2 J\cos^2(\phi)}}
 \left(\begin{array}{c} 0 \\ \omega -\sigma_1 \\
 \sigma_2 \sqrt{2J}\sin(\phi)  
    \end{array}\right)  \; , 
\label{eq:7.3}
\end{eqnarray}
and ${  \mathfrak v}^1(\phi)$ by ${  \mathfrak v}^1 := 
{  \mathfrak v}^2 \times {  \mathfrak v}^3$. 
Thus
${\cal V}:=\left[{  \mathfrak v}^1, ~{  \mathfrak v}^2, ~{  \mathfrak v}^3 
\right]$ is smooth and $2\pi$--periodic.
It is thus an IFF.
By (\ref{eq:6.4}),(\ref{eq:7.2}),(\ref{eq:7.3}) we obtain
\begin{eqnarray}
&& C_{\cal V}(\phi)\; =  
  \Omega (\phi) \cdot {  \mathfrak v}^3 (\phi) 
+ {  \mathfrak v}^1 (\phi) \cdot ({\cal D} {  \mathfrak v}^2 (\phi)) 
= \sigma + \frac{\sigma\omega(\sigma_1-\omega)}
{\sigma^2-2\sigma_2^2 J\cos^2(\phi)} \; ,
\label{eq:7.20}
\end{eqnarray}
so that
\begin{eqnarray}
&& \bar{c}_{\cal V}(\phi_0)
 = \lim_{T\rightarrow\infty}\,
        (1/T)\,
           \int_0^T \; d\theta'
         ~C_{\cal V}(\omega\theta' + \phi_0)
= (1/2\pi)\,
           \int_0^{2\pi} \; d\phi
         ~C_{\cal V}(\phi) \; ,
\label{eq:7.21}
\end{eqnarray}
where we use the $2\pi$--periodicity of $C_{\cal V}$.
Thus $\bar{c}_{\cal V}(\phi_0)$ is independent of $\phi_0$.
Since \cite[p.379]{Grad}
\begin{eqnarray}
\int_0^{\pi/2}\;  \frac{d\phi}
{(\sigma_1-\omega)^2\sin^2(\phi)+\sigma^2\cos^2(\phi)} = 
\frac{\pi}{2\sigma |\sigma_1-\omega|} \; ,
\nonumber
\end{eqnarray}
and since by (\ref{eq:7.20}) and (\ref{eq:7.21}) 
\begin{eqnarray}
&& \bar{c}_{\cal V} =  \sigma + \frac{\sigma\omega(\sigma_1-\omega)}
{2\pi}\int_0^{2\pi}\;  \frac{d\phi}
{(\sigma_1-\omega)^2\cos^2(\phi)+\sigma^2\sin^2(\phi)} \nonumber\\
&&
 =  \sigma + \frac{\sigma\omega(\sigma_1-\omega)}
{\pi}\int_0^{\pi}\;  \frac{d\phi}
{(\sigma_1-\omega)^2\cos^2(\phi)+\sigma^2\sin^2(\phi)}
 \nonumber\\
&&
 =  \sigma + \frac{2\sigma\omega(\sigma_1-\omega)}
{\pi}\int_0^{\pi/2}\;  \frac{d\phi}
{(\sigma_1-\omega)^2\sin^2(\phi)+\sigma^2\cos^2(\phi)} \; ,
\label{eq:7.0}
\end{eqnarray}
we can make the assignments
\begin{eqnarray}
&& \bar{c}_{\cal V} = 
\sigma + \omega ~{\rm sign}(\sigma_1-\omega) \; , \qquad
 \tilde{c}_{\cal V}(\phi) = \frac{\sigma\omega(\sigma_1-\omega)}
{\sigma^2-2\sigma_2^2 J\cos^2(\phi)}  - 
\omega ~{\rm sign}(\sigma_1-\omega) \; .
\label{eq:7.5}
\end{eqnarray}
Because $\tilde{c}_{\cal V}$ does not depend on $\theta$ one finds
that (\ref{eq:6.0}) has smooth solutions $\alpha(\phi)$
which do not depend on $\theta$ and that one of those solutions reads as
\begin{equation}
\alpha(\phi) =
\frac{1}{\omega} \int^{\phi}_0  d \phi' \, \tilde{c}_{\cal V}(\phi') \; .
\label{eq:7.6}
\end{equation}
Since $0=\int_0^{2\pi} \; d\phi\tilde{c}_{\cal V}(\phi)$ 
and by the $2\pi$--periodicity of $\tilde{c}_{\cal V}(\phi)$ 
it follows by (\ref{eq:7.6})
that $\alpha(\phi)$ is $2\pi$--periodic in $\phi$.
Thus Theorem 6.5a applies and one concludes that
$U(\cdot;\phi_0)$, defined by
\begin{equation}
U(\theta;\phi_0) = 
{\cal V}(\phi_0+\omega\theta) \exp \left({\cal J}(
\frac{1}{\omega} \int_{\phi_0}^{\phi_0+\omega\theta}  
d \phi' \, \tilde{c}_{\cal V}(\phi')
+ k_{\cal V} \theta) \right) \; ,
\label{eq:7.7}
\end{equation}
is a proper UPF at $\phi_0$ with UPR $\nu_{\cal V}$.
Here $\nu_{\cal V}$ and the integer $k_{\cal V}$ 
are uniquely determined by $\bar{c}_{\cal V}$
via (\ref{eq:6.109}) and (\ref{eq:7.5}). 
Because $\nu_{\cal V}$ is contained 
in each $\Xi(\phi_0)$ we conclude (recall the comment after Definition 5.4)
that the tori are well--tuned when
$\sigma_1\neq\omega$ and $\sigma_2\sqrt{2J}\neq 0$.

We now discuss the general case using the machinery of Section 6.

\vspace*{.15in}

\noindent{\bf Proposition 7.1}
{\em The single resonance model
has a uniform IFF for every value $J$
of the orbital action variable. Hence
the corresponding torus is well--tuned. }

\vspace*{.15in}
\noindent{\em Proof:}\quad 
~For ${\cal Y}(\theta,\phi):=\exp(\phi{\cal J})$ one obtains
\begin{eqnarray}
{\cal Y}^T({\cal A}{\cal Y}- {\cal D}{\cal Y})= E \; ,
\label{eq:7.001}
\end{eqnarray}
where
\begin{eqnarray}
&& E:= \left(\begin{array}{ccc} 0 & -\sigma_1+\omega  & 0 \\
 \sigma_1-\omega & 0 & -\sigma_2\sqrt{2J}  \\
 0 & \sigma_2\sqrt{2J} & 0
\end{array}\right) \; .
\nonumber
\end{eqnarray}
This has eigenvalues $\lambda(E)=\lbrace i\sigma,-i\sigma,0\rbrace$.
If $\sigma=0$, then $E=0$ and, due to (\ref{eq:6.0005}) and (\ref{eq:7.001}),
${\cal Y}$ is a uniform
IFF and ${\cal C}_{\cal Y}=0$.
Thus by Theorem 6.3a the tori are well--tuned and $0$ is a spin tune.
If $\sigma\neq 0$, then by Lemma 2.1b we can choose $W \in SO(3)$
such that
\begin{eqnarray}
&& W^{-1} E W = \sigma {\cal J} =:(\nu+k){\cal J}
\; ,
\nonumber
\end{eqnarray}
where the integer $k$ is chosen such that
$\nu\in[0,1)$. We now define
\begin{eqnarray}
&& {\cal U}(\theta;\phi):= {\cal Y}(\theta;\phi)W
\exp(k\theta{\cal J})
\; ,
\nonumber
\end{eqnarray}
and obtain
\begin{eqnarray}
{\cal U}^T({\cal A}{\cal U}- {\cal D}{\cal U})= \nu{\cal J} \; .
\nonumber
\end{eqnarray}
It follows by (\ref{eq:6.0005})
that ${\cal U}$ is a uniform
IFF and ${\cal C}_{\cal U}=\nu=\sigma-k$.
Thus by Theorem 6.3a the tori are well--tuned and $\sigma-k$ is a spin tune.
\hfill $\Box$

\vspace*{.15in}

\noindent{\bf Remarks:}
\begin{itemize}
\item [(1)] The proof of Proposition 7.1  and Definition 5.5 show that the 
spin tunes
$\omega_s$ of the single resonance model
have the form
\begin{eqnarray}
&&  \omega_s = \varepsilon \sigma+
j\omega +k \; ,
\label{eq:7.17}
\end{eqnarray}
where $\varepsilon\in\lbrace -1,1\rbrace;j,k\in{\mathbb Z}$. Conversely every
constant of the form (\ref{eq:7.17}) is a spin tune of the 
single resonance model, if it is
in $[0,1)$. In particular the set $\Xi(\phi_0)$ is 
independent of $\phi_0$. 
Note that for the single resonance model, 
spin tunes exist also on orbital resonance, i.e.\ 
for rational $\omega$. See Remark 4 too.
\item [(2)]
The case $\sigma_2\sqrt{2J}=0$ represents the absence of betatron
motion, i.e.\ motion on the design orbit. In this case 
$\varepsilon,j,k$ can be chosen in (\ref{eq:7.17}) such that
the spin tune reduces to $\sigma_1$. 
\item [(3)]
~From the expression for $\sigma$ it is clear that 
during variation of $\sigma_1$, the spin tune in (\ref{eq:7.17}) comes closest 
to the spin--orbit resonance 
$\omega_s = k + j \omega$ when $\sigma_1 = \omega$. 
However for the case $\sigma_2\sqrt{2J}\neq 0$ of principle interest,
the resonance condition is not reached.
\item [(4)]
If $\omega$ is an integer, the matrix $A$ is one--turn periodic and the 
ISF can be 
obtained as the  eigenvector of length 1 of the one--turn 
principal solution matrix.
Moreover, every proper UPF is one--turn periodic 
and its
UPR can be obtained from the complex eigenvalues of the one--turn 
principal solution matrix
just as in Section 3. Of course, this UPR is a spin tune.
If $\omega$ is rational, the  ISF can be 
obtained as the  eigenvector of length 1 of the appropriate multi--turn 
principal solution matrix. 
See, e.g.\, \cite{bar2002}.
Every proper UPF is then multi--turn periodic and its
UPR is extracted from the corresponding complex eigenvalues
and is again a spin tune.
Note that this circumstance that a spin tune exists even on orbital
resonance  has its origin in the facts that the single resonance model
has only one
orbital frequency and that $\cal A$ is independent of $\theta$.
\item [(5)]
If the orbital tune is rational then with
Proposition 7.1 we have  an example of a torus which is on orbital
resonance but is nevertheless well--tuned.
This is an example for which the torus may be on orbital resonance but still
satisfy the conditions of Theorem 6.3a.

\end{itemize}
\section{The Moser-Siegel Model}
\setcounter{equation}{0}
In this section we construct  and study an illustrative but unphysical 
model which, for reasons which will become clear, we will call
the ``Moser-Siegel model''. This model has $d=1$, 
can be solved exactly, and has two real parameters
$\sigma_1,\sigma_2$, where $\sigma_1>1$.
~For certain choices of the orbital tune $\omega$
and of $\sigma_1,\sigma_2$ the Moser-Siegel model
provides an example of
spin motion which is nonquasiperiodic.
The matrix $\cal A$ of the Moser-Siegel model
takes the form:
\begin{eqnarray}
\!\star\!\star\!\star &&  \qquad  \qquad  \qquad  {\cal A}:= 
{\cal A}_{21}{\cal J} \; , \qquad
{\cal A}_{21}(\theta,\phi):= \sum_{k,l=1}^\infty\; \sigma_1^{-k-l}\eta_{kl}
 \sin(k\phi-l\theta) \; , \qquad  \qquad \qquad  
\label{eq:8.3}
\end{eqnarray}
where
\begin{eqnarray}
&&  \eta_{kl}:=
          \left\{ \begin{array}{ll}
 1 \; , & \qquad{\rm if} \; k\sigma_2-l\geq 0  \; , \\
 -1\; , & \qquad {\rm if} \; k\sigma_2-l< 0
  \; ,
     \end{array}
                  \right.
\nonumber
\end{eqnarray}
and by definition $\cal A$ is smooth (in fact $C^\infty$)
and $2\pi$--periodic. The latter follows from the convergence
of $\sum_{k=1}^\infty\; k^n\sigma_1^{-k}$ for a nonnegative integer $n$,
which follows from the ratio
test and  which implies that the series in (\ref{eq:8.3})
and the series of all its derivatives converge uniformly (see
\cite[8.6.3]{Di60},\cite[p.117]{Lang}).

~For $\phi_0=0$, 
\begin{eqnarray}
&& A(\theta;0) =f(\theta){\cal J} \; , \qquad
f(\theta)=\sum_{k,l=1}^\infty\; \sigma_1^{-k-l}\eta_{kl}
 \sin((k\omega-l)\theta) \; .
\label{eq:8.04}
\end{eqnarray}
Clearly $f\in{\cal Q}(1,\omega;2)$ and for an irrational $\omega$, $\bar{f}=0$
by Lemma 4.3c, so that 
\begin{equation}
\Phi(\theta;0) = \exp \left({\cal J}g(\theta) \right) 
\; ,
\label{eq:8.07}
\end{equation}
where
\begin{equation}
g(\theta):=\int\limits^{\theta}_0 \tilde{f}(\theta')d \theta' \; . 
\label{eq:8.08}
\end{equation}
Because the function $g$ has been used in \cite[Paragraph 36]{Sieg}, 
we call our model the Moser--Siegel model.

\vspace*{.15in}

\noindent{\bf Proposition 8.1}
{\em For some $\omega\not\in{\mathbb Q}$ there exist values of
$\sigma_1,\sigma_2$ such that $g$ is unbounded.
~For these values, $\exp(ig)$ is not quasiperiodic whence
$\Phi(\cdot;0)$ is not quasiperiodic.}

\vspace*{.15in}
\noindent{\em Proof:} It is shown in \cite[Paragraph 36]{Sieg}
that values of
$\sigma_1,\sigma_2$ and of $\omega\not\in{\mathbb Q}$ exist such that 
$g$ is unbounded. Thus, for these values,
$\tilde{f}$ is an almost periodic function whose integral, $g$,
is unbounded. It then follows (see \cite[Chapter 6]{Fink}) that
$\exp(ig)$ is not almost periodic, whence at least one of
$\cos(g)$ and $\sin(g)$ is not almost periodic.
\hfill $\Box$

\vspace*{.15in}

\noindent{\bf Remarks:}
\begin{itemize}
\item[(1)] Proposition 8.1 shows that
for certain values of $\sigma_1,\sigma_2$ and of $\omega$ the 
principal solution matrix
at $\phi_0=0$ is not quasiperiodic, so that
$\Xi(\phi_0=0)=\emptyset$. In particular, for those
values the torus is ill--tuned so that, by Theorem 6.3a, it has no uniform IFF.
\item[(2)] The unit matrix obviously provides an IFF for the 
Moser-Siegel model.
Thus Proposition 8.1 demonstrates that the existence of an IFF is neither
sufficient for having a uniform IFF nor for having a well--tuned torus.
In the context of Theorem 6.5a this means that the existence
of an IFF does not necessarily admit a smooth solution $\alpha$ of
(\ref{eq:6.0}), which is $2\pi$-periodic.
\end{itemize}
\section{Spectral Analysis of Quasiperiodic Spin Motion}
\setcounter{equation}{0}
\setcounter{theorem}{0}
In the code SPRINT, ISF{\small s} are calculated 
nonperturbatively in three ways, 
namely by stroboscopic
averaging, by the SODOM2 algorithm and by adiabatic antidamping  
\cite{hh96,epac98,spin98,gh2000,mv2000,spin2000,hvb99}.
Perturbative algorithms for obtaining 
ISF{\small s} are listed in the Introduction.
Moreover the spin tune is calculated 
nonperturbatively in SPRINT by logging the spin
precession around the ISF or by the SODOM2 algorithm. We will make further comments
on these simulations in Section 10.

The fact that under appropriate conditions the set of generalized Floquet 
frequencies is the set of spin tunes, suggests a further way to obtain  
the 
spin tune, namely by spectral analysis of quasiperiodic functions. 
The way forward is contained in the following theorem.
As we shall see, we can also use spectral analysis to construct the ISF.

\vspace*{.15in}

\begin{theorem} 
Consider a uniform IFF ${\cal V}$ on a fixed torus.
Then the following holds.

\noindent
a) Let $S$ be a spin trajectory at $\phi_0$. If 
$0\in\lbrack {\cal C}_{\cal V}\rbrack$, where the equivalence class 
$\lbrack {\cal C}_{\cal V}\rbrack$ 
is defined in Section 5,
then $S$ is in ${\cal Q}(1,\omega;d+1)$.  

\noindent
b) Let $S$ be a spin trajectory at $\phi_0$ and let 
the torus be off orbital resonance. Also, let
$0\not\in\lbrack {\cal C}_{\cal V}\rbrack$. 
Then $\hat{S}$, defined by
\begin{eqnarray}
&& \hat{S}(\theta):= \lim_{N\rightarrow\infty} 
\sum_{m \in {\mathbb Z}^{d+1}\atop ||m||\leq N}\;  
A_{N,m} \exp(i\theta m\cdot(1,\omega))
a\biggl( S,m\cdot(1,\omega)\biggr) \; ,
\label{eq:9.01}
\end{eqnarray}
with $A_{N,m}:=\prod_{n=1}^{d+1}\;\frac{N+1-|m_n|}{N+1}$,
satisfies the relation
\begin{eqnarray}
&& \hat{S}(\theta)=n(\theta,\phi_0+\omega\theta)
\hat{S}(0)\cdot n(0,\phi_0) 
\; ,
\label{eq:9.08c}
\end{eqnarray}
where $n$ is the ISF $n:={\cal V}(0,0,1)$.
Moreover, $\hat{S}$ is a spin trajectory in ${\cal Q}(1,\omega;d+1)$.

\noindent c) For arbitrary $\phi_0$,
\begin{eqnarray}
\!\star\!\star\!\star &&  \qquad  \qquad \Lambda(\Phi(\cdot;\phi_0))\subset
\lbrace
\varepsilon{\cal C}_{\cal V}+m\cdot(1,\omega):
(\varepsilon,m)\in \lbrace 0,1,-1\rbrace\times{\mathbb Z}^{d+1}\rbrace.
\qquad  \qquad \qquad
\nonumber
\end{eqnarray}
\end{theorem}
{\em Proof of Theorem 9.1a:} For fixed $\phi_0$, define
$U(\theta):={\cal V}(\theta,\phi_0+\omega\theta)$ so that by Theorem 6.3a
$U$ is a proper UPF at $\phi_0$ and we have
\begin{eqnarray}
&& \Phi(\theta;\phi_0)=
U(\theta) \exp ({\cal J}{\cal C}_{\cal V}\theta)U^T(0) \; .
\label{eq:9.02}
\end{eqnarray}
Thus if $0\in\lbrack{\cal C}_{\cal V}\rbrack$,
then $\Phi(\cdot;\phi_0)$, and therefore $S$, is in ${\cal Q}(1,\omega;d+1)$.
\hfill $\Box$

\noindent {\em Proof of Theorem 9.1b:} 
Because the $U$, defined in the proof of Theorem 9.1a,
is in ${\cal Q}(1,\omega;d+1)$ and because the torus is off orbital resonance 
we can use Lemma 4.3b-c to write
\begin{eqnarray}
U(\theta) =
\lim_{N \rightarrow \infty}  \sum_{m \in {\mathbb Z}^{d+1}\atop ||m||\leq N} 
A_{N,m} \exp(i\theta m\cdot(1,\omega))
a\biggl( U,m\cdot(1,\omega)\biggr) \; .
\label{eq:9.03}
\end{eqnarray}
With 
\begin{eqnarray}
\Delta_\pm:=\frac{1}{2}\left( \begin{array}{ccc} 
1 & \pm i & 0\\ 
\mp i & 1 & 0\\ 
0 & 0 & 0 \end{array} \right) \; , \qquad
\Delta_0:=
\left( \begin{array}{ccc} 
0 & 0 & 0\\ 
0 & 0 & 0\\ 
0 & 0 & 1 \end{array} \right) \; ,
\label{eq:9.00}
\end{eqnarray}
and recalling  (\ref{eq:2.12}), we have
\begin{eqnarray}
&& \exp ({\cal J}{\cal C}_{\cal V}\theta) = \Delta_+
\exp (i{\cal C}_{\cal V}\theta)
+\Delta_-\exp (-i{\cal C}_{\cal V}\theta)+\Delta_0 \; ,
\label{eq:9.04}
\end{eqnarray}
so that 
\begin{eqnarray}
&& a\biggl( U\exp ({\cal J}{\cal C}_{\cal V} \, \cdot),m\cdot(1,\omega)\biggr) 
=a\biggl( U \Delta_+,m\cdot(1,\omega)-{\cal C}_{\cal V}\biggr) 
+ a\biggl( U \Delta_-,m\cdot(1,\omega)+{\cal C}_{\cal V}\biggr) 
\nonumber\\
&&\qquad
+a\biggl( U \Delta_0,m\cdot(1,\omega)\biggr) \; .
\label{eq:9.05}
\end{eqnarray}
Since $0\not\in\lbrack{\cal C}_{\cal V}\rbrack$, then 
$\pm{\cal C}_{\cal V}\not\in\lbrace m\cdot(1,\omega):m\in{\mathbb Z}^{d+1}
\rbrace$. It follows from Lemma 4.3e that 
$\pm{\cal C}_{\cal V}\not\in\Lambda(U)$. Then
$0 = a\biggl( U ,m\cdot(1,\omega)\mp {\cal C}_{\cal V}\biggr)$
so that (\ref{eq:9.05}) leads to
\begin{eqnarray}
&& a\biggl( U\exp ({\cal J}{\cal C}_{\cal V} \, \cdot),m\cdot(1,\omega)\biggr) 
=a\biggl( U \Delta_0,m\cdot(1,\omega)\biggr) \; .
\label{eq:9.06}
\end{eqnarray}
Combining (\ref{eq:9.02}), (\ref{eq:9.03}), (\ref{eq:9.06}) we obtain
\begin{eqnarray}
&& U(\theta)\Delta_0 
=\lim_{N \rightarrow \infty}  \sum_{m \in {\mathbb Z}^{d+1}\atop ||m||\leq N} 
A_{N,m} 
\exp(i\theta m\cdot(1,\omega))
a\biggl( U \Delta_0,m\cdot(1,\omega)\biggr) 
\nonumber\\
&& 
=\lim_{N \rightarrow \infty}  \sum_{m \in {\mathbb Z}^{d+1}\atop ||m||\leq N} 
A_{N,m} 
\exp(i\theta m\cdot(1,\omega))
 a\biggl( U\exp ({\cal J}{\cal C}_{\cal V} \,\cdot),m\cdot(1,\omega)\biggr)
\nonumber\\
&& 
=\lim_{N \rightarrow \infty}  
\sum_{m \in {\mathbb Z}^{d+1}\atop ||m||\leq N} 
A_{N,m} \exp(i\theta m\cdot(1,\omega))
 a\biggl( \Phi(\cdot;\phi_0)U(0) ,m\cdot(1,\omega)\biggr)\; .
\nonumber
\end{eqnarray}
Hence
\begin{eqnarray}
&& U(\theta)\Delta_0 U^T(0) =\lim_{N \rightarrow \infty}  
\sum_{m \in {\mathbb Z}^{d+1}\atop ||m||\leq N} 
A_{N,m} \exp(i\theta m\cdot(1,\omega))
 a\biggl( \Phi(\cdot;\phi_0) ,m\cdot(1,\omega)\biggr)\; .
\label{eq:9.07}
\end{eqnarray}
Then, because $S$ is a spin trajectory at $\phi_0$, 
\begin{eqnarray}
&& U(\theta)\Delta_0 U^T(0) S(0)
=\lim_{N \rightarrow \infty}  
\sum_{m \in {\mathbb Z}^{d+1}\atop ||m||\leq N} 
A_{N,m} \exp(i\theta m\cdot(1,\omega))
 a\biggl( \Phi(\cdot;\phi_0)S(0),m\cdot(1,\omega)\biggr)
\nonumber\\
&& =\lim_{N \rightarrow \infty}  
\sum_{m \in {\mathbb Z}^{d+1}\atop ||m||\leq N} 
A_{N,m} \exp(i\theta m\cdot(1,\omega))
 a\biggl(S,m\cdot(1,\omega)\biggr)
=\hat{S}(\theta) \; , \qquad
\label{eq:9.08} 
\end{eqnarray}
where the last equality follows from (\ref{eq:9.01}).
~From (\ref{eq:2.32}), (\ref{eq:9.02}), (\ref{eq:9.00}), (\ref{eq:9.08})
it follows that
\begin{eqnarray}
&& \dot{\hat{S}}(\theta)=
\dot{U}(\theta)\Delta_0 U^T(0)S(0) 
=\biggl(A(\theta;\phi_0)U(\theta)-{\cal C}_{\cal V}U(\theta){\cal J}\biggr)
\Delta_0 U^T(0)S(0) 
\nonumber\\
&&
=A(\theta;\phi_0)  
U(\theta)\Delta_0 U^T(0)S(0) =A(\theta;\phi_0)\hat{S}(\theta)
\; .
\label{eq:9.09}
\end{eqnarray}
Thus $\hat{S}$ is a spin trajectory at $\phi_0$. By (\ref{eq:9.08}) and 
because  $U$ is proper, $\hat{S}$ is in
${\cal Q}(1,\omega;d+1)$.

If $S$ is in ${\cal Q}(1,\omega;d+1)$, then by Lemmas 4.3b and 4.3c  
\begin{eqnarray}
&& S(\theta) = \lim_{N\rightarrow\infty} 
\sum_{m \in {\mathbb Z}^{d+1}\atop ||m||\leq N}\;  
A_{N,m} \exp(i\theta m\cdot(1,\omega))
a\biggl(S,m\cdot(1,\omega)\biggr) \; ,
\label{eq:9.01a}
\end{eqnarray}
so that in this special situation, $\hat S = S$. Thus, for an arbitrary
spin trajectory $S$ at $\phi_0$, the double transform $\hat{\hat{S}}$
of $S$ from the double application of 
(\ref{eq:9.01})  
is equal to the single transform $\hat{S}$ so that (\ref{eq:9.08}) yields
\begin{eqnarray}
&& U(\theta)\Delta_0 U^T(0) \hat{S}(0)=\hat{\hat{S}}(\theta)
=\hat{S}(\theta) \; .
\label{eq:9.08a} 
\end{eqnarray}
If $\theta=0$,  (\ref{eq:9.08a}) becomes an eigenproblem
for $\hat{S}(0)$. The solution is $\hat{S}(0)=\pm|\hat{S}(0)| U(0)(0,0,1)$.
Inserting this into (\ref{eq:9.08a}) yields
\begin{eqnarray}
&& \hat{S}(\theta)=\pm|\hat{S}(0)|n(\theta,\phi_0+\omega\theta) 
=n(\theta,\phi_0+\omega\theta) \hat{S}(0)\cdot n(0,\phi_0) 
\; ,
\label{eq:9.08b} 
\end{eqnarray}
where $n$ denotes the third column of ${\cal V}$ and where 
$S$ is an arbitrary spin trajectory at $\phi_0$.
\hfill $\Box$

\noindent {\em Proof of Theorem 9.1c:} 
By (\ref{eq:9.02}) and (\ref{eq:9.04})
we have
\begin{eqnarray}
&& \Phi(\theta;\phi_0)=
U(\theta) \biggl( \Delta_+\exp (i{\cal C}_{\cal V}\theta)
+\Delta_-\exp (-i{\cal C}_{\cal V}\theta)+\Delta_0\biggr)U^T(0) \; .
\label{eq:9.001}
\end{eqnarray}
If $\lambda\in \Lambda\biggl(U \Delta_+
\exp (i{\cal C}_{\cal V} \, \cdot)U^T(0)\biggr)$, then
\begin{eqnarray}
&& \lambda-{\cal C}_{\cal V}
\in \Lambda\biggl(U \Delta_+ U^T(0)\biggr) \; .
\label{eq:9.002}
\end{eqnarray}
Because $U$ is proper and with Lemma 4.3e, 
$\Lambda(U \Delta_+ U^T(0))\subset
\lbrace m\cdot(1,\omega):m\in{\mathbb Z}^{d+1}\rbrace$. Thus with 
(\ref{eq:9.002}),
$\lambda\in \lbrace {\cal C}_{\cal V}+m\cdot(1,\omega):
m\in{\mathbb Z}^{d+1}\rbrace$. Then
\begin{eqnarray}
&& \Lambda\biggl(U \Delta_+\exp (i{\cal C}_{\cal V} \, \cdot)U^T(0)\biggr) 
\subset
\lbrace{\cal C}_{\cal V}+m\cdot(1,\omega):m\in{\mathbb Z}^{d+1}\rbrace \; .
\label{eq:9.003}
\end{eqnarray}
In an analogous way
\begin{eqnarray}
&& \Lambda\biggl(U \Delta_-\exp (-i{\cal C}_{\cal V} \, \cdot)U^T(0)\biggr) 
\subset
\lbrace-{\cal C}_{\cal V}+m\cdot(1,\omega):m\in{\mathbb Z}^{d+1}\rbrace \; .
\label{eq:9.004a}
\end{eqnarray}
Moreover,
\begin{eqnarray}
&& \Lambda\biggl(U \Delta_0  U^T(0)\biggr) \subset
\lbrace m\cdot(1,\omega):m\in {\mathbb Z}^{d+1}\rbrace \; .
\label{eq:9.004b}
\end{eqnarray}
Combining (\ref{eq:9.001}), (\ref{eq:9.003}), 
(\ref{eq:9.004a}), (\ref{eq:9.004b}) gives

\vspace*{.15in}

$\Lambda(\Phi(\cdot;\phi_0))$\\
\hspace*{5mm}~~$\subset
\Biggl(
\Lambda\biggl(U \Delta_+\exp (i{\cal C}_{\cal V} \, \cdot)U^T(0)\biggr)\cup 
\Lambda\biggl(U \Delta_-\exp (-i{\cal C}_{\cal V} \, \cdot)U^T(0)\biggr)\cup
\Lambda\biggl(U \Delta_0  U^T(0)\biggr) \Biggr)$\\
\hspace*{5mm}~~$\subset 
\lbrace \varepsilon{\cal C}_{\cal V}+m\cdot(1,\omega):
(\varepsilon,m)\in \lbrace 0,1,-1\rbrace\times{\mathbb Z}^{d+1}\rbrace$.
\hfill $\Box$



\noindent{\bf Remarks:}
\begin{itemize}
\item[(1)] Let the conditions of Theorem 9.1b hold and let
the spin trajectory $S$ at $\phi_0$ be in ${\cal Q}(1,\omega;d+1)$.
In this special situation (\ref{eq:9.01a}) holds, i.e.\
(\ref{eq:9.01}) becomes the spectral expansion of $S$. However, in general 
$S$ is not in ${\cal Q}(1,\omega;d+1)$, i.e.\ in general (\ref{eq:9.01})
is {\em not} the spectral expansion of $S$ because only the tunes 
$m\cdot(1, \omega)$ appear in (\ref{eq:9.01}). \\
If the conditions of Theorem 9.1b hold then 
$\hat{S}$ is parallel to an ISF
and with (\ref{eq:9.01}) and (\ref{eq:9.08c})
one could, at least in principle, attempt to compute the ISF by
doing numerical spectral analysis on an arbitrary spin trajectory. 
Of course, by Theorem 6.4, the ISF $n$ is unique up to a sign and, by
Theorem 6.3a,
the torus is well--tuned.

\item[(2)] Consider a fixed torus and assume that a uniform IFF exists so that,
due to Theorem 6.3a, the torus is well--tuned.
Then, due to Remark 8 in Section 5,
the set of generalized Floquet frequencies is the same at every $\phi_0$
and is identical with the set $\Xi$ of spin tunes.
Therefore Theorem 9.1c implies, for arbitrary $\phi_0$,
that for every $\lambda$ in $\Lambda(\Phi(\cdot;\phi_0)) 
\setminus \lbrace m\cdot(1,\omega): m \in {\mathbb Z}^{d+1}\rbrace$ 
the fractional part of $\lambda$ is 
a generalized Floquet frequency  at $\phi_0$, and in particular a spin tune.
Thus as conjectured earlier, spin tunes can indeed be obtained by spectral
analysis. 
\end{itemize}

Theorem 9.1c addresses the spectrum, $\Lambda(\Phi(\cdot;\phi_0))$,   
for the conditions stated but gives no information on its dependence
on $\phi_0$.
However, under certain conditions, the special parameter dependence of $A$ 
given in  (\ref{eq:5.04}) guarantees that $\Lambda(\Phi(\cdot;\phi_0))$
is independent of  $\phi_0$ as we show in the next theorem.

Consider a fixed torus  and denote
$\Phi(\theta+2\pi N;\phi_0) \Phi^T(2\pi N;\phi_0)$ by 
$\Pi(\theta;\phi_0)$, where $N$ is an integer.
We conclude from (\ref{eq:5.02}) and 
(\ref{eq:5.04}) that $\Pi$ satisfies the initial value problem
\begin{eqnarray}
\frac{\partial \Pi(\theta; \phi_0)}{\partial \theta}
&=& A(\theta; 2\pi N\omega+\phi_0)\Pi(\theta; \phi_0) \; , \qquad
\Pi(0;\phi_0) = I .
\label{eq:5.05}
\end{eqnarray}
Because $\Phi(\theta; 2\pi N\omega+\phi_0)$ is the unique
solution of this initial value problem (see also Section 2), we obtain
\begin{eqnarray}
&& \Phi(\theta; 2\pi N\omega+\phi_0) =
\Phi(\theta+2\pi N;\phi_0) \Phi^T(2\pi N;\phi_0)\; ,
\label{eq:9.06n}
\end{eqnarray}
valid for arbitrary $\phi_0$ and arbitrary integer $N$.
Thus the basic property of $A$
in (\ref{eq:5.04}) leads to the basic property of
$\Phi$ as manifested in (\ref{eq:9.06n}).
It follows that 
if $\Phi(\cdot;\phi)$ is known at a fixed $\phi_0$ then it is known
for all $\phi=\phi_0+2\pi N\omega + 2\pi M$ with $M \in {\mathbb Z}^d$. Then if in addition  $(1, \omega)$ is 
nonresonant
it follows by continuity that it is known for all $\phi$ on the torus.

\vspace*{.15in}

\setcounter{theorem}{1}
\begin{theorem} 
Consider a fixed torus $J_0$ off orbital resonance 
and
assume that a $\phi_0$ exists such that $\Xi(\phi_0)\neq\emptyset$. 
Then for every real $\lambda$ and all $\phi\in{\mathbb R}^d$, 
$a(\Phi(\cdot;\phi),\lambda)$ exists and is continuous in $\phi$.
Moreover, for all $\phi$, $\Lambda(\Phi(\cdot;\phi))=
\Lambda(\Phi(\cdot;\phi_0))$.
\end{theorem}

\noindent{\em Proof:} Since $\Phi(\cdot+2\pi N;\phi_0)$ is quasiperiodic, it is
easy to see from 
Definition 4.1 that
\begin{eqnarray}
&& a(\Phi(\cdot+2\pi N;\phi_0),\lambda)=
\exp(i\lambda 2\pi N)a(\Phi(\cdot;\phi_0),\lambda)
\label{eq:9.001n} 
\end{eqnarray}
and the basic identity (\ref{eq:9.06n}) gives
\begin{eqnarray}
&& a(\Phi(\cdot;\phi_0+2\pi N\omega),\lambda)=
\exp(i\lambda 2\pi N)a(\Phi(\cdot;\phi_0),\lambda)\Phi^T(2\pi N;\phi_0) \; .
\label{eq:9.002n} 
\end{eqnarray}
Therefore
\begin{eqnarray}
&& \Lambda(\Phi(\cdot;\phi_0))=\Lambda(\Phi(\cdot;
\phi_0+2\pi N\omega + 2\pi M)) \; ,
\label{eq:9.003n} 
\end{eqnarray}
for all $N \in {\mathbb Z}, M \in {\mathbb Z}^d$, where we also used the
fact that $\Phi(\theta;\phi_0)$ is $2\pi$--periodic in $\phi_0$.
Thus the spectrum of the principal solution matrix
on the set 
\begin{eqnarray}
&& D':=\lbrace \phi_0 + 2\pi N\omega + 
2\pi M: N \in {\mathbb Z}, M \in {\mathbb Z}^d \rbrace 
\label{eq:9.011} 
\end{eqnarray}
is the same as the spectrum of the 
principal solution matrix at $\phi_0$.
Because $(1,\omega)$ is nonresonant, $D'$
is dense in  ${\mathbb R}^d$.
Now fix $\lambda$ and let $h(\phi):=a(\Phi(\cdot;\phi),\lambda)$
for $\phi\in D'$ and assume $h(\phi_0)\neq 0$ so that
$\lambda\in \Lambda(\Phi(\cdot;\phi_0))$. Then
\begin{eqnarray}
&& 0 < |h(\phi_0)|=|h(\phi_0+2\pi N\omega)\Phi(2\pi N;\phi_0)|
\leq \sqrt{3}|h(\phi_0+2\pi N\omega)| \nonumber \\
&&=\sqrt{3}|h(\phi_0+2\pi N\omega + 2\pi M)| \; ,
\label{eq:9.004} 
\end{eqnarray}
where we used (\ref{eq:9.002n}) for the first equality
and where for the second inequality we used the fact that
\begin{eqnarray}
&& |X\Phi(2\pi N;\phi_0)| \leq \sqrt{3}|X| \; ,
\label{eq:9.005n} 
\end{eqnarray}
which follows from the $SO(3)$ nature of $\Phi$.
As always, $|\cdot|$ denotes the Euclidean norm, i.e.\
$|X|:=\sqrt{X_{11}X_{11}^*+X_{12}X_{12}^*+...+X_{33}X_{33}^*}$.
If $h$ is continuous and $2\pi$--periodic and defined all over
${\mathbb R}^d$ then, since $D'$ is dense in ${\mathbb R}^d$,
$|h(\phi)|>0$ for all $\phi$ and thus $\Lambda(\Phi(\cdot;\phi_0)) \subset
\Lambda(\Phi(\cdot;\phi))$ for all $\phi$.
Conversely if, for a given $\phi$,
$|h(\phi)|>0$ 
then, by exchanging the roles of $\phi$ and $\phi_0$, we obtain 
$|h(\phi_0)|>0$ so that
$\Lambda(\Phi(\cdot;\phi_0)) \supset
\Lambda(\Phi(\cdot;\phi))$ for all $\phi$.
Thus $\Lambda(\Phi(\cdot;\phi_0)) = 
\Lambda(\Phi(\cdot;\phi))$ for all $\phi$
and it remains to show that $h$ is defined and continuous on 
${\mathbb R}^d$. The $2\pi$--periodicity of $h$ then follows immediately. 
We first state the following lemma.

\vspace*{.15in}

\setcounter{lemma}{2}
\begin{lemma} 
~For fixed $\lambda$ and nonresonant $(1,\omega)$, let
\begin{eqnarray}
&& \Phi_{q,\lambda}(\phi):=\frac{1}{q}\int_0^q\; 
\Phi(\theta;\phi)\exp(-i\lambda\theta)d\theta 
\label{eq:9.006} 
\end{eqnarray}
converge uniformly for all $\phi$ in the set $D'$ of {\rm (\ref{eq:9.011})} 
as the positive integer $q\rightarrow \infty$.
Then $\Phi_{q,\lambda}$ converges uniformly on ${\mathbb R}^d$. In particular
$a(\Phi(\cdot;\phi),\lambda)$ exists for all $\phi$ in ${\mathbb R}^d$
and is continuous in $\phi$.
\end{lemma}
{\em Proof of Lemma 9.3:} The space $Y$ of bounded (w.r.t. Euclidean norm)
functions $g:{\mathbb R}^d\rightarrow{\mathbb C}^9$ is a complex normed space 
w.r.t.
the norm $\sup_{\phi\in {\mathbb R}^d} |g(\phi)|$ and obviously
$\Phi_{q,\lambda}$ is a sequence in $Y$. Because ${\mathbb C}^9$ is complete,
so is $Y$ (see for example \cite[Section 7.1]{Di60}). Thus to show that
$\Phi_{q,\lambda}$ converges uniformly on ${\mathbb R}^d$ it suffices to show
that it is a Cauchy sequence in $Y$, i.e.\
for all positive $\delta$ there is a positive integer $m$ such that 
for all integers with $j,k\geq m$ we have
\begin{eqnarray}
&&  
\sup_{\phi\in {\mathbb R}^d} |\Phi_{j,\lambda}(\phi)-\Phi_{k,\lambda}(\phi)|<
\delta  \; .
\label{eq:9.008} 
\end{eqnarray}
We also observe that
\begin{eqnarray}
&& \sup_{\phi\in {\mathbb R}^d} |\Phi_{j,\lambda}(\phi)
-\Phi_{k,\lambda}(\phi)| 
= \sup_{\phi\in D'} |\Phi_{j,\lambda}(\phi)
-\Phi_{k,\lambda}(\phi)|
\; ,
\label{eq:9.009} 
\end{eqnarray}
which follows from the continuity of
$|\Phi_{j,\lambda}(\phi)
-\Phi_{k,\lambda}(\phi)|$ in $\phi$ and from $D'$ being dense in 
${\mathbb R}^d$.
Equation (\ref{eq:9.009}) implies that (\ref{eq:9.008}) is equivalent to
the statement
\begin{eqnarray}
&& 
\sup_{\phi\in D'} |\Phi_{j,\lambda}(\phi)-\Phi_{k,\lambda}(\phi)|<
\delta \; .
\label{eq:9.010} 
\end{eqnarray}
Clearly (\ref{eq:9.010}) holds because, by assumption,  
$\Phi_{q,\lambda}(\phi)$  converges uniformly on $D'$. Thus
$\Phi_{q,\lambda}$ converges uniformly on ${\mathbb R}^d$, so that
$a(\Phi(\cdot;\phi),\lambda)$ exists for all $\phi$ in ${\mathbb R}^d$
and is continuous in $\phi$. 
\hspace*{16.2cm} $\Box$

\vspace*{.15in}

To complete the proof of Theorem 9.2 we now show that 
the conditions of Lemma 9.3 are fulfilled, i.e.\ that
$\Phi_{q,\lambda}$ converges uniformly on $D'$, for every $\lambda$.
Clearly we have
\begin{eqnarray}
&& a(\Phi(\cdot;\phi_0+2\pi N\omega),\lambda)=
\lim_{q\rightarrow\infty}\Phi_{q,\lambda}(\phi_0+2\pi N\omega)
\nonumber\\
&&=
\exp(i\lambda 2\pi N)a(\Phi(\cdot;\phi_0),\lambda)\Phi^T(2\pi N;\phi_0) 
\label{eq:9.007} 
\end{eqnarray}
and to show that the limit in (\ref{eq:9.007}) is uniform on $D'$ 
we estimate
\begin{eqnarray}
&&|\Phi_{q,\lambda}(\phi_0+2\pi N\omega + 2\pi M)
-\exp(i\lambda 2\pi N)a(\Phi(\cdot;\phi_0),\lambda)\Phi^T(2\pi N;\phi_0)| 
\nonumber \\
&& = |\Phi_{q,\lambda}(\phi_0+2\pi N\omega)
-\exp(i\lambda 2\pi N)a(\Phi(\cdot;\phi_0),\lambda)\Phi^T(2\pi N;\phi_0)| 
\nonumber\\
&&= | \frac{1}{q}\int_{2\pi N}^{2\pi N+q}\; 
\Phi(\theta;\phi_0)\Phi^T(2\pi N;\phi_0)
\exp(-i\lambda(\theta-2\pi N))d\theta
\nonumber\\
&&\qquad
-\exp(i\lambda 2\pi N)a(\Phi(\cdot;\phi_0),\lambda)\Phi^T(2\pi N;\phi_0)|
\nonumber\\
&&
\leq \sqrt{3} | \frac{1}{q}\int_{2\pi N}^{2\pi N+q}\; 
\Phi(\theta;\phi_0)
\exp(-i\lambda\theta)d\theta
- a(\Phi(\cdot;\phi_0),\lambda)|
\; .
\label{eq:9.13} 
\end{eqnarray}
By using the first equality of (\ref{eq:9.007}) with $N=0$ and by noting that
$\Phi(\cdot;\phi_0)$ is bounded we have, for all $N$,
\begin{eqnarray}
&& \lim_{q\rightarrow\infty}|
 \frac{1}{q}\int_{2\pi N}^{2\pi N+q}
\Phi(\theta;\phi_0)
\exp(-i\lambda\theta)d\theta
- a(\Phi(\cdot;\phi_0),\lambda)| = 0 \; .
\label{eq:9.15} 
\end{eqnarray}
Moreover because $\Phi(\cdot;\phi_0)
\exp(-i\lambda \, \cdot)$ is almost periodic it follows (see
\cite[Chapter 3]{Fink})
that the convergence in (\ref{eq:9.15}) is uniform on the domain 
$\mathbb{Z}$ of $N$. Hence
(\ref{eq:9.13}) implies
that $\Phi_{q,\lambda}$ converges uniformly on $D'$.
\hfill $\Box$

\vspace*{.15in}

Thus we have proved, under the conditions of Theorem 9.2, 
that  
$\Lambda(\Phi(\cdot;\phi_0))$ is independent of $\phi_0$.



\noindent{\bf Remark:}
\begin{itemize}
\item[(3)] Remark 2 shows that under the conditions of Theorem 9.1 the spin tune 
can indeed be discovered  from a spectral analysis of the spin 
flow for arbitrary $\phi_0$. In particular,
since the spin motion is quasiperiodic and the torus is well tuned, the spectrum has at most 
countably many elements and
$\Lambda(\Phi(\cdot;\phi_0))$ will consist of sharp ``lines'' which can then be ``measured''.
Moreover, off orbital resonance and under the conditions of Theorem 9.2, all 
$\Lambda(\Phi(\cdot;\phi_0))$ are equal.
However, if the torus is ill--tuned  but the spin motion is quasiperiodic (so that the spectrum 
$\Lambda(\Phi(\cdot;\phi_0))$ will
consist of sharp lines), we expect that the union of the spectra 
$\Lambda(\Phi(\cdot;\phi_0))$ over the torus will  contain
uncountably many elements. The models in  Remarks 13 and 14 in Section 6 provide
examples. Note that in the absence of 
quasiperiodicity or even almost--periodicity, as for example in the Moser-Siegel model of 
Section 8, there may be difficulties in even computing the spectrum.
In practice, the spectrum can be obtained by tracking three mutually orthogonal spins
along an orbit and storing their values at each of a very large number of 
turns before applying a discrete Fourier transform to the data (a
well-known way of doing numerical Fourier analysis can also be found in
\cite{Las}).
But since this  spectrum can be very dense, it can be difficult to identify 
the  spin tunes. Thus it would be useful
to begin with small amplitudes, i.e.\ tori with small $J$, and look for a 
$\omega_s$ close to that for the closed orbit, which can be calculated as in 
Section 3. The spin tune at higher amplitudes could then be identified by
continuing away from the closed orbit. We return to this theme in Section 10.
Finally, tracking a spin trajectory which is parallel to an ISF 
would give a spectrum without $\omega_s$.
\item[(4)]
Of course, there might be cases where the dependence of
$\Lambda(\Phi(\cdot;\phi_0))$ on $\phi_0$ leading to the ill--tuning
mentioned in Remark 3, is very weak so that the sharp lines are just
broadened.  Intuition suggests that in such cases, spin--orbit
resonance--like phenemena might still be expected. But, of course,
the definitive conditions under which such phenomena could occur,
will only become clear by careful analysis.
\end{itemize}
\section{Discussion and Conclusion}
\setcounter{theorem}{0}
\setcounter{equation}{0}
In the foregoing sections we have presented a thorough step by step 
account of the circumstances under which spin motion may be quasiperiodic
on integrable particle orbits and have thereby put previous studies 
of the concept of spin tune onto a  rigorous basis.  In particular,
we considered integrable orbits in 
${\cal Q}(1,\omega_1,...,\omega_d;d+1)$ and by 
introducing certain conditions (e.g.\ Diophantine conditions) and assuming 
the existence of an ISF we obtained conditions under which 
the spin motion is in ${\cal Q}(1,\omega_1,...,\omega_d,\omega_s;d+2)$
where $\omega_s$ is a spin tune.
We have also shown how, by introducing UPF{\small s},  the spin motion can 
be represented in terms of generalized Floquet forms.

The scenarios covered by our treatment and the relationships between them are 
summarized by the Venn diagram of Figure 1. 
\begin{figure}[htbp]
\begin{center}
\epsfig{figure=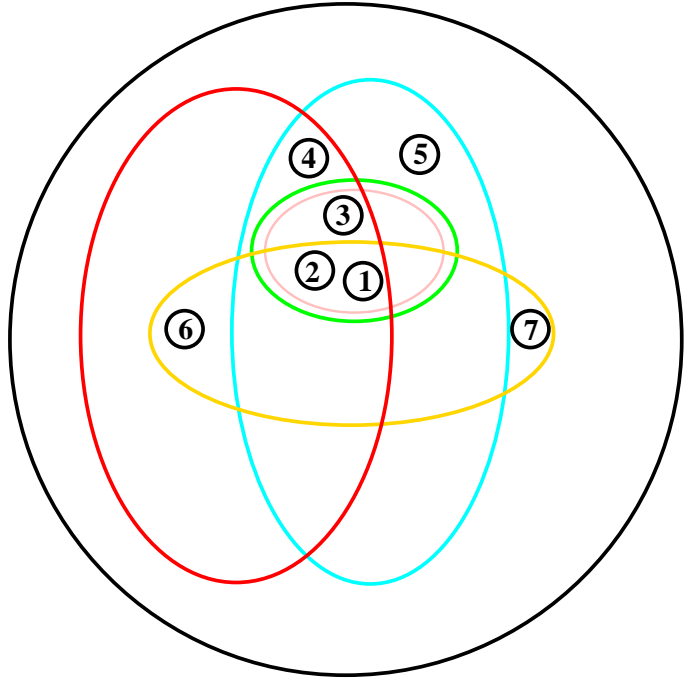,width=9.0cm,angle=-0}
\caption{\footnotesize{
The logical connections between the various scenarios.
}}
\end{center}
\end{figure}

The meanings of the domains in Figure 1 are as follows:
\begin{itemize}
\item
Inside the black circle: all tori,
i.e.\ for arbitrary real skew--symmetric $3\times 3$
matrices ${\cal A}(\cdot,\cdot,J_0)$ and arbitrary orbital tunes $\omega(J_0)$,
which are smooth and 
$2\pi$--periodic in $\theta$ and $\phi$.
\item
Inside the red ellipse: tori which have an ISF
\item
Inside the blue ellipse: tori, which at every $\phi_0$ have a proper UPF
(note that for those tori every spin trajectory is 
quasiperiodic)
\item
Inside the green ellipse: tori which are well--tuned 
(see Section 5)
\item
Inside the yellow ellipse: tori which are off orbital resonance
\item
Inside the pink ellipse:   tori with a uniform IFF
\end{itemize}
The numbered circles label specific examples, namely:
\begin{itemize}
\item
Example 1: the single resonance model
off orbital resonance (see Section 7)
\item
Example 2: the torus $J$ defined in Remark 13 in Section 6, 
           for the case of irrational $\omega (J)$
\item
Example 3: the single resonance model
on orbital resonance (see Section 7)
\item
Example 4: the torus defined in Remark 13 in Section 6, 
           for the case $\omega = 1$ 
\item
Example 5: a torus defined in Remark 14 in Section 6 
\item
Example 6: the Moser-Siegel model
(for certain choices of the parameters $\omega,\sigma_1, 
\sigma_2$) - see Section 8
\item 
Example 7: see below
\end{itemize}

Analytical solutions for spin motion or the ISF  for the 
${\cal A}$\,{\small s} arising in
real storage rings cannot be
obtained and it is not even known whether an ISF exists in general.
Nevertheless it seems that it usually exists for storage rings of
interest. 
This is supported by a large amount of numerical work in which, 
for the cases studied, it was possible to construct at least a very good 
approximation to an ISF 
\cite{hh96,epac98,spin98,hv2004,gh2000,mv2000,spin2000,hvb99}.
Note that in these simulations hard edge and some thin lens representations
of fields were used so that the ${\cal A}$\,{\small s} were not smooth. See 
Remark 16 
in Section 6. 

Those studies also included calculations of the spin tune using
SPRINT, either beginning with the 
pseudo--$u^1,u^2$--axes or  using the SODOM2 algorithm
whereby the spin tune is obtained by solving an eigenproblem for Fourier
coefficients in an SU(2) formalism. Numerical simulations with both methods 
suggest that
a spin tune normally exists if the torus is off orbital resonance.
This is the situation considered by Theorem 6.5d which implies that if we
have a ISF ${\cal S}$ and
if a $2\pi$--periodic unit vector exists which is nowhere parallel to ${\cal S}$,
then we almost always have a spin tune
(see Remark 5 in Section 6).
~For this and the other results in this paper it has been convenient to 
prescribe that the ${\cal A}$\,{\small s} are smooth.  But the demonstration 
of ISF{\small s} and spin tunes in 
\cite{hh96,epac98,spin98,hv2004,gh2000,mv2000,spin2000,hvb99}
indicates that the smoothness condition can often be relaxed.
For convenience, in the remainder of this section we will use the term ISF
in this spirit. We hope in the future to be able to present a 
treatment of the 
ISF and the spin tune in which the requirement of smoothness is relaxed.

An example of a situation where an  ISF might not exist even in 
crude   approximation, namely near 803.5 GeV in the HERA ring, 
is given in \cite{mv2000}. If the spin motion is nonquasiperiodic, this 
case  would  correspond to example 7 in Figure 1.

The spin tune is a crucial quantity for characterizing the stability of
spin motion. Spins behave somewhat like driven oscillators where
the driver is the magnetic and electric fields along particle orbits.
Near spin--orbit resonance, there is potential for marked qualitative changes in the spin 
motion which may then be quite erratic. The special 
nature of spin--orbit resonance is already clear from the fact that 
the  ISF need not be unique at resonance. See  Theorem 6.4.
Moreover, it is clear that it would make little sense to define a spin--orbit
resonance condition in terms of a UPR $\nu_s$ depending on $\phi_0$: 
such a $\nu_s$ would in general take different values for different
particles on a torus so that it would be impossible for the particles 
on a torus to be simultaneously on resonance and there would be no enhancement
of our ability to systematize spin motion. 
This was the reason
for insisting that a torus be well--tuned before considering
spin--orbit resonance. 

As mentioned in the Introduction and in Remark 5 of Section 5, it is 
clear that $\omega_s$ can vary with the beam energy and with $J$.
This is confirmed by simulations in SPRINT 
\cite{hh96,epac98,spin98,hv2004,gh2000,mv2000,spin2000,hvb99}, hence the name
``amplitude dependent spin tune''. 
To study the dependence of $\Xi_J$ on $J$ (recall Remark 5 of Section 5), 
it is necessary to choose a ``preferred'' member of $\Xi_J$. 
For the calculations with SPRINT the choice is made as follows. 

The spin tune $\nu_0 \in[0,1)$ and the corresponding UPF are
found on the closed orbit using the method outlined in Section 3.
Normally $\nu_0$ is set significantly different from zero 
in order to ensure that the direction of ${\bf n}_0$ is close to the  
``design'' direction in spite of the presence of the usual misalignments
of the ring \cite{mont98,br99,bmrr}.  
Then as in Remark 3 in Section 9, the preferred spin tune at nonzero
amplitudes is selected by requiring that it and the corresponding UPF
vary continuously with amplitude and reduce continuously to the spin
tune and UPF on the closed orbit. 

With this procedure in place, simulations in which the ISF and the
spin tune are calculated over a range of fixed amplitudes or energies
indeed show that the spin motion can become erratic near a spin--orbit
resonance.  Moreover, in such cases there is a tendency for the ISF to
become very sensitive to the parameter being varied.
With this prescription,  one also finds
that the strongest variations of the ISF occur near low order resonances
and that high order resonance effects are usually unimportant.
~For the details of these calculations and results, the reader is referred 
to  \cite{hh96,epac98,spin98,hv2004,gh2000,mv2000,spin2000,hvb99}.
These findings are consistent with perturbation theoretical 
calculations of the ISF as in \cite{mane87}
\footnote{Note that in \cite{mane87} the perturbation expansions calculate
${\bf n} - {\bf n}_0$. Then the $\omega_s$ for the closed 
orbit appears in the resonance factors, not the $\omega_s$ for nonzero orbital amplitudes.}.
The nonperturbative calculations also show that near spin--orbit resonance 
the spin tune tends to avoid exact fulfillment of the resonance condition.
As we have already indicated in Remark 3 of Section 7, the spin tune for 
the single resonance model
(see (\ref{eq:7.17})) also avoids the spin--orbit
resonance condition as $\sigma_1$ is varied through the 
condition $\sigma_1 = \omega$.  

In a ring without so--called Siberian Snakes (see below) the closed orbit 
spin tune $\nu_0$ usually varies with the beam energy \cite{mont98, chao81}.
It is then sometimes implied in the literature that in a beam with a large 
energy spread, the synchrotron motion causes the particles to oscillate  
to and fro across spin--orbit 
resonances as the spin tune $\omega_s$ oscillates. This crossing of 
resonances is then 
supposed to be the source of the low beam polarization that would be seen.
However, we have seen that a spin tune $\omega_s$ on a torus 
$(J_1, J_2, J_3)$ is a constant. As usual we assign $J_3$ to synchrotron 
motion. Thus $\omega_s$ does {\em not} oscillate and there is no resonance
crossing. Nevertheless, we do expect that a large energy spread
can lead to small beam polarization. This is explained by the fact  
that for particles of large enough
$J_3$, $\bf n$ varies strongly with the $\phi_3$. Then the maximum permissible
equilibrium beam polarization can indeed be small \cite{mont98}.

Studies of the effect on polarization of real resonance crossing
can be found in \cite{hv2004,gh2000,mv2000,spin2000}. The consequences for the
polarization of crossing first order resonances are usually quantified 
using the Froissart--Stora formula \cite{fs60}. 
But in \cite{hv2004,gh2000,mv2000,spin2000} it is shown that the
Froissart--Stora approach can be generalized to describe the change 
of the polarization at the crossing of higher order resonances too.
This is a further illustration of the value of using a wisely
defined spin tune for identifying resonances and for understanding
their properties

The spin tune can be obtained in SPRINT via the
pseudo--$u^1,u^2$--axes or by using the SODOM2 algorithm.
However, as we have seen in Section
9 the spin tune might also be obtained (``measured'') by a spectral analysis of the spin
motion.  This offers an attractive alternative. As in the case
of the other two methods the preferred spin tune would have to be
identified among the many spectral lines by matching onto the spin
tune of the closed orbit.
We have also seen in Section 9 that the ISF might be obtainable, if it exists,
by spectral analysis of spin motion. 
Spectral analysis may thus lead to a criterion for deciding whether an 
invariant spin field exists.
Other criteria for the existence of the ISF  are already available for
stroboscopic averaging and the SODOM2 algorithm.

We now complete this discussion by mentioning other quantities
that have been used in attempts to define a spin precession frequencies. 

As explained in Remark 3 in Section 3, for motion on the closed orbit the 
spin tune can
be obtained trivially from the complex eigenvalues of the 1--turn 
principal solution matrix.
This does not normally work off the closed orbit since 
$\Phi(2\pi;\phi_0) =U(2\pi) \exp \left({\cal J}2\pi
\nu_s(U) \right) U^T(0)$ 
and for this the $\exp(\pm 2\pi i\nu_s(U))$
are generally not the complex eigenvalues.
So the reader will agree that on synchro--betatron
orbits
the spin tune usually {\em cannot} be 
obtained from the complex eigenvalues of the 1--turn 
principal solution matrix.
In fact in general the real eigenvector of the general 1--turn 
principal solution matrix
$\Phi(\theta+2\pi;\phi_0)\Phi^T(\theta;\phi_0)$
starting at $\theta$ is not even a spin trajectory and is not
parallel to an ISF \cite[p.27, for example]{mont84}:  the term 
``spin closed orbit'' which is 
sometimes used for the ISF, is inappropriate. 
Moreover in the simulations with SPRINT the 
sensitivity of the ISF 
to variation of parameters shows no correlation with the spin precession 
rate extracted from the  complex eigenvalues of the 1--turn 
principal solution matrix.
Of course eigenvalues of 1--turn 
principal solution matrices
are easy to calculate but for 
$J \ne 0$ they usually have no useful function. 

The model described in  \cite[equation 21]{mane20031} provides an example.
It concerns orbits which are said in \cite{mane20032} to be exceptional. It is
stated there that exceptional orbits are characterized by the feature
that the spin tune depends on orbital phase and this is 
discussed in close conjunction with Stern--Gerlach forces. 
As we point out in the Introduction, S--G 
forces can have no practical relevance for understanding  spin resonance.
The model in \cite{mane20031} involves the 
single resonance model (see Section 7) and a single
thin lens Siberian Snake.  A Siberian
Snake \cite{dk76,dk78} of the kind used here is a magnet system that
rotates a spin by the angle $\pi$ around an axis in the plane of the
ring \cite[p.71]{mont84}.  
In the notation of Section 7 the parameters in \cite{mane20031} 
are $\sigma_1 = \omega$ and $\sigma_2 \sqrt{2 J} = 3/2$.
At all $\phi_0$ the 1--turn principal solution matrix
$\Phi(2\pi; \phi_0)$ starting at the snake 
represents a rotation around the vertical by an angle depending linearly on 
$\phi_0$. 
Then the ISF is vertical at the snake at all $\phi_0$ on this torus  
and for this zero--measure range of $J$,
the ISF at the snake {\em is} the same as the real unit--length eigenvector of 
the  1--turn principal solution matrix.
\footnote{Note that in this discussion we allow the ISF to be discontinuous in $\theta$.}
The orbital tune $\omega$ is arbitrary. 
The eigentunes extracted from the complex eigenvalues of 
$\Phi(2\pi; \phi_0)$, vary linearly with $\phi_0$.
Therefore if a particle is followed along an orbit, these eigentunes normally 
change abruptly between one turn and the next.
Other 1--turn eigentunes would be obtained if other positions around the ring
were chosen for $\theta = 0$.
Nevertheless, it is claimed in \cite{mane20032,mane20031} that the 
eigentune for $\Phi(2\pi; \phi_0)$ is $2 \pi\,\times\,$spin tune.
Obviously this is not a spin tune in the sense of our treatment. 
We have taken account of the fact that an eigentune in the SU(2)
formalism used in \cite{mane20032,mane20031} is one half of a corresponding 
eigentune in our formalism. 
Although this $\Phi(2\pi; \phi_0)$ and its eigentune are discussed in 
association with S--G forces, the inclusion of S--G forces is not
necessary to obtain either. In fact by exploiting techniques  additional to 
those used in this paper one can show that these parameters simply provide an example of 
ill--tuning and that this case would be entered next to example 6 in the diagram of Figure 1.
The inclusion of S--G efffects clouds the issue. 
It is not clear from \cite{mane20032,mane20031} whether these phase dependent  
eigentunes serve some useful function such as 
indicating the stability of spin motion.
We give another example of the use of the eigentune
of a 1--turn principal solution matrix
below. 

The situation with regard to the utility of eigenvectors and 
eigentunes is more subtle for tori on orbital
resonance where $\omega_1, \omega_2,...$ are rational. The orbit 
and $A$ are
then periodic over an appropriate number, $M$, of whole turns.  Thus 
in analogy with the method in Remark 6 in 
Section 3, there is a possibility of calculating ${\bf n}$ at $\theta = 0$ as
the unit--length real  eigenvector 
of the $M$--turn principal solution matrix
and in general it would be a 
function of $\phi_0$.  The imaginary part of the exponent of a complex 
eigenvalue of this principal solution matrix would
provide the $M$--turn advance of 
the phase of spin rotation around $\bf n$ and this could be used to obtain the 
average 1--turn  spin phase advance.   
This would usually  depend on $\phi_0$ and in such cases it  could
{\em not} be used to define a spin tune.  But it could  be used to define
a $\nu_{\cal V}$ (see (\ref{eq:6.109})).

So, as is usual at orbital resonance, an ISF can exist in general but
normally there is no spin tune. 
We note in passing that the $M$--turn complex eigentunes are $2\pi$--periodic in 
$\theta$ and $\phi$, just like  ${\cal A}$. Moreover, because eigenvalues
of matrices are invariant under similarity transformations, the eigentunes
are invariant when
the starting point for the eigenanalysis is shifted {\em along the orbit}.
Examples of an unwise use of the
term spin tune can be found in \cite{mane20021} where the dependence 
of the multi--turn eigentune on $\phi_0$ is made explicit. 
See \cite{mane20032,mane20031} too. 
Again, a ``tune'' depending on $\phi_0$ cannot be used for studying 
spin--orbit resonance. Calling such a quantity a spin tune can create
confusion. See Remark 4 in Section 9 also.

If, on orbital resonance with rational tunes, the torus is ill--tuned,
then one can expect that either there are more  or fewer  proper
UPR{\small s} than one would have on a well--tuned torus.  Thus a
spectral analysis along the lines of that in Section 9, applied to
every $\phi_0$ on the torus, could be a useful diagnostic tool to
signal these two cases of ill--tuning.  Note that in the examples in
Remarks 13 and 14 in Section 6 there are too many UPR{\small s}.
Therefore one might expect that on orbital resonance with rational
tunes an ill--tuned torus had too many proper UPR{\small s}.

Although an $M$--turn eigentune is in general $\phi_0$ dependent, we might  
expect that an approximation to the spin tune $\omega_s$ on a well--tuned 
torus off orbital resonance could be obtained by setting the orbital tunes
to rational values near to the actual tunes but such that $M$ were very large.
Indeed this is the essence of a popular perception. In effect, although usually
not clearly stated, the underlying hope is that for the smooth guide fields 
of real rings
and for large enough $M$, the average 1--turn spin phase  advance is only very
weakly dependent on $\phi_0$ and that therefore a good approximation 
to $\omega_s$ can be obtained.
That would be consistent with the heuristic expectation that for large 
enough $M$ the influence of the initial $\phi_0$ become diluted.
Such behavior might also be expected if the high order Fourier
coefficients in the $\tilde{c}_{\cal V}(\theta,\phi-\omega\theta)$ 
in (\ref{eq:6.6}) are very small owing to the smoothness of the fields.
See \cite[p.66]{bhr92} for a hint of how the Fourier coefficients
come in here. We hope in the future to be able to present
a rigorous treatment of this approximation.
This approach has been adopted in \cite{mane20021} to indicate  
that off orbital resonance the spin tune is a half integer 
at most $J$ in the model to be discussed in the next paragraph.
However, any attempt to find an approximate 
value for a spin tune by using $M$--turn maps for nearby rational tunes
should at least be checked for convergence and consistency.
See Remark 4 in Section 9 too.

~Further discussion around the topic of rational tunes can be found in 
\cite{bar2002}
in which the nature of the so--called ``snake resonances'' is studied. 
These refer to a large loss of beam polarization during acceleration in 
a model in which the spin motion in most of the ring is 
approximated by the single resonance model
and the spin motion is stabilized by 
pairs of idealized, i.e.\ thin lens, Siberian Snakes \cite{syl98}. 
The Siberian Snakes have the effect of 
 fixing $\nu_0$ at $1/2$ independently of the beam energy 
\cite[p.70]{mont84}.
Snake ``resonances'' occur at rational orbital tunes $\omega$ for which,
in the notation of (\ref{eq:5.00}), $1/2 = m_{0} + m_{1} ~\omega$ with
odd $m_1$. 
In \cite{bar2002} it is made evident that at these tunes and for most $J \ne 0$ 
there is  no amplitude dependent spin tune
so that one is not dealing with spin--orbit resonances. 
Moreover at most orbital amplitudes the ISF defined there (i.e.\ without
insistence  on smoothness) is irreducibly discontinuous at 
some orbital phases \cite{bar2002}. 
Of course, since rational tunes correspond
to orbital resonance, it should come as no surprise, given the content of
our paper, that there is no amplitude dependent spin tune in this case. 
We have explicitly chosen our nomenclature to be consistent with earlier
usage \cite{ky99} and thereby contribute clarity to the classification of phenomena.
It could be that the pathological behavior in this model, namely the large loss of 
polarization, is due to the use of  
the simplified but singular representation for the snake fields.
In this connection we note with interest that according to simulations for RHIC, the 
loss of polarization during acceleration is less severe when 
the simulations are carried out with the magnetic fields of real snakes 
than with the singular fields of thin lens snakes \cite{xk2003}.
This indicates that predictions from simplified, mathematically singular models should be 
treated with some caution. For recent experimental work around vertical orbital tunes 
corresponding to ``snake resonance'', see \cite{mbei2004}.
In any case the use of the term ``snake resonance''  is a good 
illustration  of the confusion that arises from an 
imprecise use of the concept of spin--orbit resonance. 

A  model involving the 
single resonance model and two thin snakes provides the second
example of the use of the eigentune from the 1--turn 
principal solution matrix.  Simulations
described in \cite[p.98--100]{syl98}
show some loss of polarization for all nonzero $2 \sigma_2^2 J$
during acceleration, even away from the rational orbital tunes
associated with snake ``resonances''.  It is implied  there
that this loss stems from the fact that during the
acceleration, the eigentune from the 1--turn principal solution matrix, 
which in \cite{syl98}
is called the ``perturbed spin tune'',
oscillates to and fro across a spin--orbit resonance as
the orbital phase advances from turn to turn.  But in this example the 1--turn
eigenvector of the principal solution matrix
is usually not even parallel to the ISF.  So it
is even more difficult to imagine that the ``perturbed spin tune'' can
characterize long term spin motion.  
Thus an alternative picture for the loss of polarization must be sought.

As mentioned in Remarks 1 and 4 in Section 7, for the straightforward 
single resonance model a 
spin tune {\em does} exist on orbital resonance.

This completes our discussion of notions of spin precession frequency.

We now conclude by summarizing the main message. This is, that, by just employing
the Lorentz force and the T-BMT equation and with the help of
the concept of quasiperiodicity, we are able to provide a rigorous and
broad treatment of the concepts of proper uniform precession rate,
spin tune and the invariant spin field by common methods used for
ordinary differential equations. 
This allows us to focus on the main 
phenomena without the distraction of perturbations such as noise, collective effects
and synchrotron
radiation. In principle they can be included by using perturbation theory.
We have discussed putative Stern--Gerlach forces in the Introduction.
The advantages of a clear, universally accepted and {\em useful} definition of spin tune 
has been made evident 
by the examples in the foregoing paragraphs.
\section*{Acknowledgements}
We thank Georg Hoffstaetter, Helmut Mais, Mathias Vogt, Kaoru Yokoya and the late Gerhard Ripken for 
important and fruitful discussions on this and related topics.
In addition, JAE gratefully acknowledges the 
support from DOE grant DE-FG03-99ER41104 and from
DESY during a sabbatical in 1997--1998 when the core elements of this work were first assembled
into a manuscript.

\renewcommand\refname


{\large References}
\begin{thebibliography}{99}
\renewcommand{\baselinestretch}{1}
\large
\normalsize
%
%
\bibitem{jackson}
J.D. Jackson, ``Classical Electrodynamics'', 3rd edition, Wiley (1998). 
%
%
\bibitem{bhr1}
D.P. Barber, K. Heinemann and G. Ripken, Z. f. Physik  {\bf C64}, 117  (1994).
%
%
\bibitem{eh2004}
J.A.Ellison and K. Heinemann, ``Periodic Spin Fields and Phase Space Densities:
 Stroboscopic Averaging and the Ergodic Theorem'', submitted for publication 
(2004).
%
%
\bibitem{gottfried}
K. Gottfried, ``Quantum Mechanics: Fundamentals'', Addison Wesley  (1989).
%
%
\bibitem{dbkh98}
K. Heinemann and D.P. Barber, Nucl. Instr. Meth. {\bf A463}, 62
and {\bf A469}, 294 (2001).
%
%
\bibitem{hh96}
K. Heinemann and G.H. Hoffstaetter, Phys. Rev. {\bf E 54} (4), 4240 
(1996).
%
%
\bibitem{mont98}
D.P. Barber et al., five  articles in
proceedings of ICFA workshop ``Quantum Aspects of
Beam Physics'', Monterey, U.S.A., 1998, edited by
P. Chen, World Scientific (1999). 
Also in extended form as DESY Technical Report 98-096 (1998) 
and e--print archive: physics/9901038, 9901041, 9901042, 9901043, 9901044.
%
%
\bibitem{epac98}
D.P. Barber, G.H. Hoffstaetter and M. Vogt, 
Proc. 1998 European Part. Acc. Conf. (EPAC98), Stockholm, Sweden, June 1998.
Available electronically at: http://epac.web.cern.ch/EPAC/Welcome.html.
%
%
\bibitem{spin98}
D.P. Barber, G.H. Hoffstaetter and M. Vogt, 
Proc. 13th Int. Symp. High Energy Spin Physics,
Protvino, Russia, September 1998, World Scientific (1999).
%
%
\bibitem{mane87}
S.R. Mane, Phys. Rev. {\bf A36}, 105 (1987).
%
%
\bibitem{hv2004}
G.H. Hoffstaetter and M. Vogt, Phys. Rev. {\bf E 70}, 056501 (2004).
%
%
\bibitem{gh2000}
G.H. Hoffstaetter, ``A modern view of high energy polarized proton beams''. 
To be published as a Springer Tract in Modern Physics.
%
%
\bibitem{mv2000}
M. Vogt, Ph.D. Thesis, University of Hamburg, DESY Technical Report DESY--THESIS--2000--054 
(2000).
%
%
\bibitem{chao81}
A.W. Chao, Nucl. Instr. Meth. {\bf 180}, 29 (1981).
Modern notation: replace $\bf n$ by ${\bf n}_0$.
%
%
\bibitem{br99}
D. P. Barber and G. Ripken, 
``Radiative Polarization, Computer Algorithms and 
Spin Matching in  Electron  Storage Rings'',
in the Handbook of Accelerator Physics  
and Engineering, Edited by
A.W. Chao and M. Tigner, World Scientific, 2nd edition (2002). 
%
%
\bibitem{bg98}
V.V.  Balandin and N.I. Golubeva, DESY Technical Report 98-16 (1998).
%
%
\bibitem{ey95}
Yu.~Eidelman and V.~Yakimenko,
Particle ~Accelerators {\bf 50}, 261 (1995).
%
%
\bibitem{ky99}
K. Yokoya, DESY Technical Report 99-006 (1999) and 
e--print archive: physics/9902068.
%
%
\bibitem{mane20022}
S.R. Mane, Nucl. Instr. Meth. {\bf A498}, 1  (2003).
%
%
\bibitem{dk72}
Ya.S. Derbenev and A.M. Kondratenko, Sov. Phys. JETP {\bf 35}, 230 (1972).
%
%
\bibitem{dk73}
Ya.S. Derbenev and A.M. Kondratenko, Sov. Phys. JETP {\bf 37}, 968 (1973).
%
%
\bibitem{ky86}
K. Yokoya, DESY Technical Report 86-57 (1986).
%
%
\bibitem{spin2000}
D.P. Barber, G.H. Hoffstaetter and M. Vogt, 
Proc. 14th Int. Spin Physics Symposium,
Osaka, Japan, October 2000, AIP proceedings 570, (2001).
%
%
\bibitem{bhr92}
D.P. Barber, K. Heinemann and G. Ripken, DESY Technical Report M-92-04 (1992),
second revised version, September 1999.
%
%
\bibitem{hvb99}
G.H. Hoffstaetter, M. Vogt and D.P. Barber, Phys. Rev. ST Accel. Beams
{\bf 2}, 114001 (1999).
%
%
\bibitem{si89}Y.G. Sinai (Ed.), Encycl. of math.sciences V.
Dynamical Systems II, Springer, New York (1989).
%
%
\bibitem{CFS}
I.P. Cornfeld, S.V. Fomin and Y.G. Sinai, ``Ergodic Theory'', Springer, 
New York (1982).
%
%
\bibitem{h96}
K. Heinemann, DESY Technical Report 96-229 (1996) and
e--print archive: physics/9611001.
%
%
\bibitem{khrip2000}
A.A. Pomeransky, R.A. Senkov and I.B. Khriplovich,
Phys. Usp. {\bf 43} (10), 1055 (2000). \\ See also \\
I.B. Khriplovich and A.A. Pomeransky, Surveys High Energy Phys. {\bf 14},
145 (1999)
and e--print archive gr-qc/9809069.
%
%
\bibitem{derb90}
Ya.S. Derbenev, University of Michigan--Ann Arbor, Technical Report UM--HE--90--30 (1990).
%
%
\bibitem{mane20032}
S.R. Mane, Nucl. Instr. Meth. {\bf A498}, 52  (2003).
%
%
\bibitem{mane20031}
S.R. Mane, 
Proc. 15th Int. 
Spin Physics Symposium, Brookhaven National Laboratory, 
Long Island, U.S.A., September 2002. AIP proceedings 675 (2003).
%
%
\bibitem{spin2_2000}
D.P. Barber, J.A. Ellison and K. Heinemann,
Proc. 14th Int. Spin Physics Symposium,
Osaka, Japan, October 2000, AIP proceedings 570, (2001).
%
%
\bibitem{Amann}
H. Amann, ``Ordinary Differential Equations: Introduction to Nonlinear 
Analysis'', de Gruyter, (1990). 
%
%
\bibitem{Hale}
J. K. Hale, ``Ordinary Differential Equations'', 2nd ed., Krieger,
Malabar, Florida (1980).
%
\bibitem{AA}
Y. Aharonov and J. Anandan, Phys. Rev. Letts. {\bf 58} (16), 1593 (1987).


%
\bibitem{Frank}
J. N. Franklin, ``Matrix Theory'', Prentice Hall, New Jersey (1968).
%
%
\bibitem{Gold}
H. Goldstein, ``Classical Mechanics'', Addison Wesley, 2nd ed.,New York (1982).
%
%
\bibitem{bmrr}
D.P. Barber  et al., DESY Technical Report 85-44 (1985).
Modern notation: replace $\bf n$ by ${\bf n}_0$.
%
%
\bibitem{Loch}
P. Lochak and C. Meunier, ``Multiphase Averaging for Classical Systems'', 
Springer, New York (1988).
%
%
\bibitem{Koe}
T.W. Koerner, ``Fourier Analysis'', Cambridge University Press, Cambridge 
(1988). 
%
%
\bibitem{Maa}
W. Maak, ``Fastperiodische Funktionen'', 2nd ed.,
Springer, Berlin (1967).
%
%
\bibitem{Arn}
V.I. Arnold, ``Mathematische Methoden der klassischen Mechanik'', 
Birkhaeuser, Basel (1988). 
%
%
\bibitem{Fink}
A. M. Fink, ``Almost Periodic Differential Equations'', Lecture Notes in
Math., Vol. 377, Springer (1974).
%
%
\bibitem{Dumas}
H. S. Dumas, J. A. Ellison and M. Vogt, SIAM J. App. Dynam. Syst. {\bf 3}, 409 (2004).
%
%
\bibitem{Di60}
J. Dieudonne, ``Foundations of Modern Analysis'', Academic Press, New York
(1960). 
%
%
\bibitem{Lang}
S. Lang, ``Real Analysis'', Addison Wesley, Reading
(1973). 
%
%
\bibitem{Yos}
T. Yoshizawa, ``Stability Theory and the Existence of Periodic 
Solutions and Almost Periodic Solutions'', 
Springer, New York (1975). 
%
%
\bibitem{mane92}
S.R. Mane, Nucl. Instr. Meth. {\bf A321}, 21  (1992). 
%
%
\bibitem{Grad}
I.S. Gradstein and I.M. Ryshik, ``Tables of Integrals, Series, and Products'',
Academic Press, New York (1965).
%
\bibitem{bar2002}
D.P. Barber et al., 
Proc. 15th Int. 
Spin Physics Symposium, Brookhaven National Laboratory, 
Long Island, U.S.A., September 2002. AIP proceedings 675 (2003).
%
%
\bibitem{Sieg}
C. L. Siegel and J. K. Moser, ``Lectures on Celestial Mechanics'', 
Springer, New York (1971).
%
%
\bibitem{Las}
J. Laskar, Physica {\bf D67}, 257 (1993).
%
%
\bibitem{fs60}
M. Froissart and R. Stora, Nucl. Instr. Meth. {\bf 7}, 297 (1960).
%
%
\bibitem{mont84}
B. Montague, Physics Reports {\bf 113}, 1 (1984).
%
%
\bibitem{dk76}
Ya.~S.~Derbenev and A.~Kondratenko, 
Soviet Physics Doklady {\bf 20}, 562 (1976).
%
%
\bibitem{dk78}                                              
Ya.~S.~Derbenev et al. Particle Accelerators {\bf 8}, 115 (1978).
%
%
\bibitem{mane20021}
S.R. Mane, Nucl. Instr. Meth. {\bf A480}, 328 (2002).
%
%
\bibitem{syl98}
S.Y. Lee, ``Spin Dynamics and Snakes in Synchrotrons'', World Scientific
(1997).
%
%
\bibitem{xk2003}
M.Xiao and T. Katayama, University of Tokyo Technical Report CNS-REP-51 (2003).
%
%
\bibitem{mbei2004}
M. Bei et al., Proc. 16th Int. 
Spin Physics Symposium, Trieste, Italy, October (2004). To be published by World Scientific.


\end{thebibliography}
\end{document}